\documentclass[twocolumn,showpacs,preprintnumbers,amsmath,amssymb,floatfix,aps,unsortedaddress]{revtex4}
\usepackage[dvips]{graphicx}

\begin{document}

\title{Theory of ultracold atomic Fermi gases}
\author{Stefano Giorgini}
\affiliation{Dipartimento di Fisica, Universit\`a di Trento and CNR-INFM BEC Center, I-38050 Povo, Trento, Italy}

\author{Lev P. Pitaevskii}
\affiliation{Dipartimento di Fisica, Universit\`a di Trento and CNR-INFM BEC Center, I-38050 Povo, Trento, Italy}
\affiliation{Kapitza Institute for Physical Problems, ul. Kosygina 2, 117334 Moscow, Russia}

\author{Sandro Stringari}
\affiliation{Dipartimento di Fisica, Universit\`a di Trento and CNR-INFM BEC Center, I-38050 Povo, Trento, Italy}

\begin{abstract}
The physics of quantum degenerate atomic Fermi gases in uniform as well as in harmonically trapped configurations is reviewed from a theoretical perspective. Emphasis is given to the effect of interactions which play a crucial role, bringing the gas into a superfluid phase at low temperature. In these dilute systems interactions are characterized by a single parameter, the $s$-wave scattering length, whose value can be tuned using an external magnetic field near a broad Feshbach resonance. The BCS limit of 
ordinary Fermi superfluidity, the Bose-Einstein condensation (BEC) of dimers and the unitary limit of large scattering length are important regimes exhibited by interacting Fermi gases. In particular the BEC and the unitary regimes are characterized by a high value of the superfluid critical temperature, of the order of the Fermi temperature. Different physical properties are discussed, including the density profiles and the energy of the ground-state configurations, the momentum distribution, the fraction of condensed pairs, collective oscillations and pair breaking effects, the expansion of the gas, the main thermodynamic properties, the 
behavior in the presence of optical lattices and the signatures of superfluidity, such as the existence of quantized vortices, the quenching of the moment of inertia and the consequences of spin polarization. Various theoretical approaches are considered, ranging from the mean-field description of the BCS-BEC crossover to non-perturbative methods based on quantum Monte Carlo techniques. A major goal of the review is to compare the theoretical predictions with the available experimental results.
\end{abstract}

\maketitle
\tableofcontents

\begin{figure*}
\scalebox{.75}{\includegraphics[width=1.0\textwidth]{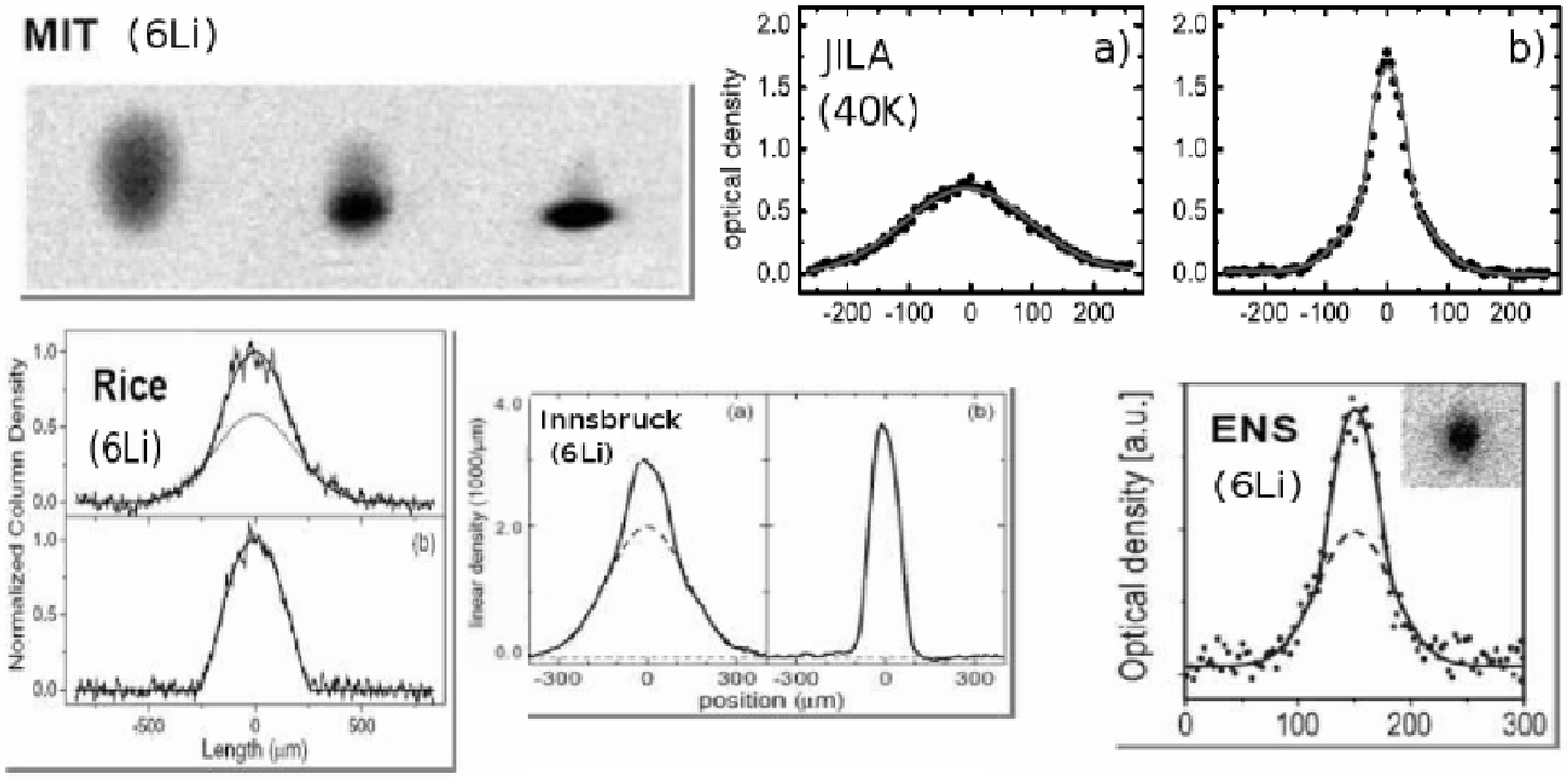}}
\caption{Gallery of molecular BEC experiments. Bimodal spatial distributions were observed for expanding gases at JILA (Greiner, Regal and Jin, 2003) with $^{40}$K, at MIT (Zwierlein {\it et al.}, 2003) and at ENS (Bourdel {\it et al.}, 2004) with $^6$Li. They were instead measured {\it in situ} at Innsbruck (Bartenstein {\it et al.}, 2004a) and at Rice University (Partridge {\it et al.}, 2005) with $^6$Li.}
\label{fig1.1}
\end{figure*}

\begin{figure}[b]
\begin{center}
\includegraphics[width=8.5cm]{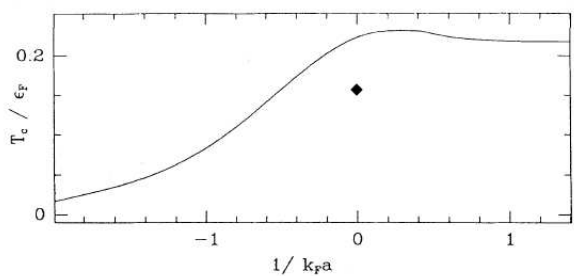}
\caption{Transition temperature in units of the Fermi energy $E_F$ as a function of the interaction strength along the BCS-BEC crossover, calculated using BCS mean-field theory (from S\'a de Melo, Randeria and Engelbrecht, 1993). The diamond corresponds to the theoretical prediction by Burovski {\it et al.} (2006a)  based on a Quantum Monte Carlo simulation at unitarity. } 
\label{fig1.2}
\end{center}
\end{figure}

\section{INTRODUCTION} \label{Chap1}

An impressive amount of experimental and theoretical activity  has characterized the last ten years of ultracold atom physics. The first realization of Bose-Einstein condensation (BEC) in dilute vapours of alkali atoms (Anderson {\it et al.}, 1995; Bradley {\it et al.}, 1995; Davis {\it et al.}, 1995) has in fact opened  new stimulating perspectives in this area of research. Most of the studies in the first years have been devoted to quantum gases of bosonic nature and were aimed to investigate the important consequences of  Bose-Einstein condensation which, before 1995, had remained an elusive and inaccessible phenomenon. 

Major achievements of these studies have been, among others, the investigation of superfluid features, including the hydrodynamic nature of the collective oscillations (Jin {\it et al.}, 1996; Mewes {\it et al.}, 1996), Josephson-like effects (Cataliotti {\it et al.}, 2001; Albiez {\it et al.}, 2005) and the realization of quantized vortices (Matthews {\it et al.}, 1999; Madison {\it et al.}, 2000; Abo-Shaeer {\it et al.}, 2001); the observation of interference of matter waves (Andrews {\it et al.}, 1997); the study of coherence phenomena in atom laser configurations (Mewes {\it et al.}, 1997; Anderson and Kasevich, 1998; Bloch, H\"ansch and Esslinger, 1999; Hagley {\it et al.}, 1999), the observation of four-wave mixing (Deng {\it et al.}, 1999) and of the Hanbury-Brown$-$Twiss effect (Schellekens {\it et al.}, 2005); the realization of spinor condensates (Stenger {\it et al.}, 1998); the propagation of solitons (Burger {\it et al.}, 1999; Denschlag {\it et al.}, 2000; Strecker {\it et al.}, 2002; Khaykovich {\it et al.}, 2002) and the observation of dispersive schock waves (Dutton {\it et al.}, 2001; Hoefer {\it et al.}, 2006); the transition to the Mott-insulator phase (Greiner {\it et al.}, 2002), the observation of interaction effects in the Bloch oscillations (Morsch {\it et al.}, 2001) and of dynamic instabilities in the presence of moving optical lattices (Fallani {\it et al.}, 2004); the realization of low dimensional configurations, including the 1D Tonks-Girardeau gas (Kinoshita, Wenger and Weiss, 2004; Paredes {\it et al.}, 2004) and the Berezinskii-Kosterlitz-Thouless phase transition in 2D configurations (Hadzibabic {\it et al.}, 2006; Schweikhard, Tung and Cornell, 2007).

On the theoretical side the first efforts were devoted to implement the Gross-Pitaevskii theory of weakly interacting Bose gases in the presence of the trapping conditions of experimental interest. This non-linear, mean-field theory  has proven capable to account for most of the relevant experimentally measured quantities in Bose-Einstein condensed gases like density profiles, collective oscillations, structure of vortices etc. The attention of theorists was later focused also on phenomena which can not be accounted for by the mean-field description like, for example, the role of correlations in low dimensional and in fast rotating configurations as well as in deep optical lattices (for general reviews on Bose-Einstein condensed gases see Inguscio, Stringari and Wieman, 1999; Dalfovo {\it et al.}, 1999; Leggett, 2001; Pethick and Smith, 2002; Pitaevskii and Stringari, 2003).

Very soon the attention of  experimentalists and theorists was also oriented towards the study of Fermi gases. The main motivations for studying fermionic systems are in many respects complementary to the bosonic case. Quantum statistics plays a major role at low temperature. Although the relevant temperature scale providing the onset of quantum degeneracy is the same  in both cases, being of the order of $k_BT_{deg}\sim \hbar^2 n^{2/3}/m$ where $n$ is the gas density and $m$ is the mass of the atoms, the physical consequences of quantum degeneracy are different. In the Bose case quantum statistical effects are associated with the occurrence of a phase transition to the Bose-Einstein condensed phase. Conversely, in a non-interacting Fermi gas the quantum degeneracy temperature only corresponds to a smooth crossover between a classical and a quantum behavior. Differently from the Bose case the occurrence of a superfluid phase in a Fermi gas can be only due to the presence of interactions. From the many-body point of view the study of Fermi superfluidity opens a different and richer class of questions which will be discussed in this review paper. Another important difference between Bose and Fermi gases concerns the collisional processes. In particular, in a single component Fermi gas $s$-wave scattering is inhibited due to the Pauli exclusion principle. This effect has dramatic consequences on the cooling mechanisms based on evaporation, where thermalization plays a crucial role. This has made the achievement of low temperatures in Fermi gases a difficult goal that was eventually realized with the use of sympathetic cooling techniques either employing two different spin components of the same Fermi gas or adding a Bose gas component as a refrigerant. 

First important achievements of quantum degeneracy in trapped Fermi gases were obtained by the group at JILA (De Marco and Jin, 1999). In these experiments temperatures of the order of fractions of the Fermi temperatures were reached by working with two spin components of $^{40}$K atoms interacting with negative scattering length. According to the Bardeen-Cooper-Schrieffer (BCS) theory this gas should exhibit superfluidity at sufficiently low temperature. However, due to the extreme diluteness of the gas, the critical temperature required to enter the superfluid phase was too small in these experiments. Quantum degeneracy effects were later observed in $^6$Li Fermi gases (Truscott et al., 2001; Schreck et al., 2001) using sympathetic cooling between $^6$Li and the bosonic $^7$Li isotope. Fermion cooling using different bosonic species has also proved very efficient for instance in the case of   $^{40}$K-$^{87}$Rb (Roati {\it et al.}, 2002) as well as $^6$Li-$^{23}$Na (Hadzibabic {\it et al.}, 2003).

It was soon realized that a crucial tool to achieve superfluidity is provided by the availability of Feshbach resonances. These resonances characterize the two body interaction and permit one to change the value and even the sign of the scattering length by simply tuning an external magnetic field. Feshbach resonances were first investigated in bosonic sytems (Courteille {\it et al.}, 1998; Inouye {\it et al.}, 1998). However, inelastic processes severely limit the possibility of tuning the interaction in Bose condensates (Stenger {\it et al.}, 1999). Strongly interacting regimes of fermionic atoms were achieved by O'Hara {\it et al.} (2002) and Bourdel {\it et al.} (2003) working at the resonance where the scattering length takes a divergent value. 
In this case three-body losses are inhibited by the Pauli exclusion principle, leading to a greater stability of the gas (Petrov, Salomon and Shlyapnikov, 2004). The resonant regime, also called unitary regime, is very peculiar since the gas is at the same time dilute (in the sense that the range of the interatomic potential is much smaller than the interparticle distance) and strongly interacting (in the sense that the scattering length is much larger than the interparticle distance). All the length scales associated with interactions disappear from the problem and the system is expected to exhibit a universal behavior, independent of the details of the interatomic potential. Bertsch (1999) and Baker (1999) first discussed the unitary regime as a model for neutron matter based on resonance effects in the neutron-neutron scattering amplitude (for a recent comparison between cold atoms and neutron matter see Gezerlis and Carlson, 2007). The critical temperature of the new system is much higher than in the BCS regime, being of the order of the quantum degeneracy temperature, which makes the realization of the superfluid phase much easier. Thanks to Feshbach resonances one can also tune the scattering length to positive and small values. Here, bound dimers composed of atoms of different spin are formed and consequently the system, that was originally a Fermi gas, is transformed into a bosonic gas of molecules. The possibility of tuning the scattering length across the resonance from negative to positive values, and vice-versa, provides a continous connection between the physics of Fermi superfluidity and Bose-Einstein condensation, including the unitary gas as an intermediate regime. 

On the BEC side of the Feshbach resonance molecules can be created either by directly cooling the gas at positive values of the scattering length $a$, or by first cooling the gas on the BCS side and then tuning the value of $a$ through the resonance. At low enough temperatures, Bose-Einstein condensation of pairs of atoms was observed through the typical bimodal distribution of the molecular profiles (see Fig.~\ref{fig1.1}) (Greiner, Regal and Jin, 2003; Jochim {\it et al.} 2003; Zwierlein {\it et al.}, 2003; Bourdel {\it et al.}, 2004; Partridge {\it et al.}, 2005). Condensation of pairs was later measured also on the fermionic side of the resonance (Regal, Greiner and Jin, 2004b; Zwierlein {\it et al.}, 2004). Other important experiments have investigated the surprisingly long lifetime of these interacting Fermi gases (Strecker, Partridge and Hulet, 2003; Cubizolles {\it et al.}, 2003), the release energy (Bourdel {\it et al.}, 2004) and the density profiles (Bartenstein {\it et al.}, 2004a) along the crossover.

Many relevant experiments have also focused on the dynamic behavior of these interacting systems, with the main motivation of exploiting their superfluid nature. The first observation of anisotropic expansion (O'Hara {\it et al.}, 2002) and the measurements of the collective oscillations (Bartenstein {\it et al.}, 2004b; Kinast {\it et al.}, 2004), although nicely confirming at low temperatures the predictions of the hydrodynamic theory of superfluids, can not be however considered a proof of superfluidity since a similar behavior is prediced also in the collisional regime of a normal gas above the critical temperature. The measurement of the pairing gap observed in radio-frequency excitation spectra (Chin {\it et al.}, 2004) was an important step toward the experimental evidence of superfluidity, even though it was not conclusive since pairing correlations are present also in the normal phase. A convincing proof of superfluidity was actually provided by the observation of quantized vortices which were realized on both sides of the Feshbach resonance (Zwierlein {\it et al.}, 2005b). 

\begin{table}
\centering
\caption{Ratio $T_c/T_F$ of the transition temperature to the Fermi temperature in various Fermi superfluids.}
\begin{tabular}{|l|c|} \hline
& $T_c/T_F$ \\ \hline
Conventional superconductors & 10$^{-5}$-10$^{-4}$ \\
Superfluid $^3$He & 10$^{-3}$ \\
High-temperature superconductors & 10$^{-2}$ \\
Fermi gases with resonant interactions & $\sim$0.2 \\ \hline
\end{tabular}
\label{Tab1}
\end{table} 

More recent experimental work (Zwierlein {\it et al.} 2006a; Partridge {\it et al.} 2006a and 2006b) has concerned the study of spin polarized configurations with an unequal number of atoms occupying two different spin states. In particular the Clogston-Chandrasekar limit, where the system looses superfluidity, has been experimentally identified at unitarity (Shin {\it et al.}, 2006). These configurations provide the unique possibility of observing the consequences of superfluidity through sudden changes in the shape of the cloud as one lowers the temperature, in analogy to the case of Bose-Einstein condensates (Zwierlein {\it et al.}, 2006b). Another rapidily growing direction of research is the study of Fermi gases in periodic potentials (Modugno {\it et al.}, 2003). First experimental results concern the effect of the periodic lattice on the binding energy of molecules across the Feshbach resonance (St\"oferle {\it et al.}, 2006) and some aspects of the superfluid to Mott insulator transition (Chin {\it et al.}, 2006). A motivation of the investigations in this field is the possibility of implementing an important model of condensed matter physics, the Hubbard hamiltonian, in analogy to the Bose-Hubbard counterpart already realized in bosonic systems (Greiner {\it et al.}, 2002). Fermi gases in periodic potentials are  of high interest also in the absence of interactions. For example they give rise to long-living Bloch oscillations that were observed in spin polarized Fermi gases (Roati {\it et al.}, 2004). 

On the theoretical side the availability of interacting Fermi gases with tunable scattering length has stimulated an impressive amount of work. Differently from the case of dilute Bose gases, where the Gross-Pitaevskii equation provides an accurate description of the many-body physics at low temperature and small densities, an analog theory for the Fermi gas along the BCS-BEC crossover is not available. The theoretical efforts started in the context of superconductors with the work by Eagles (1969), where it was pointed out that for large attraction between electrons the equations of BCS theory describe pairs of small size with a binding energy independent of density. A thorough discussion of the generalization of the BCS approach to describe the crossover in terms of the scattering length was presented in the seminal paper by Leggett (1980) (see also Leggett, 2006). This work concerned ground-state properties and was later extended to finite temperatures by Nozi\`eres and Schmitt-Rink (1985) and by S\'a de Melo, Randeria and Engelbrecht (1993) to calculate the critical temperature for the onset of superfluidity. These theories describe the properties of the many-body configurations along the BCS-BEC crossover in terms of a single parameter related to interactions, the dimensionless combination $k_Fa$ where $k_F$ is the Fermi wavevector. In Fig.~\ref{fig1.2} we report theoretical predictions for the critical temperature, showing that $T_c$ is of the order of the Fermi temperature in a wide interval of values of $k_F|a|$. For this reason one often speaks of high-$T_c$ Fermi superfluidity (see Table \ref{Tab1}). Furthermore, the results shown in Fig.~\ref{fig1.2} suggest that the transition between BCS and BEC is indeed a continous crossover. 

The first application of the concept of bound pairs to the case of Fermi gases with resonant interactions was proposed by Timmermans {\it et al.} (2001) and Holland {\it et al.} (2001). These extensions of the BCS mean-field theory  are however approximate and, even at zero temperature, the solution of the many-body problem along the crossover is still an open issue. Different methods have been developed to improve the description of the BCS-BEC crossover in uniform gases as well as in the presence of harmonic traps. These methods include the solution of the four-body problem to describe the interaction between molecules on the BEC side of the resonance (Petrov, Salomon and Shlyapnikov, 2004), applications of the BCS mean-field theory to trapped configurations with the help of the local density approximation, extensions of the mean-field approach using diagrammatic techniques (Pieri, Pisani and Strinati, 2004; Chen {\it et al.}, 2005; Haussmann {\it et al.}, 2006) and the development of theories based on the explicit inclusion of bosonic degreees of freedom in the Hamiltonian (Ohashi and Griffin, 2002; Bruun and Pethick, 2004; Romans and Stoof, 2006). At the same time more microscopic calculations based on quantum Monte Carlo (QMC) techniques have become available providing results on the equation of state at zero temperature (Carlson {\it et al.}, 2003; Astrakharchick {\it et al.}, 2004a; Juillet, 2007) and on the critical temperature for the superfluid transition (Bulgac, Drut and Magierski, 2006; Burovski {\it et al.}, 2006a; Akkineni, Ceperley and Trivedi, 2006). In addition to the above approaches, aimed to investigate the equilibrium properties of the system, a successful direction of research was devoted to the study of dynamic properties, like the expansion and the collective oscillations, by applying the hydrodynamic theory of superfluids to harmonically trapped Fermi gases (Menotti, Pedri and Stringari, 2002; Stringari, 2004), the pair-breaking excitations produced in resonant light scattering (T\"orm\"a and Zoller, 2000) and the dynamic structure factor (Minguzzi, Ferrari and Castin, 2001). A large number of theoretical papers has been recently devoted also to the study of spin polarization effects, with the aim of revealing the consequences of superfluidity on the density profiles and on the emergence of new superfluid phases.

Since the number of papers published on the subject of ultracold Fermi gases is very large we decided to limit the presentation only to some aspects of the problem which naturally reflect the main interests and motivations of the authors. In particular, we have tried to give special emphasis on the physical properties where an explicit comparison between theory and experiment is available, focusing on the effects of the interaction and on the manifestations of superfluidity exhibited by these novel trapped quantum systems. Most of the results presented in this review are relative to systems at zero temperature where the theoretical predictions are more systematic and the comparison with experiments is more reliable. A more complete review, covering all the interesting directions of theoretical research would require a much bigger effort, beyond the scope of the present paper. Some advanced topics related to the physics of ultracold Fermi gases are discussed, for example, in the Proceedings of the 2006 `Enrico Fermi' Varenna School (Ketterle, Inguscio and Salomon, 2007) and in the review article by Bloch, Dalibard and Zwerger (2007).

\section{IDEAL FERMI GAS IN HARMONIC TRAP} \label{Chap2}

\subsection{Fermi energy and thermodynamic functions} \label{Sec2.1}

The ideal Fermi gas represents a natural starting point for discussing the physics of dilute Fermi gases.
In many cases the role of interactions can in fact be neglected, as in the case of spin polarized gases 
where interactions are strongly suppressed at low temperature by the Pauli exclusion principle, or treated 
as a small perturbation.

The ideal Fermi gas in the harmonic potential
\begin{equation}
V_{ho}={1\over 2}m\omega_x^2x^2 + {1\over 2}m\omega_y^2y^2 + {1\over 2}m\omega_z^2z^2 
\label{Vho}
\end{equation} 
is a model system with many applications in different fields of physics, ranging from nuclear physics to the 
more recent studies of quantum dots. For this reason we will mainly focus on the most relevant features of 
the model, emphasizing the large $N$ behavior where many single-particle states are occupied and the semiclassical approach can be safely used. The simplest way to introduce the semiclassical description is to use a local density approximation for the Fermi distribution function of a given spin species:
\begin{equation}
f({\bf r},{\bf p}) = \frac{1 }{\exp[\beta \left( p^2/2m + V_{ho}({\bf r})-\mu \right)] +1} \;,  
\label{fF}
\end{equation}
where $\beta=1/k_BT$ and $\mu$ is the chemical potential fixed by the normalization condition
\begin{equation}
N_\sigma = \frac{1 }{(2\pi \hbar)^3}\int d{\bf r} d{\bf p} \;f({\bf r},{\bf p})= 
\int_0^{\infty} {g(\epsilon) d\epsilon \over \exp[\beta(\epsilon-\mu)] + 1} \;,
\label{NF}
\end{equation}
$N_\sigma$ being the number of atoms of the given spin species which is supposed to be sufficiently large. The semiclassical approach accounts for the Fermi pressure at low temperatures. In Eq.~(\ref{NF}) we have introduced the single-particle density of states $g(\epsilon)$ whose energy dependence is given by  $g(\epsilon) =  \epsilon^2/2(\hbar \omega_{ho})^3$ where $\omega_{ho}=(\omega_x\omega_y\omega_z)^{1/3}$ is the geometrical average of the three trapping frequencies. In terms of the density of states one can easily calculate the relevant thermodynamic functions. For example, the energy of the gas is given by the expression
\begin{equation}
E(T)= \int_0^\infty d\epsilon {\epsilon g(\epsilon) \over e^{\beta(\epsilon -\mu)}+1} \;.
\label{ET} 
\end{equation}
At $T=0$ the chemical potential $\mu$ coincides with the Fermi energy
\begin{equation}
E_F^{HO} \equiv k_BT_F^{HO}=(6N_\sigma)^{1/3}\hbar\omega_{ho} \;,
\label{EF}
\end{equation}
and the energy takes the value $E(0)=3/4E_F^{HO}N_\sigma$. Eq.~(\ref{EF}) fixes an important energy (and temperature) scale in the problem, analog to the   expression $E_F = (\hbar^2/2m)(6\pi^2 n_\sigma)^{2/3}$ of the uniform gas, where $n_\sigma$ is the density of a single spin component.

It is worth noticing that the Fermi energy (\ref{EF}) has the same dependence on the number of trapped atoms and on 
the oscillator frequency $\omega_{ho}$ as the critical temperature for Bose-Einstein condensation given by the well 
known formula $k_BT_\text{BEC}\simeq 0.94 \hbar \omega_{ho}N^{1/3}$. 

\begin{figure}[b]
\begin{center}
\includegraphics[width=8.5cm]{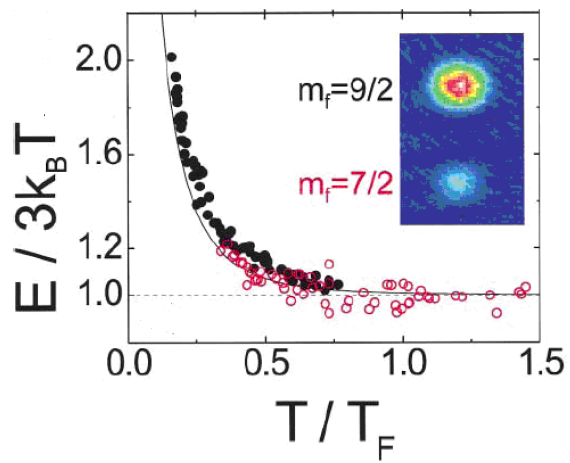}
\caption{Evidence for quantum degeneracy effects in trapped Fermi gases. The average energy per particle, 
extracted from absorption images, is shown for two-spin mixtures. In the quantum degenerate regime the data 
agree well with the ideal Fermi gas prediction (solid line). The horizontal dashed line corresponds to the 
result of a classical gas. From De Marco, Papp and Jin (2001).} 
\label{fig2.2}
\end{center}
\end{figure}

An important quantity to investigate is also the release energy $E_\text{rel}$ defined as the energy of the gas 
after a sudden switching off of the confining potential. The release energy is directly accessible in time-of-flight experiments 
and, as a consequence of the equipartition theorem applied to the ideal gas with harmonic confinement, is always 
equal to $E_\text{rel}=E/2$, where $E(T)$ is the total energy (\ref{ET}) of the gas.  At low $T$ the energy per particle deviates from the classical value $3k_BT$ due to quantum statistical effects as clearly demonstrated in the JILA experiment (De Marco and Jin, 1999; De Marco, Papp and Jin, 2001) reported in Fig.~\ref{fig2.2}.

\subsection{Density and momentum distributions} \label{Sec2.2}

The Fermi energy (\ref{EF}) can be used to define typical length and momentum scales characterizing the Fermi 
distribution in coordinate and momentum space respectively: 
\begin{equation}
R_i^0=\sqrt{2E_F^{HO}/m\omega_i^2} \;\;\;\;;\;\;\;\; p_F^0=\sqrt{2mE_F^{HO}} \;. 
\label{EFkF}
\end{equation}
The Thomas-Fermi radius ($i=x,y,z$)
\begin{equation}
R_i^0= a_{ho}\left(48 N_\sigma\right)^{1/6}\omega_{ho}  / \omega_i  \;,
\label{RFG}
\end{equation} 
gives the width of the density distribution at $T=0$, which can be calculated by integrating the distribution function in momentum space:
\begin{eqnarray}
n_\sigma({\bf r}) &=& \frac{8}{\pi ^{2}} \frac{N_\sigma}{R_x^0R_y^0R_z^0}
\label{n0r}\\
&\times& \left[1 - \left(\frac{x}{R_x^0}\right)^2-\left(\frac{y}{R_y^0}\right)^2-
\left(\frac{z}{R_z^0}\right)^2\right]^{3/2} \;.
\nonumber
\end{eqnarray}
In Eq.~(\ref{RFG}) $a_{ho}=\sqrt{\hbar/m\omega_{ho}}$ denotes the harmonic oscillator length. The Fermi wavevector 
\begin{equation}
k_F^0\equiv{p_F^0\over\hbar}={1\over a_{ho}}(48)^{1/6}N_\sigma^{1/6} \;,
\label{pFN}
\end{equation}
fixes instead the width of the momentum distribution
\begin{equation}
n_\sigma({\bf p})={\frac{8}{\pi ^{2}}}{\frac{N_\sigma}{(p_F^0)^{3}}}\left[ 
1-\left({\frac{p}{p_F^0}}\right)^2 \right]^{3/2} \;.  
\label{n0k}
\end{equation}
This result is obtained by integrating the $T=0$ distribution function in coordinate space. Equations (\ref{n0r}) and (\ref{n0k}), which are normalized to the total number of particles $N_\sigma$,  hold for positive values of 
their arguments and are often referred to as Thomas-Fermi distributions. Equation (\ref{n0k}) is the analogue of the most familiar momentum distribution $3N_\sigma/(4\pi p_{F}^{3})\Theta (1-p^{2}/p_{F}^{2})$ characterizing a uniform Fermi gas in trems of the Fermi momentum $p_F$. The broadening of the Fermi surface with respect to the uniform case is the consequence of the harmonic trapping. Notice that the value of $k_{F}^0$ defined above coincides with the Fermi wavevector $k_F^0=[6\pi ^{2}n_\sigma(0)]^{1/3}$
of a uniform gas with density $n_\sigma(0)$ calculated in the center of the trap. It is worth comparing 
Eqs.~(\ref{n0r}) and (\ref{n0k}) with the analogous results holding for a trapped Bose-Einstein condensed gas in 
the Thomas-Fermi limit (Dalfovo et al., 1999).  The shapes of Fermi and Bose profiles do not look 
very different in coordinate space. In both cases the radius of the atomic cloud increases with $N$ although 
the explicit dependence is slightly different ($N^{1/5}$ for bosons and $N^{1/6}$ for fermions). Notice however that 
the form of the density profiles has a deeply different physical origin in the two cases. For bosons it is fixed by
the repulsive two-body interactions, while in the Fermi case  is determined by the quantum pressure. 

In momentum space the Bose and Fermi distributions differ instead in a profound way. First, as a consequence of the 
semiclassical picture, the momentum distribution of the Fermi gas is isotropic even if the trapping potential is 
deformed [see Eq.~(\ref{n0k})]. This behavior differs from what happens in the BEC case where the momentum 
distribution is given by the square of the Fourier transform of the condensate wavefunction and is hence sensitive 
to the anisotropy of the confinement. Second, the width of the momentum distribution of a trapped Bose-Einstein condensed gas {\it decreases} by increasing $N$ while, according to Eqs.~(\ref{EFkF}) and (\ref{n0k}), the one of a trapped Fermi gas 
{\it increases} with the number of particles. 

It is finally useful to calculate the time evolution of the density profile after turning off the trapping potential. 
For a non-interacting gas the distribution function follows the ballistic law $f({\bf r},{\bf p},t)=f_{0}({\bf r}-{\bf p}t/m,{\bf p})$, where $f_{0}$ is the distribution function at $t=0$ given by (\ref{fF}). By integrating over ${\bf p}$ one can easily calculate the time evolution of the density and one finds the following result for the mean square radii 
\begin{equation}
\langle r_i^{2}\rangle ={\frac{E(T)}{N_\sigma}}{\frac{1}{3m\omega
_i^{2}}}(1+\omega _i^{2}t^{2}) \;.
 \label{r2}
\end{equation}
The asymptotic isotropy predicted by Eq.~(\ref{r2}) is the consequence of the absence of collisions during the 
expansion and reflects the isotropy of the momentum distribution (\ref{n0k}).

\section{TWO-BODY COLLISIONS} \label{Chap3}

\subsection{Scattering properties and binding energy} \label{Sec3.1}

Interaction effects in quantum degenerate, dilute Fermi gases can be accurately modeled by a small number of parameters characterizing the physics of two-body collisions. In the relevant regime of low temperature and large mean interparticle distance,
the spatial range $R_0$ of the interatomic potential is much smaller than both the thermal wavelength $\lambda_T=\sqrt{2\pi\hbar^2/mk_BT}$ and the inverse Fermi wavevector $k_F^{-1}$:
\begin{equation}
R_0 \ll \lambda_T \;\;\;\;\;\; R_0 \ll k_F^{-1} \;.
\label{condslow}
\end{equation}
Under the above conditions the main contribution to scattering comes from states with $\ell =0$ component of  angular momentum, i.e. $s$-wave states.   Another constraint comes from the antisymmetry of the wavefunction of identical fermions which excludes $s$-wave scattering between spin-polarized particles. As a consequence, only particles with different spin can interact.

In this Section we briefly recall some results of the theory of elastic scattering in the $s$-wave channel (see, {\it e.g.}, Landau and Lifshitz, 1987).

If one neglects small relativistic spin interactions, the problem of describing the collision process between two atoms reduces to the solution of the Schr\"{o}dinger equation for the relative motion. For positive energy $\epsilon$, the s-wave wavefunction in the asymptotic region $r\gg R_0$ can be written as $\psi_0(r)\propto\sin[kr+\delta_0(k)]/r$, where $r=|{\bf r}_{1}-{\bf r}_{2}|$ is the relative coordinate of the two atoms, $\delta_0(k)$ is the $s$-wave phase shift and $k=\sqrt{{2m_r\epsilon}}/\hbar$ is the wavevector of the scattering wave with $m_r$ the reduced mass of the pair of atoms ($m_r=m/2$ for identical atoms). The $s$-wave scattering amplitude, $f_0(k)=[{-k\cot\delta_0(k)+ik}]^{-1}$, does not depend on the scattering angle and when $k\rightarrow 0$ it tends to a constant value: $f_0(k\rightarrow 0 )=-a$. The quantity $a$ is the $s$-wave scattering length, which plays a crucial role in the scattering processes at low energy. By including terms to order $k^2$ in the expansion of the phase shift $\delta_0(k)$ at low momenta one obtains the result
\begin{equation}
f_0(k)=-\frac{1}{{a^{-1}-k^2R^\ast/2+ik}} \;,
\label{f0}
\end{equation}
defining the effective range $R^\ast$ of interactions. This length scale is usually of the same order of the range $R_0$, however in some cases, ${\it e.g.}$ close to a narrow Feshbach resonance (see Sec.~\ref{Sec3.2}), it can become much larger than $R_0$ providing a new relevant scale. In the limit $a\rightarrow\infty$, referred to as ``unitary limit'', the scattering amplitude (\ref{f0}) at wavevectors $k\ll 1/|R^\ast|$ obeys to the universal law $f_0(k)=i/k$, independent of the interaction.

For positive scattering lengths close to the resonance ($a\gg R_0$) shallow $s$-wave dimers of size $a$ exist whose binding energy $\epsilon_b$ does not depend on the short-range details of the potential and is given by
\begin{equation}
\epsilon_b=-\frac{\hbar^2}{2m_ra^2} \;.
\label{epsilonb}
\end{equation}
The binding energy of molecules of two fermionic $^{40}$K atoms formed near a Feshbach resonance was first measured by Regal {\it et al.} (2003a) using radio frequency spectroscopy.

In the many-body treatment of interactions it is convenient to use an effective potential $V_{eff}$ instead of the microscopic potential $V$. Different model potentials can be considered as the description of low-energy processes is independent of the details of $V(r)$. In many applications one introduces the regularized zero-range pseudo-potential defined as (Huang and Yang, 1957)
\begin{equation}
V_{eff}({\bf r})=g\delta({\bf r})\frac{\partial}{\partial r} r\;,
\label{pseudo}
\end{equation}
where the coupling constant $g$ is related to the scattering length by the relationship $g=2\pi\hbar^2a/m_r$. This potential has a range $R_0=0$ and results in the scattering amplitude $f(k)=-1/(a^{-1}+ik)$. For $a>0$, a bound state exists having the binding
energy (\ref{epsilonb}) and corresponding to the normalized wavefunction 
\begin{equation}
\psi_b(r)=e^{-r/a}/(\sqrt{2\pi a}r) \;.
\label{bstate}
\end{equation}
Notice that the differential operator $(\partial/\partial r) r$ in (\ref{pseudo}) eliminates the singular $1/r$ short-range behavior of the wavefunction. The use of the pseudopotential (\ref{pseudo}) in the Schr\"odinger equation is equivalent to the following contact boundary condition imposed on the wavefunction $\psi(r)$ (Bethe and Peierls, 1935) 
\begin{equation}
\left[ \frac{d(r\psi)/dr}{r\psi} \right]_{r=0}=-\frac{1}{a} \;.
\label{bcond}
\end{equation}

Another model potential which will be considered in this review in connection with quantum Monte-Carlo simulations is the attractive square-well potential defined by
\begin{equation}
V_{eff}(r)=\left\{  \begin{array}{cc}   -V_0  & (r<R_0)\;  \\
                                          0   & (r>R_0)\;.  
                    \end{array} \right.
\label{swell}
\end{equation}
The $s$-wave scattering parameters can be readily calculated in terms of the range $R_0$ and of the wavevector $K_0=\sqrt{2m_rV_0/\hbar^2}$. The scattering length is given by $a=R_0[1-\tan(K_0R_0)/(K_0R_0)]$ and the effective range by $R^\ast=R_0-R_0^3/3a^2-1/K_0^2a$.

\begin{figure}[b]
\begin{center}
\includegraphics[width=8.5cm]{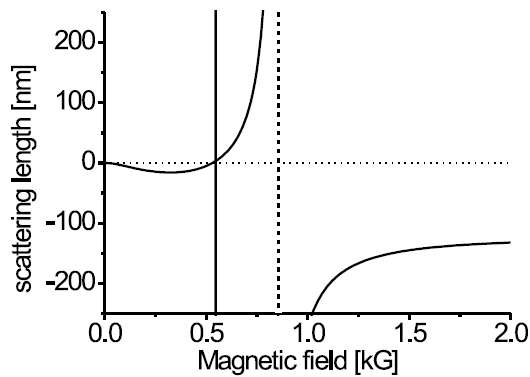}
\caption{Magnetic field dependence of the scattering length in $^6$Li, showing a broad Feshbach resonance at $B_0\simeq 834$ G and a narrow Feshbach resonance at $B_0\simeq 543$ G (can not be resolved on this scale). From Bourdel {\it et al.} (2003).}
\label{fig3.3}
\end{center}
\end{figure}

\subsection{Fano-Feshbach resonance} \label{Sec3.2}

The recent experimental achievements in the field of ultra-cold Fermi gases are based mainly on the possibility of tuning the scattering length $a$, in particular to values much larger than the mean interatomic distance, by changing an external magnetic field. This situation exists near the so called Fano-Feshbach resonances (Fano, 1961; Feshbach, 1962). These resonances take place when the energy associated with the scattering process between two particles (referred to as open channel) becomes close to the energy of a bound state of the pair in a different spin state (closed channel).

If the magnetic moments of the pairs of atoms in the two channels are different, one can go from a situation where the bound state in the closed channel is just below the threshold of the continuum spectrum in the open channel to a situation where the same bound state is just above threshold.

The transition between the two situations takes place at some value (denoted by $B_0$) of the magnetic field. In the absence of coupling, the existence of the bound state in the closed channel has no effect on the scattering in the open channel. However, in the presence of small coupling induced for example by exchange interactions, the scattering length will be large and positive if the state is below threshold and large and negative in the opposite case. As a function of the magnetic field $B$ the scattering length can be parametrized in the following form
\begin{equation}
a=a_{bg}\left( 1-\frac{\Delta_B}{B-B_0} \right) \;,
\label{Bdepend}
\end{equation}
where $\Delta_B$ is the width of the resonance and $a_{bg}$ is the background scattering length away from the resonance.

An important distinction concerns broad and narrow resonances, which in a Fermi gas involves the comparison of $k_F$ and the effective range of interactions $|R^\ast|$. Broad resonances correspond to  $k_F|R^\ast|\ll 1$. In this case, the effective range is irrelevant at the many-body level and the properties of the gas near the resonance can be described only in terms of $k_F|a|$ (Partridge {\it et al.}, 2005). On the contrary, for narrow resonances corresponding to $k_F|R^\ast|\gtrsim 1$, the effective range is negative and becomes a relevant scale of the problem (Bruun and Pethick, 2004; Bruun, 2004; De Palo {\it et al.}, 2004).

Most experiments on ultracold fermions make use of broad Feshbach resonances. This is certainly the case for the $^{40}$K resonance at $B_0\simeq 202$ G used in the experiments at JILA (Loftus {\it et al.}, 2002) and even more so for the extremely wide $^6$Li resonance at $B_0=834$ G used in the experiments at Duke (O'Hara {\it et al.}, 2002), Paris (Bourdel {\it et al.}, 2003), Innsbruck (Jochim {\it et al.}, 2003), MIT (Zwierlein {\it et al.}, 2003) and Rice (Partridge {\it et al.}, 2005). In both cases the value of $|R^\ast|$ close to the resonance is of the order of or smaller than a few nanometers and therefore $k_F|R^\ast|\lesssim 0.01$ for typical values of the density. Different is the case of the resonance in $^6$Li at $B_0\simeq 543$ G used at Rice (Strecker, Partridge and Hulet, 2003), where estimates give instead $k_F|R^\ast|\gtrsim 1$. In Fig.~\ref{fig3.3} we report the predicted behavior of the scattering length in $^6$Li as a function of the external magnetic field showing both the broad and the narrow resonance. Notice that for small values of the external magnetic field the scattering length approaches the value $a=0$, where $^6$Li atoms are expected to behave as a non-interacting gas.

\subsection{Interacting dimers} \label{Sec3.3}

The properties of shallow dimers formed near a Feshbach resonance are very important in the physics of ultracold
Fermi gases. Consisting of fermionic atoms, these dimers are bosonic molecules and interact with each other as well
as with single atoms.

The scattering between atoms and weakly-bound dimers was first investigated by Skorniakov and Ter-Martirosian (1956) in connection with neutron-deuteron scattering and, more recently, by Petrov (2003) in the context of degenerate Fermi gases (see also Brodsky {\it et al.}, 2006; Levinsen and Gurarie, 2006; Taylor, Griffin and Ohashi, 2007; Iskin and S\'a de Melo, 2007b). The solution of the three-body Schr\"odinger equation for a pair of like fermions interacting with a third particle can be obtained exactly using the contact boundary condition (\ref{bcond}) between particles with different spin. From the behavior of the scattering solution at large separation distance between the dimer and the free atom one can extract the atom-dimer scattering length, which is found to be
proportional to $a$:
\begin{equation}
a_{ad}\simeq 1.18a \;.
\label{aad}
\end{equation}
Fermi statistics plays here a crucial role since three-body bound states (Efimov states) are not permitted due to the Pauli principle.

Petrov, Salomon and Shlyapnikov (2004) have also solved the problem of the collisions between two dimers. By using again the zero-range approximation they calculated the dimer-dimer scattering length finding the value
\begin{equation}
a_{dd}\simeq 0.60a \;.
\label{add}
\end{equation}
The above result was later derived also using diagrammatic techniques by Brodsky et al. (2006) and by Levinsen and Gurarie
(2006). It is worth pointing out that by applying Born approximation one would find the results $a_{ad}=8a/3$ and $a_{dd}=2a$ (Pieri and Strinati, 2006).

The weakly-bound dimers formed near a Feshbach resonance are molecules in the highest roto-vibrational state. Due to collisions they can fall into deeper bound states of size on the order of the interaction range $R_0$. In this process a large energy of order $\hbar^2/mR_0^2$ is released and converted into kinetic energy of the colliding atoms, which then leave the system. In the case of atom-dimer collisions one can estimate the probability for the three atoms to approach each other within distances $\sim R_0$. This probability is suppressed by the Pauli principle, because two out of the three atoms have the same spin. A description of the relaxation process is provided by the equation: $\dot{n}_a=-\alpha_{ad}n_a n_d$, where $n_a$ and $n_d$ are respectively
the densities of atoms and dimers and $\dot{n}_a$ is the rate of atom losses. For the coefficent $\alpha_{ad}$ the following dependence on $a$ has been obtained (Petrov, Salomon and Shlyapnikov, 2004):
\begin{equation}
\alpha_{ad}\propto (\hbar R_{0}/m) (R_0/a)^{s} \;.
\label{alphaad}
\end{equation}
with $s=3.33$. In the case of relaxation processes caused by dimer-dimer collisions, the coefficient entering the dimer loss equation $\dot{n}_d=-\alpha_{dd}n_d^2$ satisfies the same law (\ref{alphaad}) with $s=2.55$. It is crucial that both $\alpha_{ad}$ and $\alpha_{dd}$ decrease with increasing $a$. This dependence ensures the stability of Fermi gases near a Feshbach resonance. It is the consequence of the fermionic nature of the atoms. In the case of bosons, instead, the relaxation time increases with increasing $a$ and the system becomes unstable approaching the resonance.

According to the above results,  the dimer-dimer relaxation rate should dominate over the atom-dimer one in the limit $R_0/a\to 0$. Experiments on atom losses both in potassium (Regal, Greiner and Jin, 2004a) and in lithium (Bourdel {\it et al.}, 2004) close to the Feshbach resonance give relaxation rate constants $\alpha_{dd}\propto a^{-s}$ with values of the exponent $s$ in reasonable agreement with theory. 

An interesting situation takes place in the case of heteronuclear dimers, consisting of fermionic atoms with different masses $m_1$ and $m_2$, where $m_1>m_2$ (see  Sec.~\ref{Sec9.4}). The theory (Petrov, 2003; Petrov, Salomon and Shlyapnikov, 2005) predicts that at mass ratio $m_1/m_2\simeq 12.3$ the exponent $s$ in the dimer-dimer relaxation rate $\alpha_{dd}\propto a^{-s}$ changes its sign, violating thus the stability condition of the gas near resonance. For mass ratios larger than 13.6, short-range physics dominates and the universal description in terms of the scattering length $a$ is lost.

\section{THE MANY-BODY PROBLEM AT EQUILIBRIUM: UNIFORM GAS} \label{Chap4}

\subsection{Hamiltonian and effective potential} \label{Sec4.1} 

The ideal gas model presented in Sec.~\ref{Chap2} provides a good description of a cold spin polarized 
Fermi gas. In this case interactions are in fact strongly inhibited by the Pauli exclusion principle. 
When atoms occupy different spin states interactions instead deeply affect the solution of the 
many-body problem. This is particularly true at very low temperature where, as we will discuss in this 
Section, attractive interactions give rise to pairing effects responsible for the superfluid behavior.

Let us consider a two-component system occupying two different spin states 
hereafter called, for simplicity, spin-up ($\sigma=\uparrow$) and spin-down ($\sigma=\downarrow$). 
We consider the grand canonical many-body Hamiltonian written in second quantization as
\begin{eqnarray}
\hat{H} &=& \sum_\sigma  \int d{\bf r}\;\hat{\Psi}_\sigma^\dagger({\bf r})\left(-{\hbar^2\nabla^2 
\over 2 m_\sigma}+V_{\sigma, ext}({\bf r})-\mu_{\sigma} \right)\hat{\Psi}_\sigma({\bf r}) \nonumber \\
&+&  \int d{\bf r} d{\bf r}^\prime V({\bf r}-{\bf r}^\prime)\hat{\Psi}_\uparrow^\dagger({\bf r})
\hat{\Psi}_\downarrow^\dagger({\bf r}^\prime)\hat{\Psi}_\downarrow({\bf r}^\prime)
\hat{\Psi}_\uparrow({\bf r}) \;,
\label{Hspin}
\end{eqnarray}
where the field operators obey the fermionic anticommutation relations 
$\{\hat{\Psi}_\sigma({\bf r}),\hat{\Psi}_{\sigma^\prime}^\dagger({\bf r}^\prime)\}=
\delta_{\sigma,\sigma^\prime}\delta({\bf r}-{\bf r}^\prime)$. The one-body potential $V_{\sigma,ext}$ 
and the two-body potential $V$ account for the external confinement and for the 
interaction between atoms of different spin, respectively. The number of atoms 
$N_{\sigma}=\int d{\bf r} <\hat{\Psi}_\sigma^\dagger({\bf r})\hat{\Psi}_\sigma({\bf r})>$, as well as 
the trapping potentials and the atomic masses of the two spin species can in general be different. In 
this Section we consider the uniform case ($V_{\uparrow, ext}=V_{\downarrow,ext}=0$) with 
$N_\uparrow=N_\downarrow=N/2$ and $m_\uparrow=m_\downarrow=m$. The densities of the two spin components 
are in this case uniform and the Fermi wavevector is related to the total density of the gas, $n=2n_\uparrow=2n_\downarrow$, through the expression $k_F=(3\pi^2n)^{1/3}$. The density $n$ determines 
the Fermi energy 
\begin{equation}
E_F=k_BT_F={\hbar^2 \over 2m}(3\pi^2 n)^{2/3}
\label{EF4}
\end{equation}
of the non-interacting gas. In the following we will always use the above definition of $E_F$. Notice that in the presence of interactions the above definition of $E_F$ differs from the chemical potential at $T=0$.

In our discussion we are interested in dilute gases where the range of the interatomic potential is much smaller than the interparticle distance. Furthermore, we assume that the temperature is sufficiently small so that only collisions in the $s$-wave channel are important. Under these conditions interaction effects are well described by the $s$-wave scattering length $a$ (see Sec.~\ref{Sec3.1}). In this regard one should recall that the gaseous phase corresponds to a metastable solution of the many-body problem, the true equilibrium state corresponding in general to a solid-phase configuration where the microscopic details of the  interactions are important. In Eq.~(\ref{Hspin}) we have ignored the interaction between atoms occupying the same spin state, which is expected to give rise only to minor corrections due to the quenching effect produced by the Pauli principle. 

As we have already discussed  in connection with two-body physics, a better theoretical understanding of 
the role played by the scattering length can be achieved by replacing the microscopic potential $V$ with 
an effective short-range potential $V_{eff}$. The regularized zero-range pseudo-potential has been 
introduced in Eq.~(\ref{pseudo}). Similarly to the two-body problem, the regularization accounted for by the term $(\partial / \partial r)r$ permits to cure the ultraviolet divergences in the solution of the Schr\"odinger equation that arise from the 
vanishing range of the pseudo-potential. In general, this regularization is crucial to solve the many-body 
problem beyond lowest order perturbation theory, as happens, for example, in the BCS theory of 
superfluidity (Bruun {\it et al.}, 1999). 

The effect of the zero-range pseudo-potential is accounted for by the boundary condition (\ref{bcond}) which, in the many-body problem, can be rewritten as   (Bethe and Peierls, 1935; Petrov, Salomon and Shlyapnikov, 2004)
\begin{equation}
\Psi(r_{ij}\to 0)\propto {1\over r_{ij}}-{1\over a} \;,
\label{BPbcon}
\end{equation}
where $r_{ij}=|{\bf r}_i-{\bf r}_j|$ is the distance between any pair of particles with different spin 
$(i,j)$, and the limit is taken for fixed positions of the remaining $N-2$ particles and of the center of 
mass of the pair $(i,j)$. For realistic potentials, the above short-distance behavior is expected to hold 
for length scales much larger than the effective range $|R^\ast|$ of the interaction and much smaller than the 
mean interparticle distance: $|R^\ast| \ll r_{ij}\ll k_F^{-1}$. This range of validity applies in general to 
atomic gas-like states both with repulsive ($a>0$) and attractive ($a<0$) interactions. If the many-body 
state consists instead of tight dimers of size $a\ll k_F^{-1}$ described by the wavefunction 
(\ref{bstate}) the boundary condition (\ref{BPbcon}) is valid in the reduced range: $|R^\ast|\ll r_{ij}\ll a$. 
In any case, at short distances, the physics of dilute systems is dominated by two-body effects. Under the conditions of diluteness and low temperature discussed above, the solution of the many-body problem with the full Hamiltonian (\ref{Hspin}) is completely equivalent to the solution of an effective problem where the Hamiltonian only contains the kinetic energy term and the many-body wavefunction satisfies the Bethe-Peierls boundary condition (\ref{BPbcon}).

An approach which will be frequently considered in this review is based on microscopic simulations with quantum 
Monte Carlo techniques. In this case the contact boundary conditions (\ref{BPbcon}) are difficult to implement and 
one must resort to a different effective interatomic potential. A convenient choice is the attractive 
square-well potential with range $R_0$ and depth $V_0$ defined in Eq.~(\ref{swell}) [other forms have also been 
considered in the literature (Carlson {\it et al.}, 2003)]. The interaction range $R_0$ must be taken much 
smaller than the inverse Fermi wavevector, $k_FR_0\ll 1$, in order to ensure that the many-body properties of 
the system do not depend on its value. The depth $V_0$ is instead varied so as to reproduce the actual value 
of the scattering length.   

The above approaches permit to describe the many-body features uniquely in terms of the scattering length $a$. These schemes are adequate if one can neglect the term in $k^2$ in the denominator of Eq.~(\ref{f0}). When the effective range $|R^\ast|$ of the interatomic potential becomes of the order of the inverse Fermi wavevector, as happens in the case of narrow Feshbach resonances, more complex effective potentials should be introduced in the solution of the many-body problem (see {\it e.g.} Gurarie and Radzihovsky, 2007).

\subsection{Order parameter, gap and speed of sound} \label{Sec4.2}

The phenomenon of superfluidity in 3D Fermi systems is associated with the occurrence of off-diagonal long-range 
order (ODLRO) according to the asymptotic behavior (Gorkov, 1958)   
\begin{equation}
\lim_{r\to\infty}\langle \hat{\Psi}_\uparrow^\dagger({\bf r}_2+{\bf r})
\hat{\Psi}_\downarrow^\dagger({\bf r}_1+{\bf r})\hat{\Psi}_\downarrow({\bf r}_1)
\hat{\Psi}_\uparrow({\bf r}_2)\rangle = |F({\bf r}_1,{\bf r}_2)|^2 \;, 
\label{n2}
\end{equation}
exhibited by the two-body density matrix. By assuming spontaneous breaking of gauge symmetry one can introduce the pairing field
\begin{equation}
F({\bf R},{\bf s}) = \langle \hat{\Psi}_\downarrow({\bf R}+{\bf s}/2)\hat{\Psi}_\uparrow({\bf R}-{\bf s}/2)
\rangle \;,
\label{F}
\end{equation}
[notice that ODLRO can be defined through Eq.~(\ref{n2}) also without using the symmetry breaking point of view (Yang, 1962)]. The vectors ${\bf R}=({\bf r}_1+{\bf r}_2)/2$ and ${\bf s}={\bf r}_1-{\bf r}_2$ denote, respectively, the center of mass and the relative 
coordinate of the pair of particles. In a Fermi superfluid ODLRO involves the 
expectation value of the product of two field operators instead of a single field operator as in the case of 
Bose-Einstein condensation. The pairing field in Eq.~(\ref{F}) refers to spin-singlet pairing and the spatial 
function $F$ must satisfy the even-parity symmetry requirement $F({\bf R},-{\bf s})=F({\bf R},{\bf s})$, imposed 
by the anticommutation rule of the field operators. 

The use  of the Bethe-Peierls boundary conditions (\ref{BPbcon}) for the many-body wavefunction $\Psi$ implies 
that the pairing field (\ref{F}) is proportional to $(1/s-1/a)$ for small values of the relative coordinate $s$. 
One can then write the following short-range expansion
\begin{equation}
F({\bf R},{\bf s}) = {m\over 4\pi\hbar^2}\Delta({\bf R})\left({1\over s}-{1\over a}\right) + o(s) \;,
\label{Fshortr}
\end{equation}
which defines the quantity $\Delta({\bf R})$, hereafter called the order parameter. The above asymptotic behavior 
holds in the same range of length scales as the Bethe-Peierls conditions (\ref{BPbcon}). For uniform systems the dependence on the center of mass coordinate ${\bf R}$ in Eqs.~(\ref{F})-(\ref{Fshortr}) drops and the order parameter $\Delta$ becomes constant. Moreover, in the case of $s$-wave pairing, the function (\ref{F}) becomes spherically symmetric: $F=F(s)$. The pairing field $F(s)$ can be interpreted as the wavefunction of the macroscopically occupied two-particle state. The condensate fraction of pairs is then defined according to  
\begin{equation}
n_\text{cond} = {1\over n/2} \int d{\bf s} \; |F(s)|^2 \;,
\label{n0}
\end{equation}
where $\int d{\bf s} |F(s)|^2$ is the density of condensed pairs. The 
quantity $n_\text{cond}$ is exponentially small for large, weakly bound Cooper pairs. It instead approaches the 
value $n_\text{cond}\simeq 1$ for small, tightly bound dimers that are almost fully Bose-Einstein condensed at 
$T=0$. 

Another peculiar feature characterizing superfluidity in a Fermi gas is the occurrence of a gap 
$\Delta_\text{gap}$ in the single-particle excitation spectrum. At $T=0$ this gap is related to the minimum 
energy required to add (remove) one particle starting from an unpolarized system according to the relationship 
$E(N/2\pm 1,N/2)=E(N/2,N/2)\pm\mu+\Delta_\text{gap}$. Here, $E(N_\uparrow,N_\downarrow)$ is the ground-state 
energy of the system with $N_{\uparrow (\downarrow)}$ particles of $\uparrow (\downarrow)$ spin and $\mu$ is 
the chemical potential defined by $\mu=\pm[E(N/2\pm 1,N/2\pm 1)-E(N/2,N/2)]/2=\partial E/\partial N$. By combining these two 
relations, one obtains the following expression for the gap (see, {\it e.g.}, Ring and Schuck, 1980)
\begin{eqnarray}
\Delta_\text{gap} &=& {1\over 2}\left[2E(N/2\pm 1,N/2) \right.
\label{Deltadef}\\
&-& \left. E(N/2\pm 1,N/2\pm 1) - E(N/2,N/2)\right] \;.
\nonumber
\end{eqnarray}
The gap corresponds to one half of the energy required to break a pair. The single-particle excitation spectrum, 
$\epsilon_k$, is instead defined according to the relation $E_k(N/2\pm 1,N/2)=E(N/2,N/2)\pm\mu+\epsilon_k$, 
where $E_k(N/2\pm 1,N/2)$ denotes the energy of the system with one more (less) particle with momentum  $\hbar k$. 
Since $\Delta_\text{gap}$ corresponds to the lowest of such energies $E_k$, it coincides with the minimum of 
the excitation spectrum.
In general the order parameter $\Delta$ and the gap $\Delta_\text{gap}$ are independent quantities. A direct 
relationship holds in the weakly attractive BCS regime (see Sec.~\ref{Sec4.4}) where one finds 
$\Delta_\text{gap}=\Delta$. The role of the gap in characterizing the superfluid behavior will be discussed 
in Sec.~\ref{Sec7.4}.  

Let us finally recall that a peculiar property of neutral Fermi superfluids is the occurrence of gapless density oscillations. These are the Goldstone sound modes associated to the gauge symmetry breaking and are often referred to as the Bogoliubov-Anderson modes (Bogoliubov, Tolmachev and Shirkov, 1958; Anderson, 1958). These modes are collective excitations and should not be confused with the gapped single-particle excitations discussed above. At small wavevectors they take the form of phonons propagating at $T=0$ with the velocity 
\begin{equation}
mc^2 = n  \partial \mu / \partial n
\label{soundvelocity}
\end{equation}
fixed by the compressibility of the gas. The description of these density oscillations will be presented in 
Sec.~\ref{Sec7.4}. The Bogoliubov-Anderson phonons are the only gapless excitations in the system and provide the 
main contribution to the temperature dependence of all thermodynamic functions at low temperature ($k_BT\ll\Delta_{\text gap}$). For example the 
specific heat $C$ and the entropy $S$ of the gas follow the free-phonon law $C=3S\propto T^3$.

\subsection{Repulsive gas} \label{Sec4.3}

There are important cases where the many-body problem for the interacting Fermi gas can be solved in an exact way. A first example is the dilute repulsive gas. Interactions are treated by means of the pseudo-potential (\ref{pseudo}) with a positive scattering length $a$. Standard perturbation theory can be applied with the small parameter $k_Fa\ll 1$ expressing the diluteness condition of the gas. At $T=0$, the expansion of the energy per particle up to quadratic terms in the dimensionless parameter $k_Fa$ yields the following expression (Huang and Yang, 1957; Lee and Yang, 1957)
\begin{equation}
\frac{E}{N}= \frac{3}{5}E_F \left(1+\frac{10}{9\pi}k_Fa +\frac{4(11-2\log2)}{21\pi^2}(k_Fa)^2 ...\right) \;,
\label{enexpansion}
\end{equation}
in terms of the Fermi energy (\ref{EF4}). The above result is universal as it holds for any interatomic potential with a sufficiently small effective range $|R^\ast|$ such that $n|R^{\ast}|^3\ll 1$. Higher order terms in (\ref{enexpansion}) will depend not only on the scattering length $a$, but also on the details of the potential (see Fetter and Walecka, 2003). In the case of purely repulsive potentials, such as the hard-sphere model, the expansion (\ref{enexpansion}) corresponds to the energy of the ``true'' ground state of the system (Lieb, Seiringer and Solovej, 2005; Seiringer, 2006). For more realistic potentials with an attractive tail, the above result describes instead the metastable gas-like state of repulsive atoms. This distinction is particularly important in the presence of bound states at the two-body level, since more stable many-body configurations satisfying the same condition $k_Fa\ll 1$   consist of a gas of dimers (see Sec.~\ref{Sec4.5}). The weakly repulsive gas remains normal down to extremely low temperatures when the repulsive potential produces $\ell>0$ pairing effects bringing the system into a superfluid phase (Kohn and Luttinger, 1965; Fay and Layzer, 1968; Kagan and Chubukov, 1988). In the normal phase the thermodynamic properties of the weakly repulsive gas are 
described with good accuracy by the ideal Fermi gas model.

\subsection{Weakly attractive gas} \label{Sec4.4}

A second important case   is the dilute Fermi gas interacting with negative scattering 
length ($k_F|a|\ll 1$).   In this limit the many-body problem can be solved  both at $T=0$ and at finite 
temperature and corresponds to the most famous BCS picture first introduced to describe the phenomenon of 
superconductivity (Bardeen, Cooper and Schrieffer, 1957). The main physical feature  is the 
instability of the Fermi sphere in the presence of even an extremely weak attraction and the formation of bound 
states, the Cooper pairs, with exponentially small binding energy. The many-body solution proceeds through a 
proper diagonalization of the Hamiltonian (\ref{Hspin}) by applying the Bogoliubov transformation to the Fermi 
field operators (Bogoliubov, 1958). This approach is non perturbative and predicts a second order phase transition associated with the occurrence of ODLRO. 

Exact results are available for the critical temperature and the superfluid gap (Gorkov and Melik-Barkhudarov, 1961). For the critical temperature the result is 
\begin{equation}
T_c=\left(\frac{2}{e}\right)^{7/3}\frac{\gamma}{\pi}T_F e^{\pi/2k_Fa}\approx 0.28 T_F e^{\pi/2k_Fa} \;,
\label{GMB}
\end{equation}
where $\gamma=e^C\simeq1.781$, $C$ being the Euler constant. The exponential, non analytical dependence of $T_c$ 
on the interaction strength $k_F|a|$ is typical of the BCS regime. With respect to the original treatment by Bardeen, Cooper and Schrieffer (1957),  the preexponential term in Eq.~(\ref{GMB}) is a factor $\sim$2 smaller as it accounts for the renormalization of the scattering length due to screening effects in the medium. A simple derivation of this result can be found in the book by Pethick and Smith (2002).

The spectrum of single-particle excitations close to the Fermi surface, $|k-k_F|\ll k_F$, is given by
\begin{equation}
\epsilon_k=\sqrt{\Delta_\text{gap}^2+[\hbar v_F(k-k_F)]^2} \;,
\label{elexc}
\end{equation}
where $v_F=\hbar k_F/m$ is the Fermi velocity, and is minimum at $k=k_F$.  The gap at $T=0$ is related to $T_c$ through the expression 
\begin{equation}
\Delta_\text{gap}=\frac{\pi }{\gamma}k_BT_c\approx 1.76 k_BT_c \;.
\label{Delta0}
\end{equation}
Furthermore, the ground-state energy per particle takes the form 
\begin{equation}
\frac{E}{N}=\frac{E_{normal}}{N}-\frac{3\Delta_{\text gap}^2}{8E_F} \;,
\label{E0sup}
\end{equation}
where $E_{normal}$ is the perturbation expansion (\ref{enexpansion}) with $a<0$ and the term proportional to $\Delta_{\text gap}^2$ corresponds to the exponentially small energy gain of the superfluid compared to the normal state. 

Since the transition temperature $T_c$ becomes exponentially small as one decreases the value of $k_F|a|$, the 
observability of superfluid phenomena is a difficult task in dilute gases. Actually, in the experimentally relevant 
case of harmonically trapped configurations the predicted value for the critical temperature easily becomes smaller 
than the typical values of the oscillator temperature $\hbar\omega_{ho}/k_B$.

The thermodynamic properties of the BCS gas can also be investigated. At the lowest temperatures 
$k_BT\ll\Delta_\text{gap}$ they are dominated by the Bogoliubov-Anderson phonons. However, 
already at temperatures $k_BT\sim\Delta_\text{gap}$ the main contribution to thermodynamics comes from fermionic 
excitations.  For more details 
on the thermodynamic behavior of a BCS gas see, for example, the book by Lifshitz and Pitaevskii (1980). At $T\gtrsim 
T_c$ the gas, due to the small value of $T_c$, is still strongly degenerate and the thermodynamic functions are well 
described by the ideal-gas model.

\subsection{Gas of composite bosons} \label{Sec4.5}

Thanks to the availability of Feshbach resonances it is possible to tune the value of the scattering length in 
a highly controlled way (see Sec.~\ref{Sec3.2}). For example, starting from a negative and small value of $a$ it is 
possible to increase $a$, reach the resonance where the scattering length diverges and explore the other 
side of the resonance where $a$ becomes positive and eventually small. One would naively expect to 
reach in this way the regime of the repulsive gas discussed in Sec.~\ref{Sec4.3}. This is not the case because in the 
presence of a Feshbach resonance the positive value of the scattering length is associated with the emergence of a 
bound state in the two-body problem and the formation of dimers as discussed in Sec.~\ref{Sec3.3}. The size of the 
dimers is of the order of the scattering length $a$ and their binding energy is $\epsilon_b\simeq-\hbar^2/ma^2$. 
These dimers have a bosonic nature, being composed of two fermions, and if the gas is sufficiently dilute and cold they 
consequently   give rise to the phenomenon of Bose-Einstein condensation. The size of the 
dimers can not however be too small, as it should remain large compared to the size of the deeply-bound energy levels 
of the molecule. This requires the condition $a\gg |R^\ast|$ which, according to the results of Sec.~\ref{Sec3.3}, 
ensures that the system of weakly-bound dimers is stable enough and that the transition to deeper molecular states, 
due to collisions between dimers, can be neglected.

The gas of dimers and the repulsive gas of atoms discussed in Sec.~\ref{Sec4.3} represent two different branches of the many-body problem, both corresponding to positive values of the scattering length (Pricoupenko and Castin, 2004). The atomic repulsive gas 
configuration has been experimentally achieved by ramping up adiabatically the value of the scattering length, starting from the value $a=0$ (Bourdel {\it et al.}, 2003). If one stays sufficiently away from the resonance, losses are not dramatic and the many-body state is a repulsive Fermi gas. Conversely, the gas of dimers is realized by crossing adiabatically the Feshbach resonance starting from negative values of $a$, which allows for a full conversion of pairs of atoms into molecules, or by cooling down a gas with a fixed (positive) value of the scattering length. 

The behavior of the dilute ($k_Fa\ll 1$) gas of dimers, hereafter called BEC limit, is properly described by the 
theory of Bose-Einstein condensed gases available both for uniform and harmonically trapped configurations (Dalfovo 
{\it et al.}, 1999). In particular, one can immediately evaluate the critical temperature $T_c$. In the uniform case 
this is given by the text-book relationship $T_c=(2\pi\hbar^2/k_Bm)(n_d/\zeta(3/2))^{2/3}$, where $n_d$ is the density 
of dimers (equal to the density of each spin species) and $\zeta(3/2)\simeq 2.612$. In terms of the Fermi temperature 
(\ref{EF4}) one can write 
\begin{equation}
T_c=0.218T_F \;,
\label{TCTFunif}
\end{equation}
showing that the superfluid transition, associated with the Bose-Einstein condensation of dimers, takes place at temperatures of the order of $T_F$, i.e. at temperatures much higher than the exponentially small value (\ref{GMB}) characterizing the BCS regime. The chemical potential of dimers, $\mu_d$, is defined through the relationship $2\mu=\epsilon_b+\mu_d$ involving the molecular binding energy $\epsilon_b$ and the atomic chemical potential $\mu$. 

The inclusion of interactions between molecules, fixed by the dimer-dimer scattering lenght $a_{dd}$ according to 
the relationship $a_{dd}=0.60a$ [see Eq.~(\ref{add})], is provided to lowest order in the gas parameter $n_da_{dd}^3$ by the Bogoliubov theory for bosons with mass $2m$ and density $n_d=n_\sigma$. At $T=0$, the bosonic chemical potential is given by 
$\mu_d =2\pi\hbar^{2}a_{dd}n_d/m$. Higher order corrections to the equation of state are provided by the Lee-Huang-Yang expansion (Lee, Huang and Yang, 1957) 
\begin{equation}
\frac{E}{N}=\frac{\epsilon_b}{2}+\frac{k_Fa_{dd}}{6\pi}\left[1+\frac{128}{15\sqrt{6\pi^3}}(k_Fa_{dd})^{3/2}\right]
E_F\;,
\label{enerbeyond}
\end{equation}
here expressed in units of the Fermi energy (\ref{EF4}). The validity of the expression (\ref{enerbeyond}) for a Fermi gas interacting with small and positive scattering lengths, was proven by Leyronas and Combescot (2007). At very low temperatures the thermodynamics of the gas can be calculated using the Bogoliubov gapless spectrum $\epsilon_d(k)$ of density excitations
\begin{equation}
\epsilon_d(k)=\hbar k\left(c_B^{2}+\hbar^2k^2/16m^2\right)^{1/2} \;,
\label{bogexc}
\end{equation}
where $c_B=\sqrt{\pi\hbar^2a_{dd}n_d/m^2}$ is the speed of Bogoliubov sound. The single-particle excitation spectrum is instead gapped and has a minimum at $k=0$: $\epsilon_{k\to0} = \Delta_\text{gap} +k^2/2m$ where 
\begin{equation}
\Delta_\text{gap}=\frac{|\epsilon_b|}{2}+(3a_{ad}-a_{dd})\frac{\pi\hbar^2n_d}{m} \;.
\label{gapBEC}
\end{equation}
In the above equation, the large binding-energy term $|\epsilon_b|/2=\hbar^2/2ma^2$ is corrected by a term which depends on both the dimer-dimer ($a_{dd}$) and the atom-dimer ($a_{ad}$) scattering length. This term can be derived from the definition (\ref{Deltadef}) using for $E(N_\uparrow,N_\downarrow)$ an energy functional where the interactions between unpaired particles and dimers are properly treated at the mean-field level, the coupling constant being fixed by $a_{ad}$ and by the atom-dimer reduced mass. Since $a_{ad}=1.18a$ [see Eq.~(\ref{aad})], the most important contribution to Eq.~(\ref{gapBEC}) comes from the term proportional to the atom-dimer scattering length. 

In the BEC limit the internal structure of dimers can be ignored for temperatures higher than the critical temperature since in this regime one has 
\begin{equation}
|\epsilon_b|/k_BT_c\sim 1/(k_Fa)^2\gg 1 \;.
\label{bindingtoTc}
\end{equation}
The above condition ensures that at thermodynamic equilibrium the number of free atoms is negligible, being proportional to $e^{\epsilon_b/2k_BT}$.

\subsection{Gas at unitarity} \label{Sec4.6}

A more difficult problem concerns the behavior of the many-body system when $k_F|a|\gtrsim 1$, i.e. when the scattering length becomes larger than the interparticle distance, which in turns is much larger than the range of the interatomic potential. This corresponds to the unusual situation of a gas which is dilute and strongly interacting at the same time. In this condition, it is not obvious whether the system is stable or collapses. Moreover, if the gas remains stable, does it exhibit superfluidity as in the BCS and BEC regimes? Since at present an exact solution of the many-body problem for $k_F|a|\gtrsim1$ is not available, one 
has to resort to approximate schemes or numerical simulations (see Secs.\ref{Sec5.1}-\ref{Sec5.2}). These approaches, together with experimental results, give a clear indication that the gas is indeed stable and that it is superfluid below a critical temperature. The limit $k_F|a|\to\infty$ is called the unitary regime and has been already introduced in Sec.~\ref{Sec3.1} when discussing two-body collisions. This regime is characterized by the universal behavior of the scattering amplitude $f_0(k)=i/k$ which bears important consequences at the many-body level. As the scattering length drops out of the problem, the only relevant lenght scales remain the inverse of the Fermi wavevector and the thermal wavelength. All thermodynamic quantities should therefore be universal functions of the Fermi energy $E_F$ and of the ratio $T/T_F$.

An important example of this universal behavior is provided by the $T=0$ value of the chemical potential:
\begin{equation}
\mu = (1+\beta)E_F \;,
\label{muunitarity}
\end{equation}
where $\beta$ is a dimensionless parameter. This relation fixes the density dependence of the equation of state, with non-trivial consequences on the density profiles and on the collective frequencies of harmonically trapped superfluids, as we will discuss in Secs.~\ref{Chap6}-\ref{Chap7}. The value of $\beta$ in Eq.~(\ref{muunitarity}) has been calculated using fixed-node quantum Monte-Carlo techniques giving the result $\beta=-0.58\pm0.01$ (Carlson {\it et al.}, 2003; Astrakharchik {\it et al.}, 2004a; Carlson and Reddy, 2005;). The most recent experimental determinations are in good agreement with this value (see Table \ref{Tab2} in Sec.~\ref{Chap6}). The negative value of $\beta$ implies that at unitarity interactions are attractive. By integrating Eq.~(\ref{muunitarity}) one finds that the same proportionality coefficient $(1+\beta)$ also relates the ground-state energy per particle $E/N$ and the pressure $P$ to the corresponding ideal gas values: $E/N=(1+\beta)3E_F/5$ and $P=(1+\beta)2nE_F/5$, respectively. As a consequence the speed of sound (\ref{soundvelocity}) is given by    
\begin{equation}
c=(1+\beta)^{1/2} v_F/\sqrt{3} \;,  
\label{soundunitarity}
\end{equation}
where $v_F/\sqrt{3}$ is the  ideal Fermi gas value. The superfluid gap at $T=0$ should also scale with the Fermi energy. Fixed-node quantum Monte Carlo simulations yield the result $\Delta_\text{gap}=(0.50\pm0.03)E_F$ (Carlson {\it et al.}, 2003; Carlson and Reddy, 2005). A more recent QMC study based on lattice calculations (Juillet, 2007) gives for $\beta$ a result consistent with the one reported above and a slightly smaller value for $\Delta_\text{gap}$ (see also Carlson and Reddy, 2007).

At finite temperature the most relevant problem, both theoretically and experimentally, is the determination of the transition temperature $T_c$ which is expected to depend on density through the Fermi temperature
\begin{equation}
T_c = \alpha T_F \;,
\label{Tcunitarity}
\end{equation}
$\alpha$ being a dimensionless universal parameter. Quantum Monte-Carlo methods have been recently used to determine the value of $\alpha$. Bulgac, Drut and Magierski (2006) and Burovski {\it et al.} (2006a) carried out simulations of fermions on a lattice where the sign problem, typical of fermionic quantum Monte-Carlo methods, can be avoided. These in principle ``exact'' studies require an extrapolation to zero filling factor in order to simulate correctly the continuum system and the reported value of the critical temperature corresponds to $\alpha=0.157\pm0.007$ (Burovski {\it et al.}, 2006a). Path integral Monte-Carlo simulations, which work directly in the continuum, have also been performed by Akkinen, Ceperley and Trivedi (2006) using the restricted path approximation to overcome the sign problem. The reported value $\alpha\simeq0.25$ is significantly higher compared to the previous method. 

Since at unitarity the gas is strongly correlated, one expects a significantly large critical region near $T_c$. 
Furthermore, the phase transition should belong to the same universality class, corresponding to a complex order 
parameter, as the one in bosonic liquid $^4$He and should exhibit similar features including the characteristic 
$\lambda$ singularity of the specific heat.  

The temperature dependence of the thermodynamic quantities is expected to involve universal functions of the ratio
$T/T_F$ (Ho, 2004). For example, the pressure of the gas can be written as $P(n,T)=P_{T=0}(n)f_P(T/T_F)$
where $P_{T=0}$ is the pressure at $T=0$ and $f_P$ is a dimensionless function. Analogously the entropy per atom 
takes the form $S/Nk_{B}=f_S(T/T_F)$
involving the universal function $f_S$ related to $f_P$ by the thermodynamic relation $df_P(x)/dx=xdf_S(x)/dx$. 
The above results for pressure and entropy imply, in particular, that during adiabatic changes the ratio $T/T_F$ 
remains constant. This implies that the adiabatic processes, at unitarity,  
follow the law $Pn^{-5/3}=\text{const}$, typical of non-interacting  atomic gases.  

The scaling laws for the pressure and the entropy also hold at high temperatures, $T\gg T_F$, 
provided that the thermal wavelength is still large compared to the effective range of interaction, 
$\hbar\sqrt{m/k_BT}\gg |R^\ast|$. In this regime of temperatures the unitary gas can be described, to first 
approximation, by an ideal Maxwell-Boltzmann gas. Corrections to the equation of state can be determined by 
calculating the second virial coefficient $B(T)$ defined from the expansion of the pressure 
$P\simeq nk_BT[1+nB(T)]$. Using the method of partial waves (Beth and Uhlenbeck, 1937; Landau and Lifshitz, 1980) 
and accounting for the unitary contribution of the $s$-wave phase shift $\delta_0=\mp\pi/2$ when $a\to\pm\infty$, 
one obtains the result
\begin{equation}
B(T)=-\frac{3}{4}\left(\frac{\pi\hbar^2}{mk_{B}T}\right)^{3/2} \;.  
\label{Bvirial}
\end{equation}
It is worth noticing that the negative sign of the second virial coefficient corresponds again to attraction. Equation (\ref{Bvirial}) takes into account  effects of both Fermi statistics and interaction. The 
pure statistical contribution would be given by the same expression in brackets, but with the coefficient $+1/4$
of opposite sign.

\section{THE BCS-BEC CROSSOVER} \label{Chap5}

\subsection{Mean-field approach at $T=0$} \label{Sec5.1}

As discussed in Sec.~\ref{Sec4.6}, there is not at present an exact analytic solution of the many-body 
problem along the whole BCS-BEC crossover. A useful approximation is provided by the standard BCS mean-field 
theory of superconductivity. This approach was first introduced by Eagles (1969) and Leggett (1980) with the main motivation to explore the properties of superconductivity and superfluidity beyond the weak-coupling limit $k_F|a|\ll 1$. The main merit of this approach is that it provides a comprehensive, although approximate, description of the equation of state along the whole crossover, 
including the limit $1/k_Fa \to 0$ and the BEC regime of small and positive $a$. At finite temperature the inclusion of fluctuations around the mean field is instead crucial to provide a qualitatively correct description of the crossover (Nozi\`eres and Schmitt-Rink, 1985; S\'a de Melo, Randeria and Engelbrecht, 1993). In this Section we review the mean-field treatment of the crossover at $T=0$, while some aspects of the theory at finite temperature will be discussed in Sec.~\ref{Sec5.3}. 

The account of the BCS mean-field theory we give here is based on the use of the pseudo-potential (\ref{pseudo}) 
and follows the treatment of Bruun {\it et al.} (1999). Let us start considering a simplified Hamiltonian 
without external confinement and where, in the interaction term of Eq.~(\ref{Hspin}), only 
pairing correlations are considered and treated at the mean-field level 
\begin{eqnarray}
&\hat{H}_\text{BCS} = \sum_\sigma  \int d{\bf r}\;\hat{\Psi}_\sigma^\dagger({\bf r})
\left(-{\nabla^2 \over 2 m}-\mu \right)\hat{\Psi}_\sigma({\bf r})&  
\label{HBCS} \\
& -\int d{\bf r} \left\{\Delta({\bf r}) \left[\hat{\Psi}_\uparrow^\dagger({\bf r})
\hat{\Psi}_\downarrow^\dagger({\bf r})-{1\over 2}\langle\hat{\Psi}_\uparrow^\dagger({\bf r})
\hat{\Psi}_\downarrow^\dagger({\bf r})\rangle\right] + h.c. \right\}\;.& 
\nonumber
\end{eqnarray}
We also restrict the discussion to equal masses $m$ and to unpolarized systems: $N_\uparrow=N_\downarrow=N/2$.
The direct (Hartree) interaction term proportional to the averages $\langle\hat{\Psi}_\uparrow^\dagger({\bf r})
\hat{\Psi}_\uparrow({\bf r})\rangle$ and $\langle\hat{\Psi}_\downarrow^\dagger({\bf r})
\hat{\Psi}_\downarrow({\bf r})\rangle$, is neglected in Eq.~(\ref{HBCS}) in order to avoid the 
presence of divergent terms in the theory when applied to the unitary limit $1/a\to 0$. The order parameter
$\Delta$ is defined here as the spatial integral of the short-range potential $V({\bf s})$ weighted by the 
pairing field (\ref{F})
\begin{eqnarray}
\Delta({\bf r}) &=& -\int d{\bf s}\; V({\bf s}) \langle \hat{\Psi}_\downarrow({\bf r}+{\bf s}/2)
\hat{\Psi}_\uparrow({\bf r}-{\bf s}/2)\rangle \nonumber \\
&=&-g(sF)^\prime_{s=0} \;.
\label{DBCS}
\end{eqnarray}
The last equality, which is obtained using the regularized potential (\ref{pseudo}), is consistent with the 
definition of the order parameter given in Eq.~(\ref{Fshortr}). The $c$-number term $\int d{\bf r}\Delta({\bf r})
\langle\hat{\Psi}_\uparrow^\dagger({\bf r})\hat{\Psi}_\downarrow^\dagger({\bf r})\rangle/2+ h.c.$ in the 
Hamiltonian (\ref{HBCS}) avoids double counting in the ground-state energy, a typical feature of the mean-field 
approach.

The Hamiltonian (\ref{HBCS}) is diagonalized by the Bogoliubov transformation $\hat{\Psi}_\uparrow({\bf r})=\sum_i\left( u_i({\bf r})\hat{\alpha}_i+v_i^\ast({\bf r})\hat{\beta}_i^\dagger \right)$, $\hat{\Psi}_\downarrow({\bf r})=\sum_i\left( u_i({\bf r})\hat{\beta}_i-v_i^\ast({\bf r})\hat{\alpha}_i^\dagger\right)$ which transforms particles into quasi-particles denoted by the operators $\hat{\alpha}_i$ and 
$\hat{\beta}_i$. Since quasi-particles should also satisfy fermionic 
anti-commutation relations, $\{\hat{\alpha}_i,\hat{\alpha}_{i^\prime}^\dagger\}=\{\hat{\beta}_i,
\hat{\beta}_{i^\prime}^\dagger\}=\delta_{i,i^\prime}$, the functions $u_i$ and $v_i$ obey the orthogonality relation $\int d{\bf r}\left[u_i^\ast({\bf r})u_j({\bf r})+v_i^\ast({\bf r})v_j({\bf r})\right]=\delta_{ij}$. As a consequence of the Bogoliubov transformations   the Hamiltonian (\ref{HBCS}) can be written in the form
\begin{equation}
\hat{H}_\text{BCS}=(E_0-\mu N)+\sum_i\epsilon_i\left( \hat{\alpha}_i^{\dagger }\hat{\alpha}_i
+\hat{\beta}_i^{\dagger }\hat{\beta}_i\right) \;,  
\label{Hdiag}
\end{equation}
which describes a system of independent quasi-particles. The corresponding expressions for the amplitudes $u_i$ and 
$v_i$ are obtained by solving the matrix equation
\begin{equation}
\left( \begin{array}{cc}
H_0 & \Delta({\bf r}) \\
\Delta^\ast({\bf r}) & -H_0 \end{array} \right) \left( \begin{array}{c} u_i({\bf r}) \\ v_i({\bf r}) \end{array} \right)
=\epsilon_i \left( \begin{array}{c} u_i({\bf r}) \\ v_i({\bf r}) \end{array} \right) \;,
\label{BdGnonuniform}
\end{equation}
where $H_0=-(\hbar^2/2m)\nabla^2-\mu$ is the single-particle Hamiltonian. The order parameter $\Delta({\bf r})$ is in general a complex, position dependent function.    
Eq.~(\ref{BdGnonuniform}) is known as the Bogoliubov-de Gennes equation  (see de Gennes, 1989). It can be used to describe both uniform and nonuniform configurations like, for example, quantized vortices (see Sec.~\ref{Sec8.3})) or solitons (Antezza, et al. 2007). 

In the uniform case the solutions take the simple form of plane waves $u_i({\bf r}) \to e^{i{\bf k \cdot r}}u_k/\sqrt{V}$ and $v_i \to  e^{i{\bf k \cdot r}}v_k/\sqrt{V}$ with 
\begin{equation}
u_k^2=1-v_k^2=\frac{1}{2}\left(1+\frac{\eta_k}{\epsilon_k}\right) \;; \;\;\;\;\; 
u_kv_k=\frac{\Delta}{2\epsilon_k} \;.
\label{uv}
\end{equation}
where $\eta _{k}={\frac{\hbar^2k^{2}}{2m}}-\mu$ is the energy of a free particle calculated with respect to the chemical potential. The spectrum of quasi-particles ($\epsilon_i \to \epsilon_{\bf k}$) which are the elementary excitations of the system, has the well-known form 
\begin{equation}
\epsilon_k=\sqrt{\Delta ^{2}+\eta _{k}^{2}} \;,
\label{elemexcit}
\end{equation}
and close to the Fermi surface coincides with the result (\ref{elexc}) holding in the weak coupling limit with 
$\Delta_\text{gap}=\Delta$. In this regime, the minimum of $\epsilon_k$ corresponds to the Fermi wavevector $k_F$. The minimum is shifted towards smaller values of $k$ as one approaches the unitary limit and corresponds to $k=0$ when the chemical potential changes sign on the BEC side of the resonance. Furthermore, one should notice that while the BCS theory correctly predicts the occurrence of a gap in the single-particle excitations, it is instead unable to describe the low-lying density oscillations of the gas (Bogoliubov-Anderson phonons). These can be accounted for by a time-dependent version of the theory (see, for example, Urban and Schuck, 2006). The vacuum of quasi-particles, defined by  $\hat{\alpha}_{\bf k}|0\rangle=\hat{\beta}_{\bf k}|0\rangle=0$, corresponds to the ground state of the system whose energy is given by 
\begin{equation}
E_0=\sum_{\bf k} \left(2\frac{\hbar^2k^2}{2m}v_k^2-\frac{\Delta^2}{2\epsilon_k}\right) \;.
\label{E0BCS}
\end{equation}
The above energy consists of the sum of two terms: the first is the kinetic energy of the two spin components, while the second corresponds to the interaction energy. One should notice that both terms, if calculated separately, exhibit an ultraviolet divergence which  disappears in the sum yielding a finite total energy.

The order parameter $\Delta$ entering the above equations should satisfy a self-consistent condition determined by the short range behavior  (\ref{Fshortr}) of the pairing field (\ref{F}). This function takes the form  
\begin{equation}
F(s)  =\int  \frac{d{\bf k}}{(2\pi)^3}
u_kv_k e^{i{\bf k}\cdot{\bf s}} = \Delta\int \frac{d{\bf k}}{(2\pi)^3} \frac{e^{i{\bf k}\cdot{\bf s}}}{2\epsilon_k} \;.
\label{FDelta}
\end{equation}
 By writing $(4\pi s)^{-1} = \int d{\bf k}e^{i{\bf k}\cdot{\bf s}}/[(2\pi)^3k^2]$ and comparing 
Eq.~(\ref{Fshortr}) with Eq.~(\ref{FDelta}) one straightforwardly obtains the important equation
\begin{equation}
\frac{m}{4\pi\hbar^2 a}= \int \frac{d{\bf k}}{(2\pi)^3}\left( \frac{m}{\hbar^2k^2} - 
\frac{1}{2\epsilon_k}\right) \;,  
\label{BCS1}
\end{equation}
where one is allowed to take the limit $s\to 0$ since the integral of the difference in brackets is convergent. Eq.~(\ref{BCS1}), through the expression (\ref{elemexcit}) of the elementary excitations, provides a relationship between $\Delta$ and the chemical potential $\mu$ entering the single-particle energy $\eta _{k}={\frac{\hbar^2k^{2}}{2m}}-\mu$. A second relation is given by the normalization condition 
\begin{equation}
n={2\over V}\sum_{\bf k}v_k^2 =
\int \frac{d{\bf k}}{(2\pi)^3}\; \left(1-\frac{\eta_k}{\epsilon_k}\right) \;,
\label{BCS2}
\end{equation}
which takes the form of an equation for the density.
One can prove that the density dependence of the chemical potential arising from the solution of Eqs.~(\ref{BCS1}) and (\ref{BCS2}) is consistent with the thermodynamic relation $\mu=\partial E_0/\partial N$, with $E_0$ given by Eq.~(\ref{E0BCS}).

Result (\ref{BCS1}) can be equivalently derived starting from the contact potential $\tilde{g}\delta({\bf r})$ 
[rather than from the regularized form (\ref{pseudo})]  and using the renormalized value
\begin{equation} 
{1\over \tilde{g}} = {m\over 4\pi\hbar^2a} - \int {d{\bf k}\over (2\pi)^3}{m\over \hbar^2k^2} 
\label{gtilde}
\end{equation}
of the coupling constant, corresponding to the low-energy limit of the two-body T-matrix (Randeria, 1995). In this 
case one must introduce a cut-off in the calculation of the order parameter  $\Delta=-\tilde{g}\langle\hat{\Psi}_\downarrow\hat{\Psi}_\uparrow\rangle=-\tilde{g}\int d{\bf k}\Delta/[2\epsilon_k(2\pi)^3]$, as well as in the integral in Eq.~(\ref{gtilde}), in order to avoid the emergence of ultraviolet divergences. 

In the case of weakly attractive gases ($k_F|a|\ll 1$ with $a<0$) the chemical potential approaches the Fermi energy 
$\mu\simeq E_F$ and  Eq.~(\ref{BCS1}) reduces to the equation for the gap of standard BCS theory. In the general 
case the value of $\mu$ and $\Delta$ should be calculated by solving the coupled equations (\ref{BCS1}) and 
(\ref{BCS2}). By expressing the energy  in units of the Fermi energy $E_F$ these equations only depend on the dimensionless parameter 
$1/k_Fa$ which characterizes the interaction strength along the BCS-BEC crossover. In the following we 
will be referring to Eqs.~(\ref{BCS1}) and (\ref{BCS2}) as to the BCS mean-field equations. 

Analytical results for the energy per particle are obtained in the limiting cases $1/k_Fa \to \pm \infty$ 
corresponding, respectively, to the BEC and BCS regimes. In the BCS limit the mean field equations give the result:
\begin{equation}
{E_0\over N} = \frac{3}{5}E_F\left(1-\frac{40}{e^4} e^{\pi/k_Fa} + ...\right)  
\label{E0bcs}
\end{equation}
while in the BEC limit one finds
\begin{equation}
{E_0\over N} =  -{\hbar^2 \over 2ma^2} + \frac{3}{5}E_F\left(\frac{5k_Fa}{9\pi}-\frac{5(k_Fa)^4}{54\pi^2}
+ ...\right) \;. 
\label{E0bec}
\end{equation}

While the leading term in the energy per particle is correctly reproduced in both limits (yielding, respectively, the non-interacting energy $3E_F/5$ and half of the dimer binding energy 
$-\hbar^2/2ma^2$), the higher order terms are wrongly predicted by this approach. In fact, in the BCS limit the theory 
misses the interaction-dependent terms in the expansion (\ref{enexpansion}). This 
is due to the absence of the Hartree term in the Hamiltonian (\ref{HBCS}). In the BEC limit the theory 
correctly reproduces a repulsive gas of dimers. However, the term arising from the interaction between dimers 
corresponds, in the expansion (\ref{E0bec}), to a molecule-molecule scattering length equal to $a_{dd}=2a$ 
rather than to the correct value $a_{dd}=0.60a$ (see Sec.~\ref{Sec3.3}). Furthermore, the Lee-Huang-Yang correction in the equation of state of composite bosons [see Eq.~(\ref{enerbeyond})] is not accounted for by the expansion (\ref{E0bec}). 

Finally, at unitarity ($1/k_Fa=0$) one finds $E_0/N\simeq 0.59(3E_F/5)$ which is 40\% larger than the value predicted by quantum Monte Carlo simulations (see Sec.~\ref{Sec4.6}).

It is worth noticing that the energy per particle, as well as the chemical potential, change sign from the BCS 
to the BEC regime. This implies that there exists a value of $k_Fa$ where $\mu=0$. This fact bears important 
consequences on the gap $\Delta_\text{gap}$ characterizing the spectrum (\ref{elemexcit}) of single-particle 
excitations. If $\mu>0$, $\Delta_\text{gap}$ coincides with the order parameter $\Delta$. This is the case, in 
particular, of the BCS regime, where one finds the result
\begin{equation}
\Delta_\text{gap}=\Delta=\frac{8}{e^2}E_F\exp\left[\frac{\pi}{2k_Fa}\right] \;.
\label{DeltaBCS}
\end{equation}
Notice that result (\ref{DeltaBCS}) does not include the Gorkov$-$Melik-Barkhudarov [see Eqs.~(\ref{GMB}), (\ref{Delta0})]. At unitarity one finds (Randeria, 1995) $\Delta_\text{gap}=\Delta\simeq 0.69 E_F$. When $\mu <0$ the gap is instead given by $\Delta_\text{gap}=\sqrt{\Delta^2+\mu^2}$. In particular in the BEC limit, where $\mu\simeq-\hbar^2/2ma^2$, one finds $\Delta=(16/3\pi)^{1/2}E_F/\sqrt{k_Fa}\ll|\mu|$. In the same limit the gap is given by $\Delta_\text{gap}=\hbar^2/2ma^2+3a\pi\hbar^2n/m$.  

The momentum distribution of either spin species $n_{\bf k}=\langle\hat{a}_{{\bf k}\uparrow}^\dagger\hat{a}_{{\bf k}
\uparrow}\rangle=\langle\hat{a}_{{\bf k}\downarrow}^\dagger\hat{a}_{{\bf k}\downarrow}\rangle$ is another direct 
output of the BCS mean-field theory. For a given value of $k_Fa$ along the crossover it is readily obtained from the 
corresponding values of $\mu$ and $\Delta$ through the expression
\begin{equation}
n_{\bf k}=v_k^2=\frac{1}{2}\left(1-\frac{\eta_k}{\sqrt{\eta_k^2+\Delta^2}}\right) \;.
\label{momdis}
\end{equation}
In the BCS regime, $n_{\bf k}$ coincides approximately with the step function $\Theta(1-k/k_F)$ characteristic of 
the non-interacting gas, the order parameter $\Delta$ providing only a small broadening around the Fermi wavevector. By increasing the interaction strength the broadening of the Fermi surface becomes more and more 
significant. At unitarity, $1/k_Fa=0$, the effect is of the order of $k_F$, consistently with the size of $\Delta$ being proportional to the Fermi energy. In the BEC regime $n_{\bf k}$ takes instead the limiting form 
\begin{equation}
n_{\bf k}=\frac{4(k_Fa)^3}{3\pi(1+k^2a^2)^2} \;, 
\label{mdismol}
\end{equation}
which is proportional to the square of the Fourier transform of the molecular wavefunction (\ref{bstate}). 

It is important to notice that at large wavevectors the momentum distribution (\ref{momdis}) decays as 
$n_{\bf k}\simeq m^2\Delta^2/(\hbar^4k^4)$ for $k\gg m|\mu|/\hbar$. The large-$k$ $1/k^4$ tail has important 
consequences for the kinetic energy of the system defined as $E_{kin}=2\sum_{\bf k}n_{\bf k}\hbar^2k^2/2m$, which 
diverges in 2D and 3D. This unphysical behavior arises from the use of the zero-range pseudopotential (\ref{pseudo}) 
which describes correctly only the region of wavevectors much smaller than the inverse effective range of interactions, 
$k\ll 1/|R^\ast|$. It reflects the fact that the kinetic energy is a microscopic quantity that in general 
can not be expressed in terms of the scattering length. 

Let us finally discuss the many-body structure of the ground-state wavefunction. The BCS ground state, defined as the 
vacuum state for the quasi-particles $\hat{\alpha}_{\bf k}$ and $\hat{\beta}_{\bf k}$, can be written explicitly in 
terms of the amplitudes $u_k$, $v_k$ giving the well known result
\begin{equation}
|\text{BCS}\rangle=\prod_{\bf k}\left(u_k+v_k \hat{a}_{{\bf k}\uparrow}^\dagger
\hat{a}_{-{\bf k}\downarrow}^\dagger\right)|0\rangle \;,
\label{BCSGS}
\end{equation}
where $|0\rangle$ is the particle vacuum. It is important to stress that the BCS mean-field 
Eqs.~(\ref{BCS1})-(\ref{BCS2}) can be equivalently derived from a variational calculation applied to the state 
(\ref{BCSGS}) where the grand-canonical energy is minimized with respect to $u_k$, $v_k$, subject to the normalization 
constraint $u_k^2+v_k^2=1$. The BCS state (\ref{BCSGS})  does not correspond to a definite number of particles. In fact, 
it can be decomposed into a series of states having an even number of particles $|\text{BCS}\rangle\propto 
|0\rangle+\hat{P}^\dagger |0\rangle+(\hat{P}^\dagger)^2|0\rangle+...$, where $\hat{P}^\dagger=\sum_{\bf k}v_k/u_k
\hat{a}_{{\bf k}\uparrow}^\dagger\hat{a}_{-{\bf k}\downarrow}^\dagger$ is the pair creation operator. By projecting the 
state (\ref{BCSGS}) onto the Hilbert space corresponding to $N$ particles one can single out the component 
$|\text{BCS}\rangle_N\propto(\hat{P}^\dagger)^{N/2}|0\rangle$ of the series. In coordinate space this $N$-particle 
state can be expressed in terms of an antisymmetrized product of pair orbitals (Leggett, 1980)
\begin{equation}
\Psi_\text{BCS}\left({\bf r}_1,...,{\bf r}_N\right)=\hat{{\cal A}}\left[\phi(r_{11^\prime})\phi(r_{22^\prime})...
\phi(r_{N_\uparrow N_\downarrow})\right] \;.
\label{BCSGS1}
\end{equation}
Here $\hat{{\cal A}}$ is the antisymmetrizer operator and the function $\phi(r)=(2\pi)^{-3}\int d{\bf k}(v_k/u_k) e^{i{\bf k}
\cdot{\bf r}}$ depends on the relative coordinate $r_{ii^\prime}=
|{\bf r}_i-{\bf r}_{i^\prime}|$ of the pair of particles, $i$ ($i^\prime$) being labels for the spin-up (down) particles. It is 
worth noticing that in the deep BEC regime, corresponding to $|\mu|\simeq\hbar^2/2ma^2\gg\Delta$, the pair orbital becomes proportional to the molecular wavefunction (\ref{bstate}): $\phi(r)=
(\sqrt{n_\sigma}/2)e^{-r/a}/(\sqrt{2\pi a}r)$, and the many-body wavefunction (\ref{BCSGS1}) describes a system where 
all atoms are paired into bound molecules. Wave functions of the form (\ref{BCSGS1}) are used in order to 
implement more microscopic approaches to the many-body problem (see Sec.~\ref{Sec5.2}).

In conclusion, we have shown how BCS mean-field theory is capable of giving a comprehensive and qualitatively correct 
picture of the BCS-BEC crossover at $T=0$. The quantitative inadequacies of the model will be discussed in more
details in Sec.~\ref{Sec5.2.2}.

\subsection{Quantum Monte Carlo approach at $T=0$} \label{Sec5.2}

\subsubsection{Method} \label{Sec5.2.1}

A more microscopic approach to the theoretical investigation of the ground-state properties of the gas along the BCS-BEC crossover is provided by the fixed-node diffusion Monte Carlo (FN-DMC) technique. This method is based on the Hamiltonian (\ref{Hspin}) where, as in Sec.~\ref{Sec5.1}, we consider uniform and unpolarized configurations of particles with equal masses. A convenient choice for the effective interatomic potential $V_{eff}(r)$ consists of using the square-well model (\ref{swell}) where, in order to reduce finite-range effects, the value of $R_0$ is taken as small as $nR_0^3=10^{-6}$. The depth $V_0=\hbar^2K_0^2/m$ of the potential is varied in the range $0<K_0R_0<\pi$ to reproduce the relevant regimes along the crossover. For $K_0R_0<\pi/2$, the potential does not support a two-body bound state and the scattering length is negative. Vice-versa, for $K_0R_0>\pi/2$, the scattering length is positive and a molecular state appears with binding energy $\epsilon_b$. The value $K_0R_0=\pi/2$ corresponds to the unitary limit, where $|a|=\infty$ and $\epsilon_b=0$.  

In a diffusion Monte Carlo (DMC) simulation one introduces the function $f({\bf R},\tau)=\psi_T({\bf R})
\Psi({\bf R},\tau)$, where $\Psi({\bf R},\tau)$ denotes the wavefunction of the system and $\psi_T({\bf R})$ is a 
trial function used for importance sampling. The function $f({\bf R},\tau)$ is evolved in imaginary time, 
$\tau=it/\hbar$, according to the Schr\"odinger equation
\begin{eqnarray}
-\frac{\partial f({\bf R},\tau)}{\partial\tau}= &-&\frac{\hbar^2}{2m} \left\{ \nabla_{\bf R}^2 f({\bf R},\tau) - 
\nabla_{\bf R}[{\bf F}({\bf R})f({\bf R},\tau)] \right\} \nonumber \\
&+& [E_L({\bf R})-E_{ref}]f({\bf R},\tau) \;.
\label{FNDMC}
\end{eqnarray}
In the above equation ${\bf R}$ denotes the position vectors of the $N$ particles, ${\bf R}=
({\bf r}_1,...,{\bf r}_N)$, $E_L({\bf R})=\psi_T({\bf R})^{-1}H\psi_T({\bf R})$ denotes the local energy, 
${\bf F}({\bf R})=2\psi_T({\bf R})^{-1}\nabla_{\bf R}\psi_T({\bf R})$ is the quantum drift force, and $E_{ref}$ is a 
reference energy introduced to stabilize the numerics. The various observables of the system are calculated from 
averages over the asymptotic distribution function $f({\bf R},\tau\to\infty)$ (for more details see {\it e.g.} 
Boronat and Casulleras, 1994). As an example, the DMC estimate of the energy is obtained from
\begin{equation}
E_\text{DMC}=\frac{\int d{\bf R} E_L({\bf R}) f({\bf R},\tau\to\infty)}{\int d{\bf R} f({\bf R},\tau\to\infty)}  \;.
\label{EDMC}
\end{equation}

A crucial requirement, which allows for the solution of Eq.~(\ref{FNDMC}) as a diffusion equation, is the positive 
definiteness of the probability distribution $f({\bf R},\tau)$. The condition $f({\bf R},\tau)\ge 0$ can be easily satisfied for the ground state of bosonic systems by choosing both $\Psi$ and $\psi_T$ positive definite, corresponding to the nodeless ground-state function. In this case the asymptotic distribution converges to $f({\bf R},\tau\to\infty)=
\psi_T({\bf R})\Psi_0({\bf R})$, where $\Psi_0({\bf R})$ is the exact ground-state wavefunction, and the average 
(\ref{EDMC}) yields the exact ground-state energy $E_0$. The case of fermionic ground-states or of more general excited states is 
different, due to the appearance of nodes in the wavefunction $\Psi$. In this case an exact solution is 
in general not available. An approximate treatment is provided by the FN-DMC method which enforces the positive 
definiteness of $f({\bf R},\tau)$ through the constraint that $\psi_T$ and $\Psi$ change sign together and thus share 
the same nodal surface. The nodal constraint is kept fixed during the calculation and the function 
$f({\bf R},\tau)$, after propagation in imaginary time according to Eq.~(\ref{FNDMC}), reaches for large times the 
asymptotic distribution $f({\bf R},\tau\to\infty)=\psi_T({\bf R})\Psi_{FN}({\bf R})$, where $\Psi_{FN}({\bf R})$ is an approximation to the exact eigenfunction of the many-body Schr\"odinger equation. It can be proved that, due to the 
nodal constraint, the fixed-node energy obtained from Eq.~(\ref{EDMC}) is an upper bound to the exact eigenenergy for a given symmetry (Reynolds {\it et al.}, 1982). In particular, if the nodal surface of $\psi_T$ were exact, then $E_\text{DMC}$ would also be exact. The energy calculated in a FN-DMC simulation depends crucially on a good parametrization of the many-body nodal surface.  

The calculations are carried out using a cubic simulation box of volume $V=L^3$ with periodic boundary conditions. 
The number of particles in the system ranges typically from $N=14$ to $N=66$ and finite-size analysis are performed to 
extrapolate the results to the thermodynamic limit. The most general trial wavefunction used in studies of ultracold 
fermionic gases has the form (Carlson {\it et al.}, 2003; Chang {\it et al.}, 2004; Astrakharchik {\it et al.}, 2004a; 
Chang and Pandharipande, 2005; Astrakharchik {\it et al.}, 2005a)
\begin{equation}
\psi_T({\bf R})=\Psi_\text{J}({\bf R})\Psi_\text{BCS}({\bf R}) \;,
\label{trialwf1}
\end{equation}
where $\Psi_\text{J}$ contains Jastrow correlations between all the particles, $\Psi_\text{J}({\bf R})=\prod_{i,j}
f_{\sigma\sigma^\prime}(|{\bf r}_{i\sigma}-{\bf r}_{j\sigma^\prime})$, where ${\bf r}_{i\sigma}$ denotes the position of the $i$-th particle with spin $\sigma$, and the BCS-type wavefunction $\Psi_\text{BCS}$ is an antisymmetrized product
of pair wavefunctions of the form (\ref{BCSGS1}). The pair orbital $\phi({\bf r})$ is chosen of the general form
\begin{equation}
\phi({\bf r})=\alpha\sum_{k_\alpha\le k_{max}}e^{i{\bf k}_\alpha \cdot {\bf r}} + \phi_s(r) \;,
\label{trialwf3}
\end{equation}
where $\alpha$ is a variational parameter and the sum is performed over the plane-wave states satisfying periodic 
boundary conditions, ${\bf k}_\alpha=2\pi/L(\ell_{\alpha x}\hat{x}+\ell_{\alpha y}\hat{y}+\ell_{\alpha z}\hat{z})$ (the 
$\ell$'s are integer numbers), up to the largest closed shell $k_{max}=2\pi/L(\ell_{max x}^2+\ell_{max y}^2+
\ell_{max z}^2)^{1/2}$ occupied by the $N/2$ particles. A convenient functional form of the Jastrow correlation 
terms $f_{\sigma\sigma^\prime}(r)$ and of the $s$-wave orbital $\phi_s(r)$ is discussed by Astrakharchik {\it et al.} 
(2005a). The Jastrow function in Eq.~(\ref{trialwf1}) is chosen positive definite, $\Psi_\text{J}({\bf R})\ge 0$, and therefore the nodal surface of the trial function is determined only by $\Psi_\text{BCS}$. 

An important point concerns the wavefunction $\Psi_\text{BCS}$ which can be used to describe both the normal and the superfluid state. In fact, if one chooses in Eq.~(\ref{trialwf3}) $\phi_s(r)=0$, it can be shown (Bouchaud, Georges and Lhuillier, 1988; Bouchaud and Lhuillier, 1989) that $\Psi_\text{BCS}$ coincides with the wavefunction of a free Fermi gas, {\it i.e.}, the product of the plane-wave 
Slater determinants for spin-up and spin-down particles. In this case the trial wavefunction (\ref{trialwf1}) is 
incompatible with ODLRO and describes a normal Fermi gas similarly to the wavefunction employed in the study of liquid $^3$He 
at low temperatures (Ceperley, Chester and Kalos, 1977). On the contrary, if $\phi_s(r)\neq 0$, the wavefunction 
(\ref{trialwf1}) accounts for $s$-wave pairing and describes a superfluid gas. In particular, in the deep BEC limit 
$1/k_Fa\gg1$, the choice $\alpha=0$ and $\phi_s(r)$ given by the molecular solution of the two-body problem in Eq.~(\ref{trialwf3}) reproduces the mean-field wavefunction (\ref{BCSGS1}). As an example, at 
unitarity, the normal-state wavefunction [$\phi_s(r)=0$ in Eq.~(\ref{trialwf3})] yields the value 
$E/N=0.56(1) 3E_F/5$ for the energy per particle, to be compared with the result $E/N=0.42(1) 3E_F/5$ obtained with the superfluid-state wavefunction.

Another important remark concerns the gas-like nature of the many-body state described by the wave 
function $\Psi_{FN}({\bf R})$. In the limit of a zero-range interatomic 
potential this state corresponds to the true ground-state of the system, because bound-states with more than two 
particles are inhibited by the Pauli exclusion principle (Baker, 1999). 
The situation is different for a finite-range potential. In the case of  
the square-well potential (\ref{swell}) one can easily calculate an upper bound to the ground-state energy using 
the Hartree-Fock state $|\text{HF}\rangle=\prod_{k\le k_F}\hat{a}_{{\bf k},\uparrow}^\dagger
\hat{a}_{{\bf k},\downarrow}^\dagger|0\rangle$. One finds the result $\langle\text{HF}|H|\text{HF}\rangle/N=
3E_F/5-\pi V_0nR_0^3/3$, showing that at large $n$ the interaction energy is unbounded from below and the kinetic 
energy can not prevent the system from collapsing. However, for realistic short-range potentials having a large depth $V_0\sim\hbar^2/mR_0^2$, this instability sets in at very high densities on order $nR_0^3\sim 1$. Such large density fluctuations are extremely unlikely, so that one can safely ignore this collapsed state when performing the simulations. Thus, for small enough values of $R_0$, the gas-like state corresponding to the wavefunction $\Psi_{FN}$ is expected to describe the ground state of the system with zero-range interactions.

\begin{figure}[b]
\begin{center}
\includegraphics[width=7.5cm]{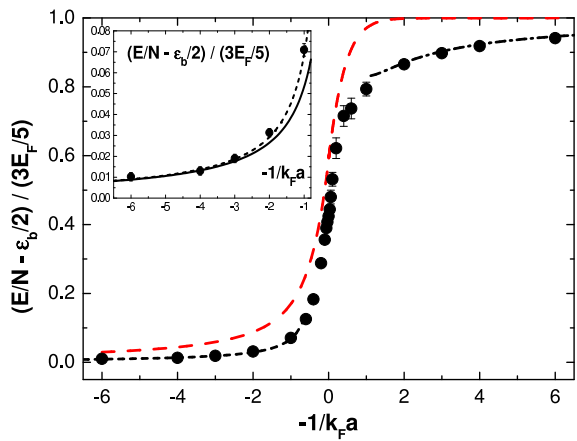}
\caption{(color online). Energy per particle along the BCS-BEC crossover with the binding-energy term subtracted from $E/N$. Symbols: quantum Monte Carlo results from Astrakharchik {\it et al.} (2004a). The dot-dashed line corresponds to the expansion (\ref{enexpansion}) and the dashed line to the expansion (\ref{enerbeyond}) holding, respectively, in the BCS and in the BEC regime. The long-dashed line (red) refers to the result of the BCS mean-field theory. Inset: enlarged view of the BEC regime $-1/k_Fa\le-1$. The solid line corresponds to the mean-field term in the expansion (\ref{enerbeyond}), the dashed line includes the Lee-Huang-Yang correction.}
\label{fig5.1}
\end{center}
\end{figure}

\begin{figure}[b]
\begin{center}
\includegraphics[width=7.5cm]{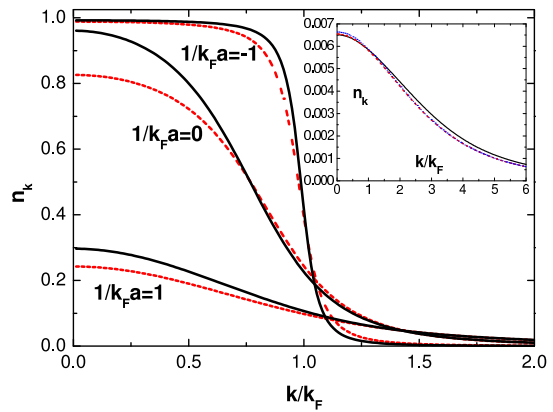}
\caption{(color online). QMC results of the momentum distribution $n_{\bf k}$ for different values of $1/k_Fa$ (solid lines). The dashed lines (red) correspond to the mean-field results from Eq.~(\ref{momdis}). Inset: $n_{\bf k}$ for $1/k_Fa=4$. The dotted line (blue) corresponds to the momentum distribution of the molecular state Eq.~(\ref{mdismol}).}
\label{fig5.3}
\end{center}
\end{figure}

\begin{figure}[b]
\begin{center}
\includegraphics[width=7.5cm]{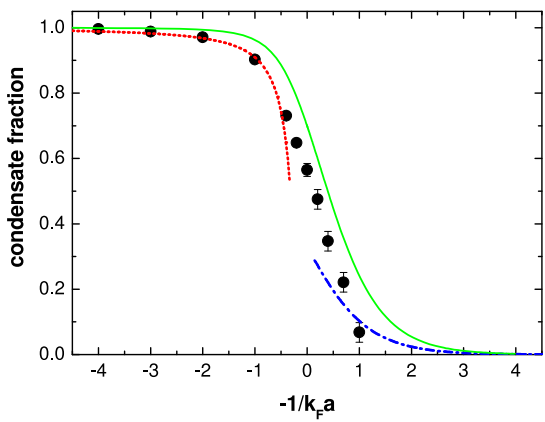}
\caption{(color online). Condensate fraction of pairs $n_\text{cond}$ [Eq.~(\ref{n0})] as a function of the 
interaction strength: FN-DMC results (symbols), Bogoliubov quantum depletion of a Bose gas with $a_{dd}=0.60a$ 
[dashed line (red)], BCS theory including the Gorkov$-$Melik-Barkhudarov correction [dot-dashed line (blue)] 
and mean-field theory using Eq.~(\ref{FDelta}) [solid line (green)].}
\label{fig5.4}
\end{center}
\end{figure}

\begin{figure}[b]
\begin{center}
\includegraphics[width=8.5cm]{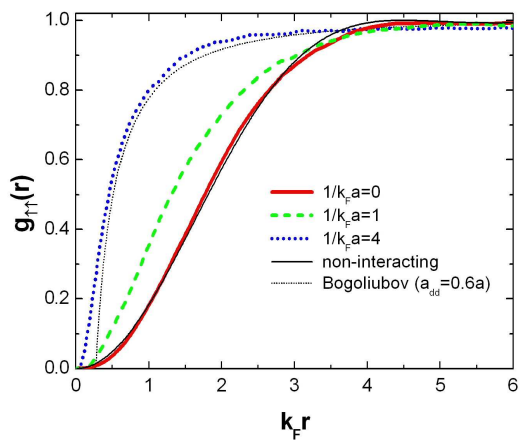}
\caption{(color online). QMC results for the pair correlation function of parallel spins, $g_{\uparrow\uparrow}(r)$, for different values of the interaction strength: solid line (red) $1/k_Fa=0$, dashed line (green) $1/k_Fa=1$, dotted line (blue) $1/k_Fa=4$. 
The thin solid line (black) refers to the non-interacting gas and the thin dotted line (black) is the pair correlation function of a Bose gas with $a_{dd}=0.60a$ and the same density as the single-component gas corresponding to $1/k_Fa=4$.}
\label{fig5.5}
\end{center}
\end{figure}

\begin{figure}[b]
\begin{center}
\includegraphics[width=8.5cm]{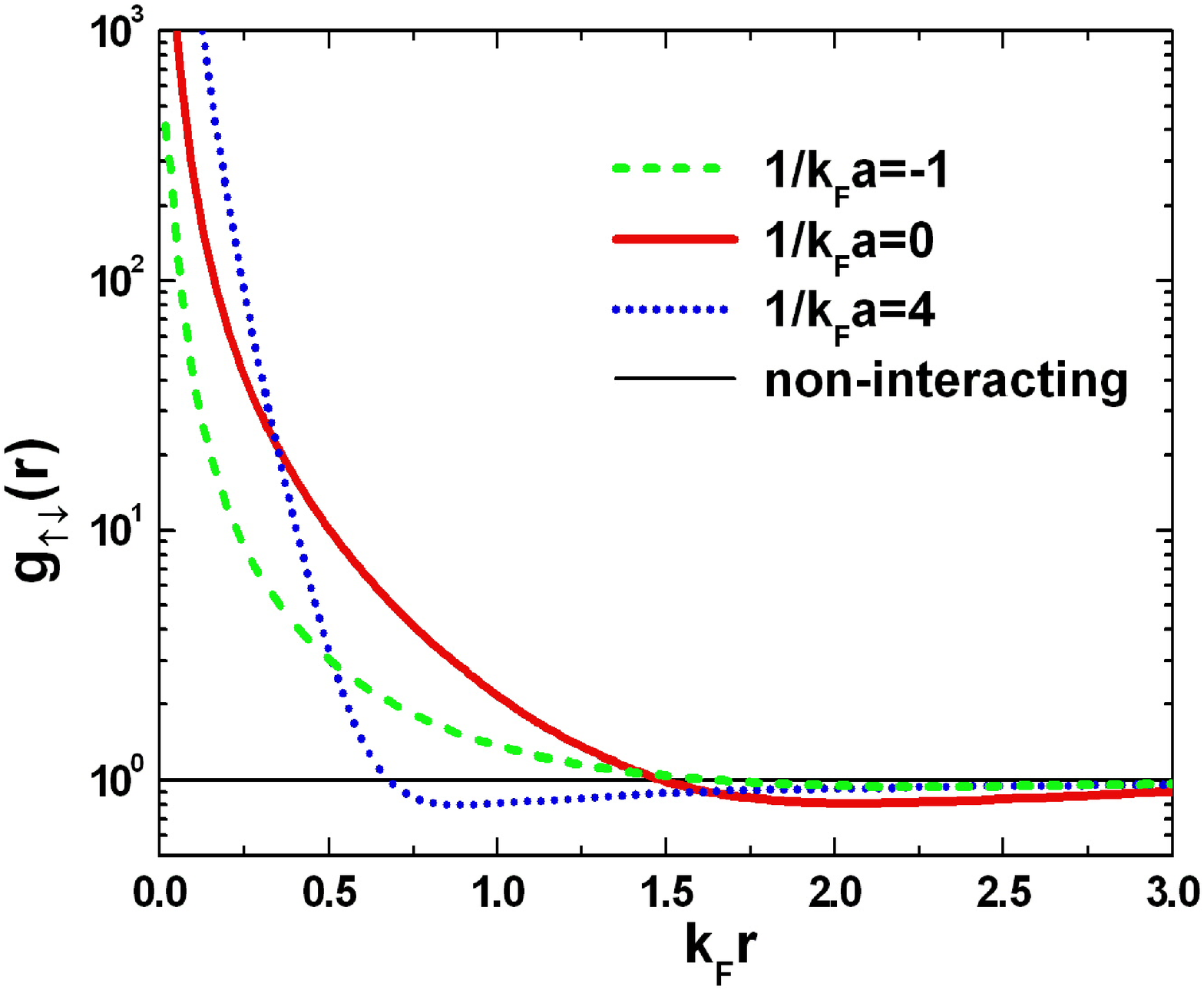}
\caption{(color online). QMC results for the pair correlation function of antiparallel spins, $g_{\uparrow\downarrow}(r)$, for different values of the interaction strength: dashed line (green) $1/k_Fa=-1$, solid line (red) $1/k_Fa=0$, dotted line (blue) 
$1/k_Fa=4$. The thin solid line (black) refers to the non-interacting gas.}
\label{fig5.6}
\end{center}
\end{figure}

\subsubsection{Results} \label{Sec5.2.2}

The results of the FN-DMC calculations are reported in Figs.~\ref{fig5.1}-\ref{fig5.6}. In Fig.~\ref{fig5.1} we show the energy per particle as a function of the interaction strength along the BCS-BEC crossover (Astrakharchik {\it et al.}, 2004a; Chang {\it et al.}, 2004). In order to emphasize many-body effects on the BEC side of the resonance we subtract from $E/N$ the two-body contribution $\epsilon_b/2$ arising from the molecules. Notice that $\epsilon_b$ refers here to the dimer binding energy in the square-well potential (\ref{swell}), which, for the largest values of $1/k_Fa$, includes appreciable finite-range corrections compared to $-\hbar^2/ma^2$. Nevertheless, no appreciable change is found in the difference $E/N-\epsilon_b/2$ if the value of $R_0$ is varied, demonstrating the irrelevance of this length scale for the many-body problem. Both in the BCS and in the BEC regime the FN-DMC energies agree respectively with the perturbation expansion (\ref{enexpansion}) and (\ref{enerbeyond}). In particular, in the inset of Fig.~\ref{fig5.1} we show an enlarged view of the results in the BEC regime indicating good agreement with the $a_{dd}=0.60a$ result for the dimer-dimer scattering length as well as evidence for beyond mean-field effects. In Fig.~\ref{fig5.1} we compare the FN-DMC results with the ones from the BCS mean-field theory of Sec.~\ref{Sec5.1}. The mean-field results reproduce the correct qualitative behavior, but are affected by important quantitative inadequacies.

The quantum Monte Carlo and mean-field results for the momentum distribution $n_{\bf k}$ and for the condensate fraction of pairs $n_\text{cond}$ are reported in Fig.~\ref{fig5.3} and \ref{fig5.4}, respectively (Astrakharchik {\it et al.}, 2005a). In both cases the mean-field predictions are in reasonable agreement with the findings of FN-DMC calculations. Significant discrepancies are found in the momentum distribution at unitarity (Fig.\ref{fig5.3}) (see also Carlson {\it et al.}, 2003), where the broadening of the 
distribution is overestimated by the mean-field theory consistently with the larger value predicted for the pairing gap. 
The momentum distribution in harmonic traps is discussed in Sec.~\ref{Sec6.3}. 

An important remark concerns the condensate fraction $n_\text{cond}$ defined in Eq.~(\ref{n0}) which, in the BEC regime, 
should coincide with the Bogoliubov quantum depletion $n_\text{cond}=1-8\sqrt{n_d a_{dd}^3}/3\sqrt{\pi}$ characterizing 
a gas of interacting
composite bosons with density $n_d=n/2$ and scattering length $a_{dd}=0.60a$. This behavior is indeed demonstrated by 
the FN-DMC results (Fig.\ref{fig5.4}), but is not recovered within the mean-field approximation. At $-1/k_Fa\sim 1$ on the 
opposite side of the resonance, the FN-DMC results agree with the condensate fraction calculated using BCS theory including  the Gorkov$-$Melik-Barkhudarov correction. This result is expected to reproduce the correct behavior of 
$n_\text{cond}$ in the deep BCS regime and is significantly lower as compared to the mean-field prediction. 

Condensation of pairs has been observed on both sides of the Feshbach resonance by detecting the emergence of a bimodal distribution in the released cloud after the conversion of pairs of atoms into tightly bound molecules using a fast 
magnetic-field ramp (Regal, Greiner and Jin, 2004b; Zwierlein {\it et al.}, 2004; Zwierlein {\it et al.}, 2005a). The magnetic-field sweep is slow enough to ensure full transfer of atomic pairs into dimers, but fast enough to act as a 
sudden 
projection of the many-body wavefunction onto the state of the gas far on the BEC side of the resonance. The resulting 
condensate fraction is an out-of-equilibrium property, whose theoretical interpretation is not straightforward (Perali, 
Pieri and Strinati, 2005; Altman and Vishwanath, 2005), but it strongly supports the existence of ODLRO in the gas at 
equilibrium also on the BCS side of the Feshbach resonance (negative $a$) where no stable molecules exist in vacuum.

Another quantity that can be calculated using the FN-DMC method is the spin-dependent pair correlation function 
(Astrakharchik {\it et al.}, 2004a; Chang and Pandharipande, 2005) defined as
\begin{eqnarray}
g_{\sigma\sigma^\prime}(|{\bf s}-{\bf s}^\prime|)=
\frac{4}{n^2}\langle\hat{\Psi}_{\sigma^\prime}^\dagger({\bf s}^\prime)\hat{\Psi}_\sigma^\dagger({\bf s})\hat{\Psi}_\sigma({\bf s})\hat{\Psi}_{\sigma^\prime}({\bf s}^\prime)\rangle \;.
\label{gofr}
\end{eqnarray}
This function measures the relative probability of finding a particle with equal or opposite spin at distance $|{\bf s}-{\bf s}^\prime|$ from a given particle. The results for $g_{\uparrow\uparrow}(r)$ are shown in Fig.~\ref{fig5.5} for different values of the interaction strength. Notice that in all cases $g_{\uparrow\uparrow}(r)$ must vanish at short distances due to the Pauli exclusion principle. This tendency of fermions to avoid each other (anti-bunching), as opposed to the bunching effect exhibited by bosons, has been recently revealed in the experiment by Jeltes {\it et al.} (2007) using helium atoms. In the BCS regime, where the effects of interaction in the $\uparrow$-$\uparrow$ spin channel are negligible, one expects that the pair correlation function is well described by the non-interacting gas result $g_{\uparrow\uparrow}(r)=1-9/(k_Fr)^4[\sin(k_Fr)/k_Fr-\cos(k_Fr)]^2$.  Quite unexpectedly this result holds true even at unitarity. Only when one approaches the BEC regime the effect of indirect coupling, mediated through interactions with the opposite spin component, becomes relevant and $g_{\uparrow\uparrow}(r)$ deviates significantly from the behavior of the non-interacting gas. In particular, for the largest value of $1/k_Fa$ reported in Fig.~\ref{fig5.5} ($1/k_Fa=4$), we show the pair distribution function of a gas of weakly interacting bosons of mass $2m$, density $n/2$ and scattering length $a_{dd}=0.60a$ calculated within Bogoliubov theory. For large distances, $r\gg a_{dd}$, where the Bogoliubov approximation is expected to hold, one finds a remarkable agreement.

Finally, in Fig.~\ref{fig5.6}, we report the results for the pair correlation function $g_{\uparrow\downarrow}(r)$. In the 
$\uparrow$-$\downarrow$ spin channel interactions are always relevant and give rise to a $1/r^2$ divergent behavior at short distances, the coefficient being determined by many-body effects on the BCS side of the 
resonance and at unitarity and by the molecular wavefunction in the deep BEC regime. In the latter case one finds $(k_Fr)^2g_{\uparrow\downarrow}(r)
\to3\pi/(k_Fa)$ while at unitarity one finds  $(k_Fr)^2
g_{\uparrow\downarrow}(r)\to 2.7$.  The divergent behavior of the pair-correlation function at short distances gives rise to a sizable bunching effect 
due to interactions in the $\uparrow$-$\downarrow$ spin channel, as opposed to the anti-bunching effect due to the Pauli 
principle in the $\uparrow$-$\uparrow$ spin channel (Lobo {\it et al.}, 2006a). The function $g_{\uparrow\downarrow}(r)$ 
has been the object of experimental studies using spectroscopic techniques (Greiner {\it et al.}, 2004; Partridge 
{\it et al.}, 2005). In particular, Partridge {\it et al.} (2005) measured the rate of molecular photoexcitation using 
an optical probe sensitive to short-range pair correlations. They found that the rate is proportional to $1/k_Fa$ in the 
BEC regime and decays exponentially in the BCS regime. 

Following the proposal by Altman, Demler and Lukin (2004), pair correlations have also been detected using the atom shot noise in absorption images (Greiner {\it et al.}, 2005). The noise, related to the fluctuations of the column integrated density, is extracted from 2D absorption images of the atom cloud by subtracting from each image pixel the azimuthal average of the signal. Crucial is the size of the effective image pixel ($\sim$15$\mu$m) which should be small enough to be sensitive to atom shot noise. Using this technique spatial $\uparrow$-$\downarrow$ pair correlations have been observed on the BEC side of the resonance by comparing pixel-to-pixel the processed noise images relative to the two spin components. These images are obtained immediately after dissociating the molecules through a rapid sweep of the scattering length across the resonance. Even more spectacular is the observation of nonlocal pair correlations between atoms that have equal but opposite momenta and are therefore detected at diametrically opposite points of the atom cloud in time-of-flight expansion. These correlations in momentum space are produced by dissociating the molecules and by allowing the gas to expand before imaging. An important requirement here, which for the moment has limited the application of this technique to the BEC regime, is that the relative momentum should be significantly larger than the center of mass motion of the pairs, since this latter would degrade the correlation signal due to image blurring. This method provides a novel, useful tool for detecting quantum correlations in many-body systems (F\"olling {\it et al.}, 2005).

\subsection{Other theoretical approaches at zero and finite temperature} \label{Sec5.3}

The extension of the BCS mean-field approach discussed in Sec.~\ref{Sec5.1} to finite temperatures requires the inclusion of 
thermal fluctuations in the formalism (see, {\it e.g.}, Randeria, 1995). This can be accomplished by expanding the effective action determining the partition function of the system to quadratic terms in the order parameter $\Delta$ (Nozi\`eres and Schmitt-Rink, 1985; S\'a de Melo, Randeria and Engelbrecht, 1993). The method goes beyond the saddle-point approximation which corresponds to the mean-field theory at $T=0$. The resulting predictions for $T_c$ are shown in Fig.~\ref{fig1.2}. At unitarity one finds the value $T_c=0.224T_F$, which is not too far from the most reliable theoretical estimate based on quantum Monte Carlo simulations (Burovski   {\it et al.}, 2006a)  discussed in Sec.~\ref{Sec4.6}. In the BEC regime the region between the transition temperature $T_c$ and the higher temperature scale $T^\ast=\hbar^2/k_Bma^2$, fixed by the molecular binding energy, corresponds to the so called pseudogap regime, where bound pairs exist but are not ``condensed'' and form a normal phase (Chen {\it et al.}, 2005). The presence of pseudogap effects near unitarity has been investigated by Stajic {\it et al.} (2004) and by Perali {\it et al.} (2004).  

Diagrammatic methods based on the $T$-matrix approach have been proposed to extend the original treatment by Nozi\`eres and Schmitt-Rink (1985) to the broken-symmetry phase below $T_c$ (Pieri, Pisani and Strinati, 2004) as well as to improve on it by including higher order diagrams (Haussmann, 1993 and 1994; Chen {\it et al.}, 2005; Combescot, Leyronas and Kagan, 2006; Haussmann {\it et al.}, 2006). In particular, in the recent approach by Haussmann {\it et al.} (2006) based on a ladder approximation, the fermionic degrees of freedom are accounted for using interacting Green's functions which are determined in a self-consistent way. This approach applies to arbitrary temperatures and interaction strengths. At unitarity it predicts the value $T_c=0.16T_F$ for the transition temperature.

Fully non-perturbative numerical techniques have also been applied to investigate the thermodynamics at finite temperature in the unitary regime. They are based on the auxiliary field (Bulgac, Drut and Magierski, 2006 and 2007) and diagrammatic determinant (Burovski {\it et al.}, 2006a and 2006b) quantum Monte Carlo method on a lattice and on the restricted path-integral Monte Carlo method in the continuum (Akkineni, Ceperley and Trivedi, 2006). The results of these approaches for the critical temperature and the thermodynamic functions have already been discussed in Sec.~\ref{Sec4.6} and the ones referring to harmonically trapped configurations are discussed in Sec.~\ref{Sec6.4}. Lattice quantum Monte Carlo methods have also been recently applied to investigate the ground-state properties at unitarity (Lee, 2006; Juillet, 2007). 

An alternative approach to the theoretical treatment of the BCS-BEC crossover is provided by the two-channel model (also called Bose-Fermi model). In this approach (Friedberg and Lee, 1989) the  Hamiltonian includes interaction terms involving both fermionic and bosonic degrees of freedom. A thorough account of the two-channel model can be found in the recent work by Gurarie and Radzihovsky (2007). This work  discusses the comparison with the single-channel Hamiltonian (\ref{Hspin}) and the predictions of the model in the case of narrow Feshbach resonances, where the many-body problem can be exactly solved using a perturbative expansion (see also Diehl and Wetterich, 2006). In the case of broad resonances this approach reduces to the single-channel Hamiltonian (\ref{Hspin}), where interactions are accounted for by a contact potential fixed  by the scattering length. The two-channel model is thus more general and can describe situations where the effective range  plays an important role.  The two-channel model was first introduced in the context of fermions with resonantly enhanced interactions by Holland {\it et al.} (2001) and by Timmermans {\it et al.} (2001) and was later developed to describe properties of the BCS-BEC crossover both at zero (Bruun and Pethick, 2004; Romans and Stoof, 2006) and at finite temperature (Ohashi and Griffin, 2002 and 2003a; Diehl and Wetterich, 2006; Falco and Stoof, 2007) as well as in trapped configurations (Ohashi and Griffin, 2003b).

Analytical methods based on an expansion around $D=4-\epsilon$ spatial dimensions have also been recently applied to the unitary Fermi gas both at $T=0$ (Nishida and Son, 2006a and 2006b) and at finite temperature (Nishida, 2006). The starting point of the method is the observation (Nussinov and Nussinov, 2006) that a unitary Fermi gas in $D=4$ behaves as an ideal Bose gas, {\it i.e.} that at $T=0$ the proportionality coefficient in Eq.~(\ref{muunitarity}) is $1+\beta=0$. Results for the physical case of $D=3$ are obtained by extrapolating the series expansion to $\epsilon=1$. A similar approach is based on a $1/N$ expansion, where $N$ is the number of fermionic species (Nikoli\'c and Sachdev, 2007; Veillette, Sheehy and Radzihovsky, 2007). For $N\to\infty$ the field-theoretical problem can be solved exactly using the mean-field theory. Corrections to the mean-field predictions can be calculated in terms of the small parameter $1/N$ and the results can be extrapolated to the relevant case of $N=2$.  

Other theoretical approaches that have been applied to the physics of the BCS-BEC crossover include the dynamical mean-field theory (Garg, Krishnamurthy and Randeria, 2005) and the renormalization group method (Nikoli\'c and Sachdev, 2007; Diehl {\it et al.}, 2007) and the generalization of the BCS mean-field theory to include effective mass and correlation terms within a density functional approach (Bulgac, 2007).

\section{INTERACTING FERMI GAS IN HARMONIC TRAP} \label{Chap6}

The solution of the many-body problem for non-uniform configurations is a difficult task involving in most cases numerical calculations which are more complex than in uniform matter (an example of this type of numerical studies will be given when discussing the structure of vortex configurations in Sec.~\ref{Sec7.4}). However, in the experimentally relevant case of quite large systems ($N\simeq$ 10$^5$-10$^7$) confined by a harmonic potential, the local density approximation (LDA) provides a reliable and relatively simple description. This approximation, which is also often referred to as semiclassical or Thomas-Fermi approximation, profits of the knowledge of the equation of state of uniform matter to infer on the behavior of the system in traps.

Let us also point out that on the BEC side of the resonance, where the interacting Fermi gas behaves like a gas of weakly interacting dimers, systematic information is available from our advanced knowledge of the physics of dilute Bose gases in traps (see, {\it e.g.}, Dalfovo {\it et al.}, 1999). Although the deep BEC regime is not easily achieved in present experiments with ultracold Fermi gases, the corresponding predictions nevertheless provide a useful reference.

\begin{figure}[b]
\begin{center}
\includegraphics[width=8.5cm]{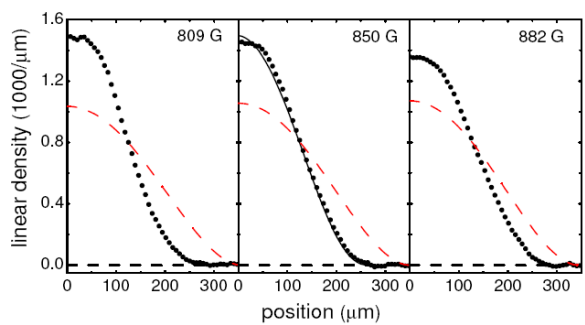}
\caption{Experimental results of the double integrated density profiles along the BCS-BEC crossover for a gas of $^6$Li atoms. The results at 850 G correspond to unitarity, while the ones at 809 G and 882 G correspond respectively to the BEC and BCS side of the resonance. The continuous curve at unitarity, is the best fit based on Eq.~(\ref{column}). The dashed lines correspond to the predictions for a non-interacting gas. From Bartenstein {\it et al.} (2004a).} 
\label{fig6.1}
\end{center}
\end{figure}

\subsection{Local density approximation at $T=0$: density profiles} \label{Sec6.1}

In this Section we will consider systems at zero temperature where the equation of state of the uniform gas is provided by the density dependence $\epsilon (n)$ of the energy density. The LDA consists of assuming that, locally, the system behaves like a uniform gas, so that its energy density can be expressed as $\epsilon(n)=n E(n)/N$
where $E(n)/N$ is the energy per atom of uniform matter. The energy of the trapped system is then written in the integral form 
\begin{equation}
E=\int d{\bf r} \left\{ \epsilon [n({\bf r})]+V_{ho}({\bf r})n({\bf r}) \right\} \;,
\label{Etot}
\end{equation}
and consists of the sum of the internal (also called release) energy 
\begin{equation}
E_{rel}=\int d{\bf r}\; \epsilon [n({\bf r})]  \;,
\label{Erel}
\end{equation}
and of the oscillator energy 
\begin{equation}
E_{ho}=\int d{\bf r}\; V_{ho}({\bf r})n({\bf r})  \;,
\label{Epot}
\end{equation}
provided by the trapping potential $V_{ho}({\bf r})$, introduced in Eq.~(\ref{Vho}), which is assumed to be the same for both spin species. Furthermore, we also assume $N_\uparrow=N_\downarrow$. In Eqs.~(\ref{Etot}-\ref{Epot}), 
$n({\bf r})=n_\uparrow({\bf r})+n_\downarrow({\bf r})$ is the total density profile. Its value at equilibrium is 
determined by the variational relation $\delta (E-\mu_0 N)/\delta n({\bf r})=0$, which yields the Thomas-Fermi equation 
\begin{equation}
\mu_0=\mu[n({\bf r})]+V_{ho}({\bf r})  \;,
\label{LDA}
\end{equation}
where $\mu(n)=\partial \epsilon(n)/\partial n$
is the local chemical potential determined by the equation of state of the uniform system and $\mu_0$ is the chemical potential of 
the trapped gas, fixed by the normalization condition $\int d{\bf r}\;n({\bf r})=N$. Eq.~(\ref{LDA}) provides an implicit 
equation for $n({\bf r})$. 

The applicability of LDA is justified if the relevant energies are much larger than the single-particle oscillator energy $\hbar\omega_i$, {\it i.e.} if $\mu _{0}\gg\hbar\omega_i$ ($i=x,y,z$). While in the case of Bose-Einstein condensed 
gases this condition is ensured by the repulsive interaction among atoms, in the Fermi case the situation is different. In fact, thanks to the quantum pressure term related to the Pauli principle, even in the non-interacting case one can apply the Thomas-Fermi relationship (\ref{LDA}) using the density dependence
\begin{equation}
\mu(n)=\left( 3\pi ^{2}\right) ^{2/3}\frac{\hbar ^{2}}{2m}n^{2/3} \;,
\label{munF}
\end{equation} 
for the local chemical potential yielding the equilibrium profile (\ref{n0r}). 

Interactions modify the shape and the size of the density profile. The effects are accounted for by Eq.(\ref{LDA}) once 
the local chemical potential $\mu(n)$ is known. A simple result is obtained at unitarity, where the equation of state has 
the same density dependence (\ref{munF}) as for the ideal gas, apart from the dimensionless renormalization factor 
$(1+\beta)$ [see Eq.~(\ref{muunitarity})]. By dividing Eq.~(\ref{LDA}) by $(1+\beta)$, one finds that the results at 
unitarity are obtained from the ones of the ideal Fermi gas by simply rescaling the trapping frequencies and the chemical potential according to $\omega_i\rightarrow \omega_i/\sqrt{1+\beta}$ and $\mu_0\rightarrow\mu_0/(1+\beta)$. In particular, 
the density profile at unitarity takes the same form (\ref{n0r}) as in the ideal gas, the Thomas-Fermi radii being given 
by the rescaled law 
\begin{equation}
R_i=(1+\beta)^{1/4}R_i^0=(1+\beta)^{1/4}a_{ho}(24N)^{1/6}\frac{\omega_{ho}}{\omega_i} \;.
\label{Runitarity}
\end{equation}
From the above results one also finds the useful expression $E_{ho}=(1+\beta)^{1/2}E_{ho}^0$ for the oscillator energy (\ref{Epot}) of the trapped gas in terms of the ideal gas value $E_{ho}^0=(3/8)NE_F^{HO}$ (see Sec.~\ref{Sec2.1}).

In the BCS regime ($a$ negative and small), the first correction to the non-interacting density profile (\ref{n0r}) can be calculated using perturbation theory. Interactions are treated at the Hartree level by adding the term $-4\pi\hbar^2|a|n/2m$ to the expression (\ref{munF}) for the local chemical potential. The resulting density profile is compressed due to the effect of the attractive interaction and the Thomas-Fermi radii reduce according to the law
\begin{equation}
R_i=\sqrt{\frac{2\mu_0}{m\omega_i^2}}=R_i^0\left(1-\frac{256}{315\pi^2}k_F^0|a|\right) \;,
\label{radiusBCS}
\end{equation}
holding if $k_F^0|a|\ll 1$, where $k_F^0$ is the Fermi wavevector (\ref{pFN}) of the non-interacting gas.

Another interesting case is the BEC limit where one treats the interaction between dimers using the mean-field term 
$\mu_d=g_dn/2$ in the equation of state. The coupling constant $g_d=2\pi\hbar^2a_{dd}/m$ is fixed by the molecule-molecule
scattering length $a_{dd}=0.60a$. In this limit  the density 
is given by the inverted parabola profile (Dalfovo {\it et al.}, 1999) and the Thomas-Fermi radii reduce to
\begin{equation}
R_i=a_{ho}\left( {15\over 2}N \frac{a_{dd}}{a_{ho}} \right)^{1/5} \frac{\omega_{ho}}{\omega_{i}} \;. 
\label{RBEC}
\end{equation}

In Fig.~\ref{fig6.1} we show the results for the density profiles measured \textit{in situ} in a harmonically trapped 
Fermi gas at low temperatures. The plotted profile is the double integrated density $n^{(1)}(z)=\int dxdy\;n({\bf r})$, corresponding to the quantity measured in the experiment (Bartenstein {\it et al.}, 2004a). Very good agreement between experiment and theory is found at unitarity where one finds 
\begin{equation}
n^{(1)}(z)=\frac{N}{R_z}\frac{16}{5\pi}\left( 1- \frac{z^{2}}{R_z^{2}} \right)^{5/2} \;,   
\label{column}
\end{equation}
with $R_z$ given by Eq.(\ref{Runitarity}). The best fit to the experimental curve yields the value $\beta =-0.73_{-0.09}^{+0.12}$ (Bartenstein {\it et al.}, 2004c). The attractive nature of interactions at unitarity is explicitly revealed in Fig.~\ref{fig6.1} through the comparison with the density profile of a non-interacting gas. For a systematic comparison between experimental and theoretical results for the density profiles, see Perali, Pieri and Strinati (2004).

A more recent experimental determination of $\beta$, also based on the \textit{in situ} measurement of the radius of the cloud, gives the value $\beta = -0.54(5)$ (Partridge {\it et al.}, 2006a). These measurements refer to a gas of $^6$Li atoms. An important result consists in the agreement found with experiments on $^{40}$K atoms, where $\beta$ was determined by extrapolating to low temperature the measured values of the oscillator energy (Stewart {\it et al.}, 2006). The fact that the value of $\beta$ does not depend on the atomic species considered is a further important proof of the universal behavior exhibited by these systems at unitarity (see Table~\ref{Tab2} for a list of available experimental and theoretical results of $\beta$).

\begin{table}
\centering
\caption{Experimental and theoretical values of the universal parameter $\beta$. The experimental
results are taken from: (1) Tarruell {\it et al.} (2007); (2) Kinast {\it et al.} (2005); (3) Partridge {\it et al.} (2006a); (4) Bartenstein {\it et al.} (2004c), and (5) Stewart {\it et al.} (2006). The theoretical results are from: (6) Carlson {\it et al.} (2003); (7) Astrakharchik {\it et al.} (2004a); (8) Carlson and Reddy (2005) and (9) Perali, Pieri and Strinati (2004).}
\begin{tabular}{|l|c|c|} \hline
& & $\beta$ \\ \hline
Exp. ($^6$Li) & ENS (1)& -0.59(15) \\
& Duke (2)& -0.49(4) \\
& Rice (3)& -0.54(5) \\
& Innsbruck (4)& -$0.73_{-0.09}^{+0.12}$ \\ \hline
Exp. ($^{40}$K) & JILA (5)& -$0.54_{-0.12}^{+0.05}$ \\ \hline
Theory & BCS mean field & -0.41 \\ 
& QMC (6,7,8)& -0.58(1) \\
& $T$-matrix (9)& -0.545 \\ \hline
\end{tabular}
\label{Tab2}
\end{table}

\subsection{Release energy and virial theorem} \label{Sec6.2}

In addition to the \textit{in situ} density profiles a valuable source of information comes from the measurement of the 
energy after switching off the confining trap (release energy). This energy consists of 
the sum of the kinetic and the interaction term
\begin{equation}
E_{rel}=E_{kin}+E_{int}  \;,
\label{release}
\end{equation}
and within LDA it is simply given by Eq.~(\ref{Erel}). 

The release energy was first measured along the crossover by Bourdel {\it et al.} (2004) on a gas of $^6$Li atoms. The most recent experimental determination at unitarity yields the estimate $\beta =-0.59(15)$ (Tarruell {\it et al.}, 2007). This value agrees with the one extracted from \textit{in situ} measurements of the radii (see Table~\ref{Tab2}).

A general relationship between the release energy (\ref{Erel}) and the potential energy (\ref{Epot}) can be derived with the help of the virial theorem. This theorem holds when the energy density $\epsilon(n)$ has a polytropic (power law) dependence on the density: $\epsilon (n)\propto n^{\gamma+1}$. The polytropic dependence characterizes the BEC limit, where $\gamma =1$, as well as the unitary limit, where $\gamma=2/3$. The theorem is derived by applying the number conserving transformation $n({\bf r})
\rightarrow(1+\alpha)^3n[(1+\alpha){\bf r}]$ to the density of the gas at equilibrium. By imposing that the total energy variation vanish to first order in $\alpha$, one easily gets the result 
\begin{equation}
3\gamma E_{rel}=2E_{ho}  \;,
\label{releaseext} 
\end{equation}
which reduces to $E_{rel}=E_{ho}$ at unitarity.

\begin{figure}[b]
\begin{center}
\includegraphics[width=7.5cm]{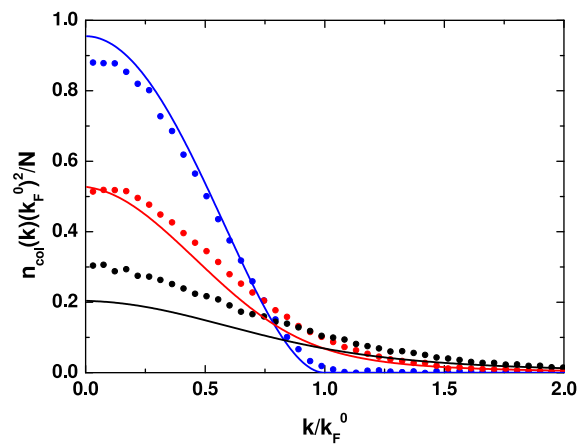}
\caption{(color online). Column integrated momentum distributions of a trapped Fermi gas. The symbols correspond to the experimental results from Regal {\it et al.} (2005) and the lines to the mean-field results based on Eq.~(\ref{nktrap}) for the same values of the interaction parameter $1/k_F^0a$. From top to bottom $1/k_F^0a=-71$ (blue), $1/k_F^0a=0$ (red) and $1/k_F^0a=0.59$ 
(black).} 
\label{fig6.4}
\end{center}
\end{figure}

\subsection{Momentum distribution} \label{Sec6.3}

The momentum distribution of ultracold Fermi gases is an important quantity carrying a wealth of information on the role 
played by interactions. A simple theoretical approach for trapped systems is based on the BCS mean-field treatment 
introduced in Sec.~\ref{Sec5.1} and on the local density approximation (Viverit {\it et al.}, 2004). The result is given by 
the spatial integral of the particle distribution function (\ref{momdis}) of the uniform gas
\begin{equation}
n({\bf k})=\int \frac{d{\bf r}}{(2\pi)^3} n_{\bf k}({\bf r}) \;,
\label{nktrap}
\end{equation}
where the ${\bf r}$-dependence enters through the chemical potential $\mu({\bf r})$ and  the order parameter $\Delta({\bf r})$ 
which are determined by the local value of the density $n({\bf r})$. The momentum distribution $n({\bf k})$ is calculated for given values of the interaction strength $k_F^0a$. Two limiting cases can be derived analitically: one corresponds to the non-interacting gas, where $n_{\bf k}({\bf r})=\Theta[1-k/k_F({\bf r})]$ depends on the local Fermi wavevector $k_F({\bf r})=[3\pi^2n({\bf r})]
^{2/3}$ and the integral (\ref{nktrap}) yields the result (\ref{n0k}). The other corresponds to the deep BEC regime, where, using Eq.~(\ref{mdismol}),  
one finds the molecular result $n({\bf k})=(a^3N/2\pi^2)/(1+k^2a^2)^2$.  A general feature emerging from these results is the broadening of the Fermi surface which, for trapped systems, is caused both by the confinement and by interaction effects. 

The momentum distribution has been measured along the crossover in a series of studies (Regal {\it et al.}, 2005;
Chen {\it et al.}, 2006 and Tarruell {\it et al.}, 2007). The accessible quantity in experiments is the column 
integrated distribution $n_{col}(k_\perp)=\int_{-\infty}^\infty dk_z n({\bf k})$, where $k_\perp=\sqrt{k_x^2+k_y^2}$. These experiments are based on the technique of time-of-flight expansion followed by absorption imaging. A crucial requirement is that the gas must expand freely without any interatomic force. To this purpose the scattering length is set to zero by a fast magnetic-field ramp immediately before the expansion. The measured column integrated distributions along the crossover from Regal {\it et al.} (2005) are shown in Fig.~\ref{fig6.4} together with the mean-field calculations of $n_{col}$ based on Eq.~(\ref{nktrap}) for the same values of the interaction strength $1/k_F^0a$ (the value of $a$ corresponds here to the scattering length before the magnetic-field ramp). There is an overall qualitative agreement between the theoretical predictions and the experimental results. However, the dynamics of the ramp produces a strong quenching of $n_{col}$ at large momenta $k\gg k_F^0$ which greatly affects the released kinetic energy $E_{rel}=2\pi\int dk_\perp k_\perp n_{col}(k_\perp) (3\hbar^2k_\perp^2/4m)$. A theoretical estimate of $E_{rel}$, based on the equilibrium distributions would predict a very large value, on the order of the energy scale $\hbar^2/mR_0^2$ associated with the interatomic potential [see discussion in Sec.~\ref{Sec5.1} after Eq.~(\ref{mdismol})]. On the contrary, the measured energies from Regal {\it et al.} (2005) do not depend on the details of the interatomic potential because the magnetic-field ramp is never fast enough to access features on order of the interaction range $R_0$. A good quantitative agreement for the measured momentum distribution and kinetic energy is provided by the approach developed by Chiofalo, Giorgini and Holland (2006), which explicitly accounts for the time dependence of the scattering length produced by the magnetic-field ramp.

\begin{figure}[b]
\begin{center}
\includegraphics[width=7.5cm]{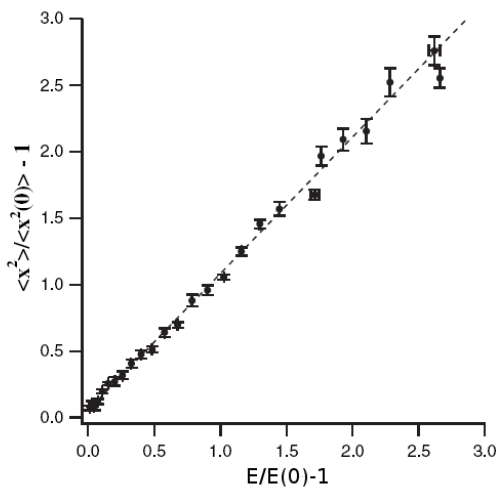}
\caption{Verifying the virial theorem at unitarity in a Fermi gas of $^6$Li: $\langle x^2\rangle/\langle x^2(0)\rangle$ versus $E/E(0)$ showing linear scaling. Here $\langle x^2\rangle$ is the transverse mean square radius, proportional to the oscillator energy. $E$ is the total energy evaluated as in Kinast {\it et al.} (2005). $E(0)$ and $\langle x^2(0)\rangle$ denote ground-state values. From Thomas, Kinast and 
Turlapov (2005).}
\label{fig6.6}
\end{center}
\end{figure}

\begin{figure}[b]
\begin{center}
\includegraphics[width=7.5cm]{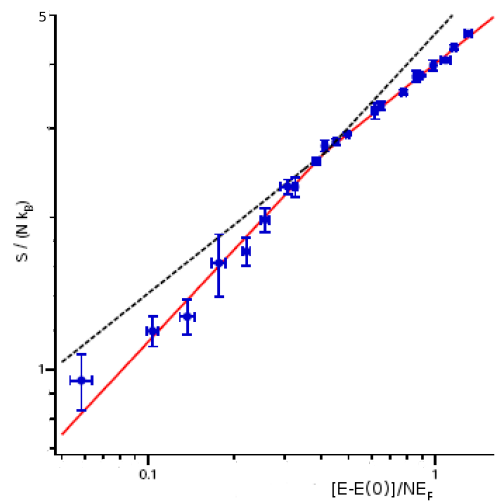}
\caption{Measured entropy in a trapped $^6$Li gas at unitarity vs. total energy. The data reveal the occurrence of a characteristic change of behavior in the slope at $[E-E(0)]/N=0.41E_F$, which is interpreted as the signature of the phase transition to the superfluid regime. The solid lines are power-law fits below and above the critical point, while the dashed lines show the extended fits. From Luo {\it et al.} (2007).}
\label{fig6.7}
\end{center}
\end{figure}

\subsection{Trapped gas at finite temperatures} \label{Sec6.4}

The local density approximation (\ref{LDA}) for the density profile can be successfully used also at finite temperature, where 
the chemical potential depends on both the density and the temperature and is defined according to the thermodynamic relationship $\mu(n,T)=(\partial \epsilon(n,T)/\partial n)_{s}$
where $s=nS(n,T)/N$ is the entropy density. Since the transition temperature decreases when the density decreases, the LDA predicts a sharp spatial boundary between a superfluid core and a normal external region. The effect is particularly evident in the 
BEC regime where it gives rise to a typical bimodal distribution characterized by a narrow Bose-Einstein condensate of molecules sorrounded by a broader cloud of thermal molecules. The experimental observation of this bimodal distribution (Greiner, Regal and Jin, 2003; Bartenstein {\it et al.}, 2004a; Zwierlein {\it et al.}, 2003; Bourdel {\it et al.}, 2004; Partridge {\it et al.}, 2005) represents the most spectacular and direct evidence of the emergence of Bose-Einstein condensation from an interacting Fermi 
gas (see Fig.~\ref{fig1.1}). In this BEC regime the critical temperature for the superfluid transition is given by the well known expression $k_BT_\text{BEC}=\hbar\omega_{ho}\left[N_d/\zeta(3)\right]^{1/3}$, where $N_d=N/2$ is the number of dimers and 
$\zeta(3)\simeq1.202$. In terms of the Fermi temperature (\ref{EF}) this result reads
\begin{equation}
T^{HO}_\text{BEC}=0.52 T_F^{HO}  \;,
\label{TBECtrap}
\end{equation}
and should be compared with the corresponding result (\ref{TCTFunif}) holding in uniform gases. By including interaction effects to lowest order using a mean-field approach, one predicts a positive shift of the transition temperature (\ref{TBECtrap}), proportional to the dimer-dimer scattering length $a_{dd}$ (Giorgini, Pitaevskii and Stringari, 1996). 
Due to large interaction effects in the condensate and in the thermal cloud, the bimodal structure becomes less and less pronounced as one approaches the resonance and  consequently it becomes more difficult to reveal the normal-superfluid phase transition by just looking at the density profile.

Once one knows the density profiles of the interacting Fermi gas as a function of temperature and interaction strength, one can calculate the various thermodynamic functions along the crossover using the LDA. For example, the release energy and the entropy 
are given, respectively, by Eq.~(\ref{Erel}) and by the integral $S(T)=\int d{\bf r}s[T,n({\bf r})]$. At unitarity the relevant thermodynamic functions can be conveniently expressed in terms of the universal functions $f_{P}(x)$ and $f_{S}(x)$ defined in Sec.~\ref{Sec4.6}, where the ratio $x=T/T_F$ is replaced by the local quantity 
$k_BT2m/\hbar^2[3\pi^2n({\bf r})]^{2/3}$ (Ho, 2004; Thomas, Kinast and Turlapov, 2005). Furthermore, a useful result at 
unitarity, which keeps its validity at finite temperature, is the virial relationship $E_{rel}=E_{ho}$ [see Eq.~(\ref{releaseext})]. Indeed, since the energy has a minimum at constant $N$ and $S$, by imposing the stationarity condition at constant $T/n^{2/3}$ one immediately finds the above identity. 

The temperature dependence of the thermodynamic functions in a trapped gas along the crossover has been calculated in a 
series of papers using self-consistent many-body approaches (Perali {\it et al.}, 2004; Kinast {\it et al.}, 2005; Chen, 
Stajic and Levin, 2005; Stajic, Chen and Levin, 2005; Hu, Liu and Drummond, 2006a and 2006b). Other 
studies are based on quantum Monte Carlo techniques (Burovski {\it et al.}, 2006b; Bulgac, Drut and Magierski, 2007). In 
particular, the transition temperature has been determined using LDA and the Monte Carlo results for uniform systems already discussed in connection with Eq.~(\ref{Tcunitarity}) of Sec.~\ref{Sec4.6}. The reported value is $T_c=0.20(2)E_F$ (Burovski 
{\it et al.}, 2006b). Furthermore, the values of the energy and entropy per particle at the transition point have also been 
determined (Bulgac, Drut and Magierski, 2007), yielding respectively the results $[E(T_c)-E(0)]/N=0.32E_F$ and $S(T_c)/Nk_B=2.15$.

From the experimental point of view a major difficulty in the study of the thermodynamic functions is the determination of 
the temperature when one cools down the system into the deeply degenerate regime. In fact, when the system is very cold the
measurement of the density profile in the tails, which in general provides a natural access to $T$, is not accurate, the 
number of thermally excited atoms being small. This is not a severe problem in the deep BCS regime, where the non-interacting 
Thomas-Fermi profile fitted to the whole spatial distribution of the expanded cloud provides a reliable thermometry. Instead, in the 
most interesting strongly coupled regime, a model-independent temperature calibration is very difficult to obtain.
Useful estimates of the temperatures achievable through adiabatic transformations along the BCS-BEC crossover can be obtained using entropy arguments. For example, starting from an initial configuration in the molecular BEC regime with temperature $T$ and changing adiabatically the value of the scattering length from positive to small negative values on the other side of 
the Feshbach resonance, one eventually reaches a final temperature in the BCS regime given by (Carr, Shlyapnikov, and Castin, 
2004)
\begin{equation}
\left( \frac{T}{T_{F}}\right)_{final}=\frac{\pi^2}{45\zeta(3)}\left( \frac{T}{T_\text{BEC}} \right)_{initial}^{3} \;.
\end{equation}
This relationship has been obtained by requiring that the entropies of the initial and final regimes, calculated using the predictions of the degenerate ideal Bose and Fermi gases respectively, be the same. The adiabatic transformation results in 
a drastic reduction of $T$. 

Many thermodynamic functions have already been measured in trapped Fermi gases. Measurements of the specific heat and a first   determination of the critical temperature $T_c$ were reported by Kinast {\it et al.} (2005) who extracted the value 
$T_c=0.27T_F$ at unitarity. The value of $T_c$ was determined by identifying a characteristic change of slope of the measured energy as a function of temperature and relied on a model-dependent temperature calibration.

In order to overcome the difficulties of the direct measurement of the temperature, recent experiments have also focused on the study of relevant thermodynamic relationships. One of these experiments concerns the verification of the virial relation at unitarity (Thomas, Kinast and Turlapov, 2005) that has been achieved by measuring independently the mean-square radius, proportional to the oscillator energy, and the total energy. The results shown in Fig.~\ref{fig6.6} demonstrate that the virial relation is verified with remarkable accuracy, confirming the universality of the equation of state at unitarity.

Another experiment concerns the measurement of the entropy of the trapped gas (Luo {\it et al.}, 2007). In this experiment one starts from a strongly interacting configuration (for example at unitarity) and switches off adiabatically the interactions  bringing the system into a final weakly interacting state, where the measurement of the radii (and hence of the oscillator energy) is expected to provide a reliable determination of the entropy. Since entropy is conserved during the transformation, the measurement of the radii of the final (weakly interacting) sample determines the entropy of the initial (strongly interacting) configuration. This procedure has been used to measure the entropy as a function of the energy at unitarity in a gas of $^6$Li atoms. The results are shown in Fig.~\ref{fig6.7}. These measurements give access to the temperature of the gas through the relationship $dE/dS=T$. They have been used to study the thermodynamic behavior near the critical point which can be identified as a change in the energy dependence $S(E)$ of the entropy. This method provides the experimental estimate $T_{c}=0.29T_{F}$ for the critical temperature in good agreement with the analysis of the specific heat (Kinast {\it et al.}, 2005). Furthermore, the extracted values of the critical entropy, $S(T_c)/Nk_B=2.7(2)$, and energy per particle, $[E(T_c)-E(0)]/N=0.41(5)E_F$, reasonably agree with the theoretical estimates reported above.

\section{DYNAMICS AND SUPERFLUIDITY} \label{Chap7}

Superfluidity is one of the most striking properties exhibited by ultracold Fermi gases, analog to superconductivity in charged Fermi systems. Among the most noticeable manifestations of superfluidity one should recall the absence of shear viscosity, the hydrodynamic nature of macroscopic dynamics even at zero temperature, the Josephson effect, the irrotational quenching of the moment of inertia, the existence of quantized vortices and the occurrence of pairing effects. The possibility of exploring these phenomena in ultracold gases is providing  a unique opportunity to complement our present knowledge of superfluidity in neutral Fermi systems, previously limited to liquid $^3$He, where pairing occurs in a $p$-wave state. In this Section we will first discuss the hydrodynamic behavior exhibited by superfluids and its implications on the dynamics of trapped Fermi gases (expansion and collective oscillations). The last part is devoted instead to a discussion of pair-breaking effects and of Landau's critical velocity.  Other manifestations of superfluidity will be discussed in Sec.~\ref{Chap8}.

\subsection{Hydrodynamic equations of superfluids at $T=0$} \label{Sec7.1}

The macroscopic behavior of a neutral superfluid is governed by the Landau equations of irrotational hydrodynamics. The condition of irrotationality is the consequence of the occurrence of off-diagonal long-range order, characterized by the order parameter $\Delta({\bf r})$ [see Eqs.~(\ref{n2})-(\ref{Fshortr})]. In fact the velocity field is proportional to the gradient of the phase $S$ of the order parameter according to 
\begin{equation}
\mathbf{v}=\frac{\hbar}{2m}\nabla S \;.
\label{vgrad}
\end{equation}

At zero temperature the  hydrodynamic equations of superfluids consist of coupled and closed equations for the density and the velocity field. Actually, due to the absence of the normal component, the superfluid density coincides with the total density and the superfluid current with the total current. The hydrodynamic picture assumes that  the densities of the two spin components are equal and move in phase [$n_\uparrow({\bf r},t)=n_\downarrow({\bf r},t)\equiv n({\bf r},t)/2$ and ${\bf v}_\uparrow({\bf r},t)={\bf v}_\downarrow({\bf r},t)\equiv {\bf v}({\bf r},t)$]. This implies, in particular, that the number of particles of each spin species be equal ($N_\uparrow=N_\downarrow$). In the same picture the current is given by ${\bf j}=n{\bf v}$. The equations take the form  
\begin{equation}
{\partial \over \partial t}n + \nabla {\bf \cdot} (n {\bf v})=0 \;,
\label{continuity}
\end{equation}
for the density (equation of continuity) and  
\begin{equation}
m{\partial \over \partial t}{\bf v} + \nabla \left({1\over 2}m{\bf v}^2 + \mu(n) + V_{ho} \right)=0 \;,
\label{euler}
\end{equation}
for the velocity field (Euler's equation). Here $\mu(n)$ is the atomic chemical potential, fixed by the equation of state of uniform matter. At equilibrium (${\bf v}=0$) Euler's equation reduces to the LDA equation (\ref{LDA}) for the ground-state density profile. The hydrodynamic equations (\ref{continuity}-\ref{euler}) can be generalized to the case 
of superfluids of unequal masses and unequal trapping potentials (see Sec.~\ref{Sec9.3}). 

Despite the quantum nature underlying the superfluid behavior, the hydrodynamic equations of motion have a classical form and do not depend explicitly on the Planck constant. This peculiarity raises the question whether the hydrodynamic behavior of a cold Fermi gas can be used to probe the superfluid regime. In fact, as we will see, Fermi gases above the critical temperature with resonantly enhanced interactions enter a collisional regime where the dynamic behavior is governed by the same equations (\ref{continuity})-(\ref{euler}). In this respect it is important to stress that collisional hydrodynamics admits the possibility of rotational components in the velocity field which are strictly absent in the superfluid. A distinction between classical and superfluid hydrodynamics is consequently possible only studying the rotational properties of the gas (see Sec.~\ref{Chap8}). Let us also emphasize that the hydrodynamic equations of superfluids have the same form both for Bose and Fermi systems, the effects of statistics entering only the form of $\mu(n)$. 

The applicability of the hydrodynamic equations is in general limited to the study of macroscopic phenomena, characterized by long wavelength excitations. In particular, the wavelengths should be larger than the so-called healing length $\xi$. The value of $\xi$ is easily estimated in the BEC limit where the Bogoliubov spectrum (\ref{bogexc}) approaches the phonon dispersion law for wavelengths larger than $\xi\sim(na_{dd})^{-1/2}$. In the opposite BCS regime the healing length is instead fixed by the pairing gap. In fact, the Bogoliubov-Anderson phonon mode is defined up to energies of the order $\hbar ck\sim\Delta_{\text gap}$, corresponding to $\xi\sim\hbar v_F/\Delta_{\text gap}$. In both the BEC and BCS limits the healing length becomes larger and larger as $k_F|a| \to 0$. Near resonance instead the only characteristic length is fixed by the average interatomic distance and the hydrodynamic theory 
can be applied for all wavelengths larger than $k_F^{-1}$. In Sec.~\ref{Sec7.4} we will relate the healing length to the critical Landau velocity and discuss its behavior along the crossover. 

\begin{figure}[b]
\begin{center}
\includegraphics[width=8.5cm]{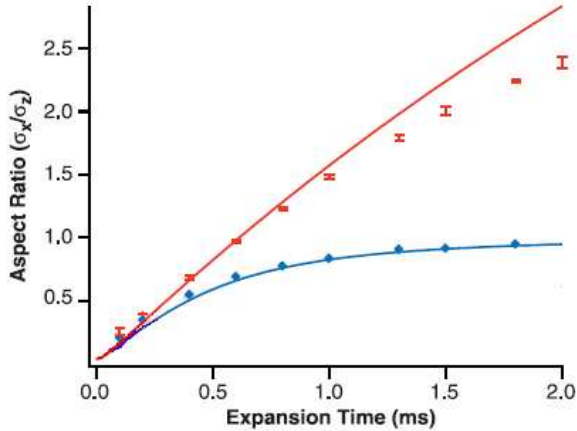}
\caption{(color online). Aspect ratio as a function of time during the expansion of an ultracold Fermi gas at unitarity. Upper (red) symbols: experiment; upper (red) line: hydrodynamic theory. For comparison the figure also shows the results in the absence of interactions. Lower (blue) symbols: experiment; lower (blue) line: ballistic expansion. From O'Hara {\it et al.} (2002).}
\label{fig7.2}
\end{center}
\end{figure}

\begin{figure}[b]
\begin{center}
\includegraphics[width=7.5cm]{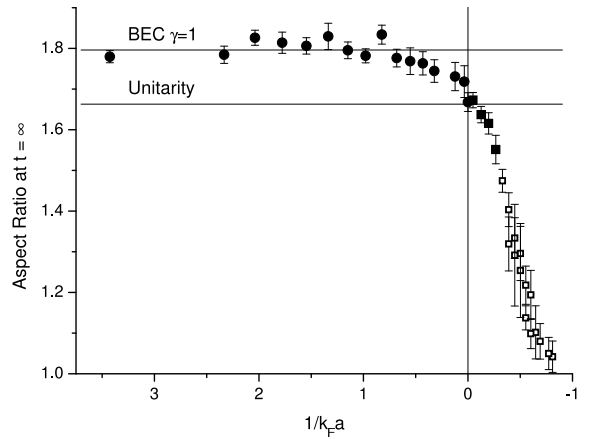}
\caption{Aspect ratio after expansion from the trap along the crossover. The solid lines are the hydrodynamic predictions at unitarity and for a gas in the deep BEC regime. The Fermi momentum $k_F$ in the figure corresponds to the value (\ref{pFN}). From Altmeyer {\it et al.} (2007c).}
\label{fig7.3}
\end{center}
\end{figure}

\subsection{Expansion of a superfluid Fermi gas} \label{Sec7.2}  

In most experiments with ultracold atomic gases images are taken after expansion of the cloud. These measurements  provide information on the state of the gas in the trap and it is consequently of crucial importance to have an accurate theory describing the expansion. As discussed in Sec.~\ref{Sec2.2}, in the absence of interactions the expansion of a Fermi gas is asymptotically isotropic also if before the expansion the gas is confined by an anisotropic potential. Deviations from isotropy are consequently an important signature of the role of interactions. In the experiment of O'Hara {\it et al.} (2002) the first clear evidence of anisotropic expansion of a cold Fermi gas at unitarity was reported (see Fig.~\ref{fig7.2}), opening an important debate in the community aimed to understand the nature of these novel many-body configurations. Hydrodynamic theory has been extensively used in the past years to analyze the expansion of Bose-Einstein condensed gases. More recently (Menotti, Pedri and Stringari, 2002) this theory was proposed to describe the expansion of a Fermi superfluid. The hydrodynamic solutions are obtained starting from the equilibrium configuration, corresponding to a Thomas-Fermi profile, and then solving Eq.~(\ref{euler}) by setting $V_{ho}=0$ for $t>0$. In general the hydrodynamic equations should be solved numerically. However, for an important class of configurations the spatial dependence can be determined analytically. In fact, if the chemical potential has the power-law dependence $\mu\propto n^{\gamma}$ on the density, then the Thomas-Fermi equilibrium profiles have the analytic form $n_0\propto(\mu_0-V_{ho})^{1/\gamma}$ and the scaling ansatz 
\begin{equation}
n(x,y,z,t) = (b_xb_yb_z)^{-1}n_0({x\over b_x},{y\over b_y},{z\over b_z}) \;,
\label{scaling}
\end{equation}
provides the exact solution. The scaling parameters $b_i$ obey the simple time-dependent equations:
\begin{equation}
\ddot{b}_i - {\omega^2_i \over b_i(b_xb_yb_z)^\gamma }=0 \;.
\label{bi}
\end{equation}
The above equation generalizes the scaling law previously introduced in the case of an interacting Bose gas ($\gamma=1$) (Castin and Dum, 1996; Kagan, Surkov and Shlyapnikov, 1996). From the solutions of Eq.~(\ref{bi}) one can calculate the aspect ratio as a function of time. For an axially symmetric trap ($\omega_x=\omega_y\equiv \omega_{\perp}$) this is defined as the ratio between the radial and axial radii. In terms of the scaling parameters $b_i$ it can be written as 
\begin{equation}
{R_{\perp}(t) \over Z(t)} = {b_\perp(t)\over b_z(t)}{\omega_z \over \omega_\perp} \;.
\label{ratio}
\end{equation}

For an ideal gas the aspect ratio tends to unity, while the hydrodynamic equations yield an asymptotic value $\neq 1$. Furthermore, hydrodynamics predicts a peculiar inversion of shape during the expansion caused by the hydrodynamic forces which are larger in the direction of larger density gradients. As a consequence an initial cigar shaped configuration is brought into a disk shaped profile at large times and vice-versa. One can easily estimate the typical time at which the inversion of shape takes place. For a highly elongated trap ($\omega_\perp \gg \omega_z$) the axial radius is practically unchanged for short times since the relevant expansion time along the $z$ axis is fixed by $1/\omega_z$. Conversely, the radial size increses fast and, for $\omega_\perp t \gg 1$ one expects $R_\perp(t)\sim R_\perp(0)\omega_\perp t$. One then finds that the aspect ratio is equal to unity when $\omega_z t\sim 1$. 

In Fig.~\ref{fig7.2} we show the predictions for the aspect ratio given by Eqs.~(\ref{bi}-\ref{ratio}) at unitarity, where $\gamma=2/3$, together with the experimental results of O'Hara {\it et al.} (2002) and the predictions of the ideal Fermi gas. The configuration shown in the figure corresponds to an initial aspect ratio equal to $R_\perp/Z=0.035$. The comparison strongly supports the hydrodynamic nature of the expansion of these ultracold Fermi gases. The experiment was repeated at higher temperatures and found to exhibit a similar hydrodynamic behavior even at temperatures of the order of the Fermi temperature, where the system can not be superfluid. One then concludes that even in the normal phase the hydrodynamic regime can be ensured by collisional dynamics. This is especially plausible close to unitarity where the scattering length is very large and the free path of atoms is of the order of the interatomic distances. The anisotropic expansion exhibited by ultracold gases shares important analogies with the expansion observed in the quark gluon plasma produced in heavy ion collisions (see, for example, Kolb {\it et al.}, 2001; Shuryak, 2004). 

In more recent experiments (Tarruell {\it et al.}, 2007; Altmeyer {\it et al.}, 2007c) the aspect ratio was measured at the coldest temperatures along the BCS-BEC crossover, at a fixed expansion time. At unitarity the results (see Fig.~\ref{fig7.3}) well agree with the hydrodynamic predictions but, as one moves towards the BCS regime, the aspect ratio becomes closer and closer to unity thereby revealing important deviations from the hydrodynamic behavior. In the deep BCS regime the hydrodynamic theory predicts indeed the same behavior for the time dependence of the aspect ratio as at unitarity, the equation of state being characterized by the same power law coefficient $\gamma= 2/3$. The deviations from the hydrodynamic picture follow from the fact that, since the density becomes smaller and smaller during the expansion, the BCS gap becomes exponentially small and the hydrodynamic picture can not longer be safely applied. Superfluidity is expected to be instead robust at unitarity since the energy gap is much larger than the typical oscillator frequency (whose inverse fixes the time scale of the expansion) and that, as a consequence, pairs can not easily break during the expansion. 

A case which deserves special attention is the expansion of the unitary gas for  isotropic harmonic trapping. In this case an exact solution of the many-body problem is available (Castin, 2004) without invoking the hydrodynamic picture. One makes use of the following scaling ansatz for the many-body wavefunction
\begin{equation} 
\Psi({\bf r}_1, ...{\bf r}_N,t)=N(t)e^{i\sum_jr^2_jm\dot{b}/2\hbar b}\Psi_0({\bf r}_1/b, ...{\bf r}_N/b) \;,
\label{castin}
\end{equation}
where $\Psi_0$ is the many-body wavefunction at $t=0$, $b(t)$ is a time-dependent scaling variable and $N(t)$ is a normalization factor. Eq.~(\ref{castin}) exactly satisfies the Bethe-Peierls boundary condition (\ref{BPbcon}) if one works at unitarity where $1/a =0$. At the same time, for distances larger than the range of the force where the Hamiltonian coincides with the kinetic energy, the Schr\"odinger equation is exactly satisfied by Eq.~(\ref{castin}) provided $\ddot{b}=\omega^2_{ho}/b^3$. One consequently finds that the expansion of the unitary gas is exactly given by the non-interacting law $b(t)=\sqrt{\omega_{ho}^2t^2+1}$, which coincides with the prediction of the hydrodynamic equations (\ref{bi}) after setting $\omega_i\equiv \omega_{ho}$.

\begin{figure}[b]
\begin{center}
\includegraphics[width=8.5cm]{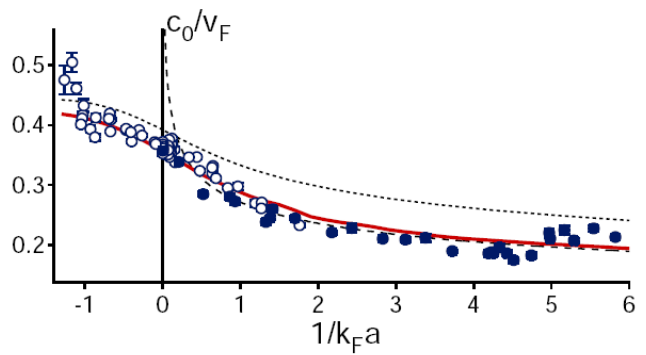}
\caption{Sound velocity in the center of the trap measured along the BCS-BEC crossover (symbols). The theory curves are obtained using Eq.~(\ref{cz}) and are based on different equations of state: BCS mean field (dotted line), quantum Monte Carlo (solid line) and Thomas-Fermi molecular BEC with $a_{dd}=0.60a$ (dashed line). The Fermi momentum $k_F$ in the figure corresponds to the value (\ref{pFN}). From Joseph {\it et al.} (2006).}
\label{fig7.4}
\end{center}
\end{figure}

\begin{figure}[b]
\begin{center}
\includegraphics[width=8.5cm]{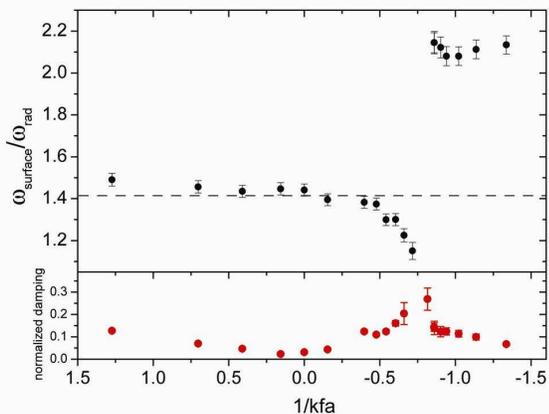}
\caption{Frequency of the radial quadrupole mode for an elongated Fermi gas in units of the radial frequency (upper panel). The dashed line is the hydrodynamic prediction $\sqrt2\omega_\perp$. The lower panel shows the damping of the collective mode. The Fermi momentum $kf$ in the figure corresponds to the value (\ref{pFN}). From Altmeyer {\it et al.} (2007b).}
\label{fig7.5}
\end{center}
\end{figure}

\begin{figure}[b]
\begin{center}
\includegraphics[width=8.5cm]{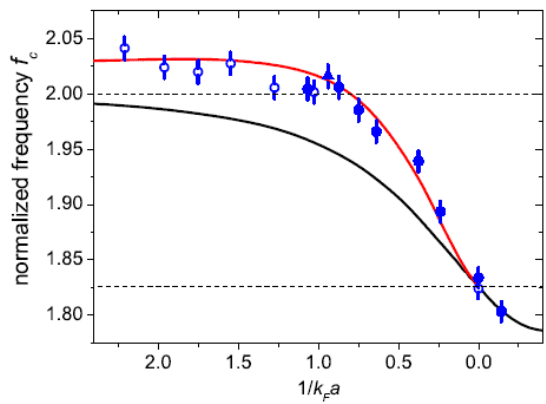}
\caption{(color online). Frequency of the radial compression mode for an elongated Fermi gas in units of the radial frequency (blue symbols). The curves refer to the equation of state based on BCS mean-field theory [lower (black) line] and on Monte Carlo simulations [upper (red) line]. The Fermi momentum $k_F$ in the figure corresponds to the value (\ref{pFN}). From Altmeyer {\it et al.} (2007a).}
\label{fig7.6}
\end{center}
\end{figure}

\subsection{Collective oscillations} \label{Sec7.3}

The collective oscillations of a superfluid gas provide a unique source of  information, being at the same time of relatively easy experimental access and of relevant theoretical importance. At zero temperature they can be studied by considering the linearized form of the time dependent hydrodynamic equations (\ref{continuity})-(\ref{euler}), corresponding to small oscillations of the density, $n=n_0+\delta n\exp(-i\omega t)$, with respect to the equilibrium profile $n_0({\bf r})$. The linearized equations take the form
\begin{equation}
-\omega^2\delta n = {1\over m}\nabla \cdot \left[n_0\nabla ({\partial \mu \over \partial n}\delta n)\right] \;.
\label{linearized}
\end{equation}

In the absence of harmonic trapping the solutions correspond to  sound waves with dispersion $\omega =c q$ and, in a Fermi superfluid, coincide with the Bogoliubov-Anderson phonons. In the BEC limit one recovers the Bogoliubov result $c=\sqrt{\pi\hbar^2a_{dd}n_d/m^2}$ for the sound velocity, where $a_{dd}$ is the dimer-dimer scattering length and $n_d=n/2$ is the molecular density. In the BCS limit one instead approaches the ideal gas value $c=v_F/\sqrt3$. Finally, at unitarity one has the result (\ref{soundunitarity}).

In the presence of harmonic trapping the propagation of sound is affected by the inhomogeneity of the medium. In a cylindrical configuration ($\omega_z=0$) the sound velocity can be easily calculated using the 1D result $mc_{1D}^2=n_{1D}\partial \mu_{1D}/\partial n_{1D}$, where $n_{1D}=\int dxdy n$ is the 1D density and the 1D chemical potential $\mu_{1D}$ is determined by the Thomas-Fermi relation $\mu(n) + V_{ho}({\bf r}_\perp)=\mu_{1D}$ for the radial dependence of the density profile. One finds (Capuzzi {\it et al.}, 2006)
\begin{equation}
c_{1D}=\left({1\over m} {\int dxdy n \over \int dxdy (\partial\mu/\partial n)^{-1}} \right)^{1/2} \; .
\label{cz}
\end{equation}
On the BEC side, where $\mu \propto n$, one recovers the result $c_{1D}=c/\sqrt{2}$ in terms of the value (\ref{soundvelocity}) of uniform systems. This result was first derived by Zaremba (1998) in the context of Bose-Einstein condensed gases. At unitarity, where $\mu \propto n^{2/3}$, one instead finds $c_{1D}=\sqrt{3/5}c$. The propagation of sound waves in very elongated traps has been recently measured by Joseph {\it et al.} (2006). The experimental results are in good agreement with the estimate of $c_{1D}$ based on the QMC equation of state and clearly differ from the prediction of the BCS mean-field theory (see Fig.~\ref{fig7.4}).

In the presence of 3D harmonic trapping the lowest frequency solutions of the hydrodynamic equations are discretized, their frequencies being of the order of the trapping frequencies. These modes correspond to wavelengths of the order of the size of the cloud. Let us first consider the case of isotropic harmonic trapping ($\omega_x=\omega_y=\omega_z\equiv \omega_{ho}$). A general class of divergency free (also called surface) solutions is available in this case. They are characterized by the velocity field ${\bf v} \propto\nabla(r^\ell Y_{\ell m})$, satisfying $\nabla\cdot{\bf v}=0$ and corresponding to density variations of the form $(\partial \mu/\partial n)\delta n \propto r^\ell Y_{\ell m}$. $Y_{\ell m}$ is here the spherical harmonic function. By using the identity $(\partial \mu/\partial n)\nabla n_0=-\nabla V_{ho}$, holding at $T=0$ for the density profile at equilibrium, one finds that these solutions obey the dispersion law $\omega(\ell)= \sqrt\ell \omega_{ho}$, independent of the form of the equation of state, as expected for the surface modes driven by an external force. This result provides a useful characterization of the hydrodynamic regime. The result in fact differs from the prediction $\omega(\ell)=\ell\omega_{ho}$ of the ideal gas model, revealing the importance of interactions accounted for by the hydrodynamic description. Only in the dipole case ($\ell=1$), corresponding to the center of mass oscillation, the interactions do not affect the frequency of the mode. 

In addition to surface modes an important solution is the $\ell=0, m=0$ breathing mode whose frequency can be found in analytic form if the equation of state is polytropic ($\mu\propto n^{\gamma}$). In this case the velocity field has the radial form ${\bf v} \propto {\bf r}$ and one finds $\omega(m=0)=\sqrt{3\gamma+2}\omega_{ho}$. For $\gamma=1$ one recovers the well known BEC result $\sqrt5\omega_{ho}$ (Stringari, 1996b) while, at unitarity, one finds $2\omega_{ho}$. It is worth stressing that the result at unitarity keeps its validity beyond the hydrodynamic approximation. The proof follows from the same arguments used for the free expansion in Sec.~\ref{Sec7.2}. In fact, the scaling ansatz (\ref{castin}) solves the Schr\"odinger equation also in the presence of isotropic harmonic trapping (Castin, 2004). The scaling function in this case obeys to the equation $\ddot{b}=\omega^2_{ho}/b^3-\omega_{ho}^2 b$, which admits undamped solutions of the form $b(t)=\sqrt{A\cos(2\omega_{ho}t+\phi)+B}$. Here, $A=\sqrt{B^2-1}$ and $B$ ($\ge 1$) is proportional to the energy of the system (at equilibrium $B=1$).  In a BEC gas a similar result holds for the radial oscillation in cylindrical geometry (Pitaevskii, 1996; Kagan, Surkov and Shlyapnikov, 1996) and the corresponding mode was experimentally investigated by Chevy {\it et al.} (2002). 

In the case of axisymmetric trapping ($\omega_x=\omega_y\equiv \omega_\perp\ne \omega_z$) the third component $\hbar m$ of the angular momentum is still a good quantum number and one also finds simple solutions of Eq.~(\ref{linearized}). The dipole modes, corresponding to the center of mass oscillation, have frequencies $\omega_\perp$ ($m=\pm1$) and $\omega_z$ ($m=0$). Another surface solution is the radial quadrupole mode ($m=\pm 2$) characterized by the velocity field ${\bf v}\propto\nabla(x\pm iy)^2$ and by the frequency 
\begin{equation}
\omega(m=\pm 2)=\sqrt2\omega_\perp \;,
\label{m=2}
\end{equation}
independent of the equation of state. This collective mode has been recently measured by Altmeyer {\it et al.} (2007b). The experimental results (see Fig.~\ref{fig7.5}) show that while the agreement with the theoretical prediction (\ref{m=2}) is very good on the BEC side, including unitarity, as soon as one enters the BCS regime strong damping and deviations from the hydrodynamic law take place, suggesting that the system is leaving the superfluid phase. For small and negative $a$ the system is actually expected to enter a collisionless regime where the frequency of the $m=\pm 2$ quadrupole mode is $2\omega_\perp$ apart from small mean-field corrections (Vichi and Stringari, 1999). Another class of surface collective oscillations, the so called scissors mode, will be discussed in Sec.~\ref{Sec8.1}, due to its relevance for the rotational properties of the system.

Differently from the surface modes, the $m=0$ compressional modes depend instead on the equation of state. For a polytropic dependence of the chemical potential ($\mu \propto n^\gamma$) the corresponding solutions can be derived analytically. They are characterized by a velocity field of the form ${\bf v}\propto\nabla[a(x^2 +y^2)+bz^2]$, resulting from the coupling between the $\ell=2$ and $\ell=0$ modes caused by the deformation of the trap (Cozzini and Stringari, 2003). In the limit of elongated traps ($\omega_z\ll \omega_\perp$) the  two solutions  reduce to the radial [$\omega=\sqrt{2(\gamma+1)}\omega_\perp$] and axial  [$\omega=\sqrt{(3\gamma+2)/(\gamma+1)}\omega_z$] breathing modes.
Experimentally both  modes have been investigated (Kinast {\it et al.}, 2004; Bartenstein {\it et al.}, 2004b; Altmeyer {\it et al.}, 2007a). In Fig.~\ref{fig7.6} we show the most recent experimental results (Altmeyer {\it et al.}, 2007a) for the radial breathing mode. At unitarity the agreement between theory ($\sqrt{10/3}\omega_\perp=1.83\omega_\perp$) and experiment is remarkably good. It is also worth noticing that the damping of the oscillation is smallest near unitarity.

When we move from unitarity the collective oscillations exhibit other interesting features. Theory predicts that in the deep BEC regime $(\gamma=1$) the frequencies of both the axial and radial modes are higher than at unitarity. Furthermore, the first corrections with respect to the mean-field prediction can be calculated analytically, by accounting for the first correction to the equation of state $\mu_d=g_dn_d$ produced by quantum fluctuations. This is the so called Lee-Huang-Yang (LHY) correction [see Eq.~(\ref{enerbeyond})] first derived in the framework of Bogoliubov theory of interacting bosons. The resulting shifts in the collective frequencies are positive (Pitaevskii and Stringari, 1998; Braaten and Pearson, 1999). As a consequence, when one moves from the BEC regime towards unitarity, the dispersion law exhibits a typical non monotonic behavior, since it first increases, due to the LHY effect, and eventually decreases to reach the lower value $\sqrt{10/3}\omega_\perp$ at unitarity (Stringari, 2004). 

In general, the collective frequencies can be calculated numerically along the  crossover by solving the hydrodynamic equations once the equation of state is known (Combescot and Leyronas, 2002; Hu {\it et al.}, 2004; Heiselberg, 2004; Kim and Zubarev, 2004; Bulgac and Bertsch, 2005; Manini and Salasnich, 2005; Astrakharchick {\it et al.}, 2005b). Fig.~\ref{fig7.6} shows the predictions (Astrakharchik {\it et al.}, 2005b) obtained using the equation of state of the Monte Carlo simulations discussed in Sec.~\ref{Sec5.2} and of the BCS mean-field theory of Sec.~\ref{Sec5.1}. The Monte Carlo equation of state accounts for the LHY effect while the mean-field BCS theory misses it, providing a monotonic behavior for the compressional frequency as one moves from the BEC regime to unitarity. The accurate measurements of the radial compression mode reported in Fig.~\ref{fig7.6} confirm the Monte Carlo predictions, providing an important test of the equation of state and first evidence for the LHY effect. Notice that the LHY effect is visible only at the lowest temperatures (Altmeyer {\it et al.}, 2007a) where quantum fluctuations dominate over thermal fluctuations (Giorgini, 2000). This explains why previous measurements of the breathing mode failed in revealing the enhancement of the collective frequency above the value $2\omega_\perp$.

The behavior of the breathing modes on the BCS side of the resonance exhibits different features. Similarly to the case of the quadrupole mode (see Fig.\ref{fig7.5}) one expects that the system soon looses superfluidity and eventually behaves like a dilute collisionless gas whose collective frequencies are given by $2\omega_\perp$ and $2\omega_z$ for the radial and axial modes respectively. Experimentally this transition is clearly observed for the radial mode (Kinast {\it et al.}, 2004, Bartenstein {\it et al.}, 2004b), where it occurs at about $k_F|a|\sim 1$. It is also associated with a strong increase of the damping of the collective oscillation.

The temperature dependence of the collective oscillations has also been the object of experimental investigations. Kinast, Turlapov and Thomas (2005) have shown that the frequency of the radial compression mode, measured at unitarity, remains practically constant when one increases the temperature, suggesting that the system is governed by the same hydrodynamic equations both at the lowest temperatures, when the gas is superfluid, and at higher temperature when it becomes normal. Conversely, the damping exhibits a significant temperature dependence, becoming smaller and smaller as one lowers the temperature. This behavior strongly supports the superfluid nature of the system in the low temperature regime. In fact a normal gas is expected to be less and less hydrodynamic as one decreases $T$ with the consequent {\it increase} of the damping of the oscillation. The accurate determination of the damping of the collective oscillations can provide useful information on the viscosity coefficients of the gas whose behavior at unitarity has been the object of recent theoretical investigations using universality arguments [see Son (2007) and references therein].

\begin{figure*}
\scalebox{.75}{\includegraphics[viewport=-30 -50 590 425,width=0.4\textwidth,height=5.3cm]{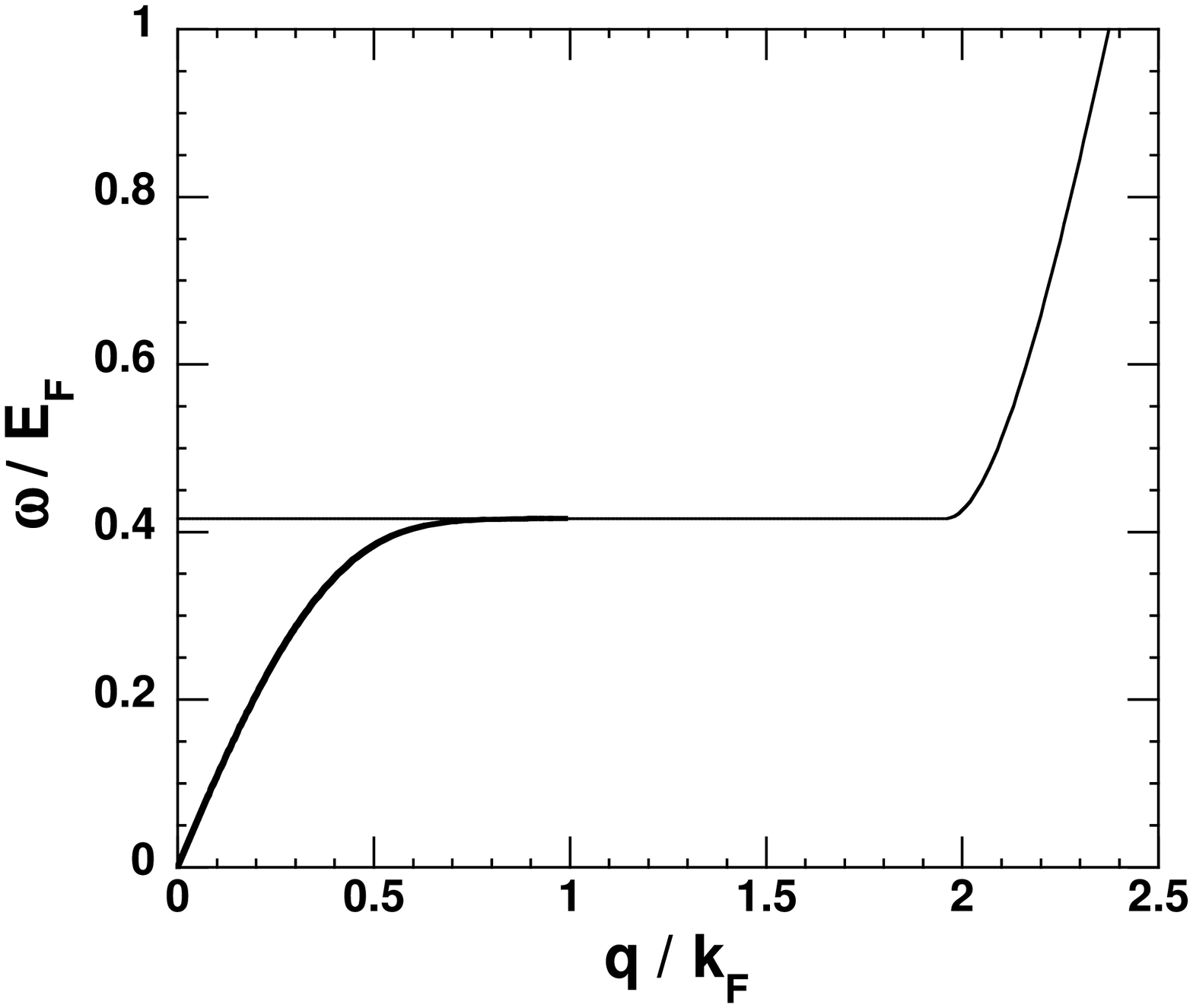}
\includegraphics[width=0.4\textwidth]{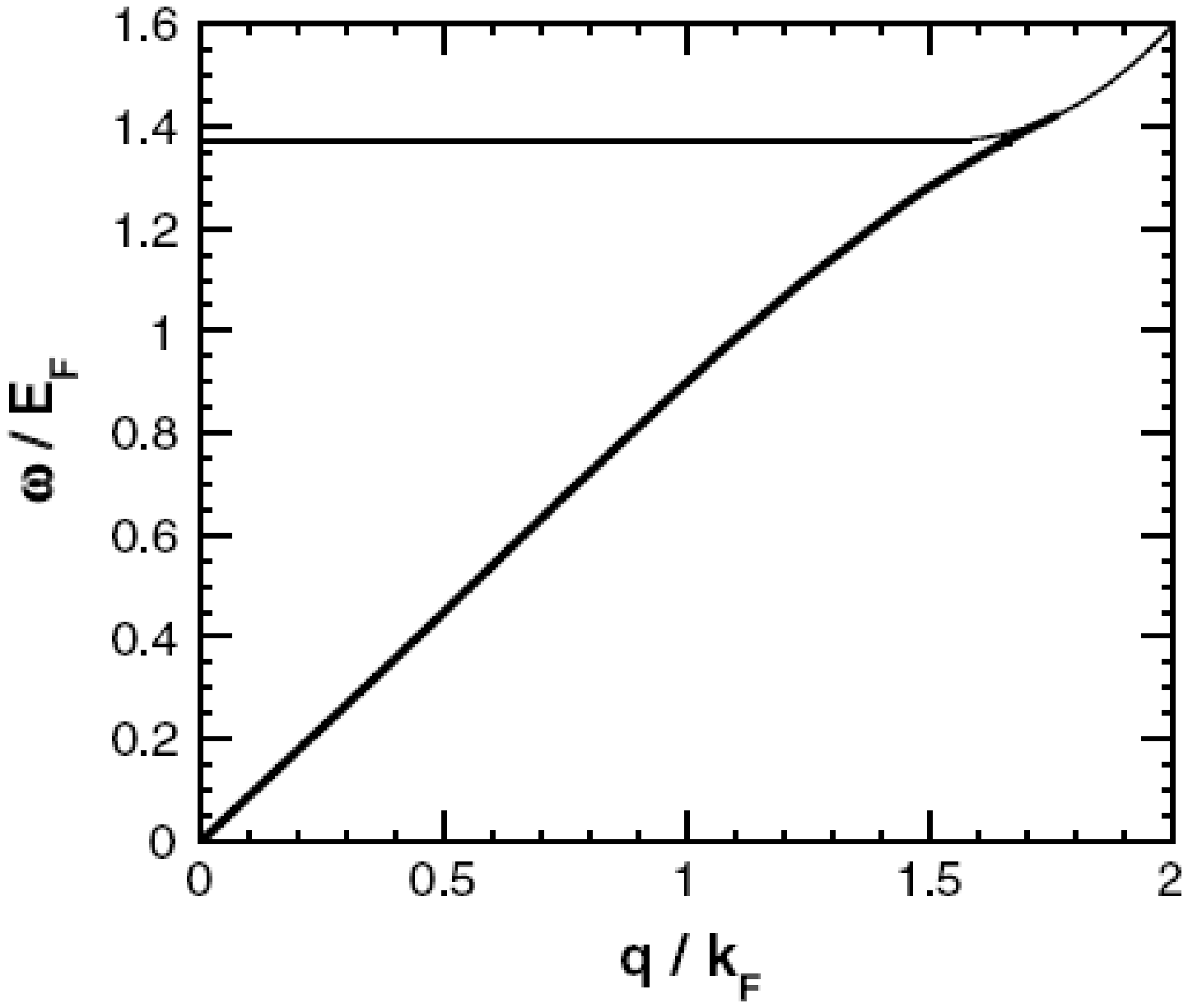}\includegraphics[width=0.4\textwidth]{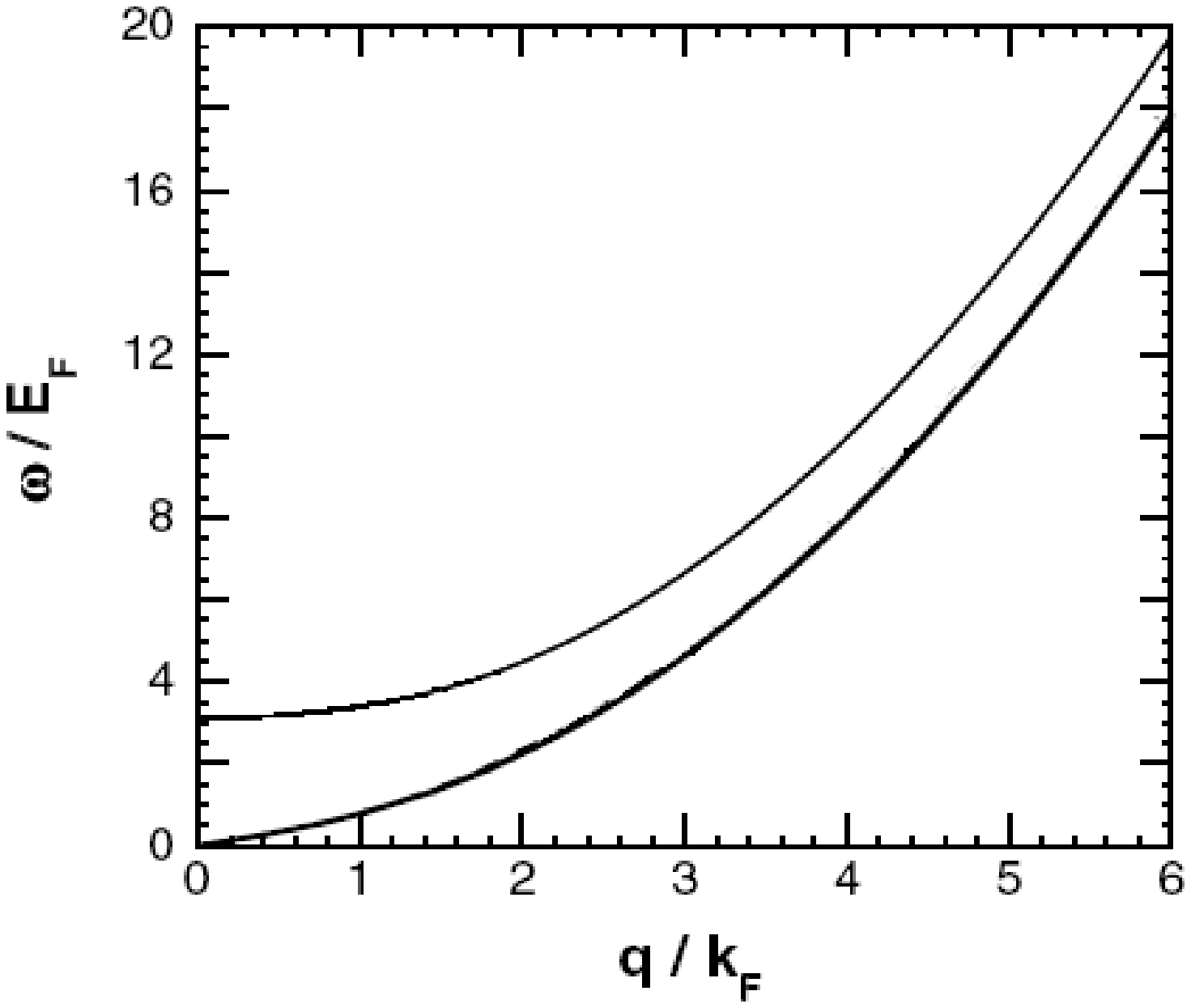}}
\caption{Spectrum of collective excitations of the superfluid Fermi gas along the BCS-BEC crossover (thick lines). Energy is given in units of the Fermi energy. Left: BCS regime ($k_Fa=-1$). Center: unitarity. Right: BEC regime ($k_Fa=+1$). The thin lines denote the threshold of single-particle excitations. From Combescot, Kagan and Stringari (2006).}
\label{fig7.7}
\end{figure*}

\begin{figure}[b]
\begin{center}
\includegraphics[width=7.5cm]{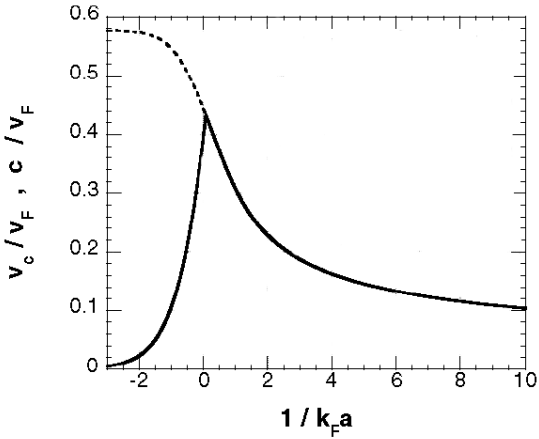}
\caption{Landau's critical velocity (in units of the Fermi velocity) calculated along the crossover using BCS mean field theory. The figure clearly shows that the critical velocity is largest near unitarity. The dashed line is the sound velocity. From Combescot, Kagan and Stringari (2006).}
\label{fig7.8}
\end{center}
\end{figure}

\begin{figure}[b]
\begin{center}
\includegraphics[width=7.0cm]{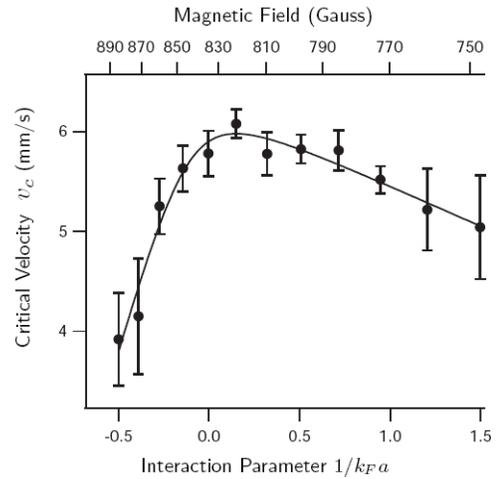}
\caption{Measured critical velocity along the BCS-BEC crossover. The solid line is a guide to the eye. From Miller {\it et al.} (2007).}
\label{fig7.9}
\end{center}
\end{figure}

\subsection{Phonons vs pair-breaking excitations and Landau's critical velocity} \label{Sec7.4}

In Sec.~\ref{Sec7.3} we have described the discretized modes predicted by hydrodynamic theory in the presence of harmonic trapping. This theory describes correctly only the low-frequency oscillations of the system, corresponding to sound waves in a uniform body. When one considers higher excitations energies the dynamic response should also include dispersive corrections to the phononic branch and the breaking of pairs into two fermionic excitations. The general picture of the excitations produced by a density probe in the superfluid state can then be summarized as follows (for simplicity we consider a uniform gas): at low frequency the system exhibits a gapless phononic branch whose slope  is fixed by the sound velocity; at high frequency one expects the emergence of a continuum of excitations starting from a given threshold, above which one can break pairs. The value of the threshold frequency depends on the momentum $\hbar{\bf q}$ carried by the perturbation. A first estimate is provided by the BCS mean-field theory discussed in Sec.~\ref{Sec5.1}, according to which the threshold is given by the line $\omega_{thr}={\text Min}_{\bf k} (\epsilon_++\epsilon_-)$, where $\epsilon_{\pm}$ is the energy (\ref{elemexcit}) of a quasi-particle excitation carrying momentum $\hbar({\bf q}/2\pm {\bf k})$. The pair then carries total momentum $\hbar{\bf q}$. It is easy to see that the minimum is obtained for ${\bf k \cdot q}=0$ which gives $\epsilon_+=\epsilon_-$. One can distinguish two cases: $\mu >0$ and $\mu < 0$. In the first case (including the unitary as well as the BCS regimes) and for $\hbar q< 2\sqrt{2m\mu}$, the minimum is at $\hbar^2 k^2/2m = \mu -\hbar^2q^2/8m$, so that $\hbar \omega_{thr} =2\Delta$. For $\mu >0$ and $\hbar q>2\sqrt{2m\mu}$, as well as for $\mu <0$ (including the BEC regime), the minimum is at $k=0$, which leads to $\hbar\omega_{thr}=2\sqrt{(\hbar^2 q^2/8m-\mu)^2+\Delta^2}$. 

The interplay between phonon and pair breaking excitations gives rise to different scenarios along the crossover. In the BCS regime the threshold occurs at low frequencies and the phonon branch very soon reaches the continuum of single-particle excitations. The behavior is quite different in the opposite BEC regime where the phonon branch extends up to high frequencies. At large momenta this branch actually looses its phononic character and approaches the dispersion $\hbar^2q^2/4m$, typical of a free molecule. In the deep BEC limit it coincides with the Bogoliubov spectrum of a dilute gas of bosonic molecules. At unitarity the system is expected to exhibit an intermediate behavior, the discretized branch surviving up to momenta of the order of the Fermi momentum. A detailed calculation of the excitation spectrum, based on a time-dependent formulation of the BCS mean-field equations is reported in Fig.~\ref{fig7.7} (Combescot, Kagan and Stringari, 2006).

The results for the excitation spectrum provide a useful insight on the superfluid behavior of the gas in terms of the Landau criterion according to which a system can not give rise to energy dissipation if its velocity, with respect to a container at rest, is smaller than Landau's critical velocity defined by the equation
\begin{equation}
v_{c} = min_q(omega_q/ q)  \;,
\label{criticalv}
\end{equation}
where $\hbar\omega_q$ is the energy of an excitation carrying momentum $\hbar{\bf q}$. According to this criterion the ideal Fermi gas is not superfluid because of the absence of a threshold for the single-particle excitations, yielding $v_{c}=0$. The interacting Fermi gas  is instead superfluid in all regimes. By inserting the results for the threshold frequency derived above into Eq.~(\ref{criticalv}) one can calculate the critical value $v_c$ due to pair breaking. The result is
\begin{equation}
m(v_{c}^{sp})^2= \sqrt{\Delta^2+\mu^2} -\mu  \;,
\label{vcrsp}
 \end{equation}
and coincides, as expected, with the critical velocity calculated by applying directly the Landau criterion (\ref{criticalv}) to the single-particle dispersion law $\epsilon_q$ of Eq.~(\ref{elemexcit}). In the deep BCS limit $k_F|a|\to 0$, corresponding to $\Delta\ll \mu$, Eq.~(\ref{vcrsp}) approaches the exponentially small value $v_{c}=\Delta/\hbar k_F$. On the BEC side of the crossover the value (\ref{vcrsp}) becomes instead larger and larger and the relevant excitations giving rise to Landau's instability are no longer single-particle excitations but phonons and the critical velocity coincides with the sound velocity: $v_{c}=c$. A simple estimate of the critical velocity along the whole crossover is then given by the expression 
\begin{equation}
v_{c}= {\text Min}\left(c,v_{c}^{sp}\right) \; .
\label{vcrcrossover}
\end{equation}
Remarkably one sees that $v_{c}$ has a maximum near unitarity (see Fig.~\ref{fig7.8}), further confirming the robustness of superfluidity in this regime.  This effect has been recently demonstrated experimentally by moving a one dimensional optical lattice in a trapped superfluid Fermi at tunable velocity (see Fig.~\ref{fig7.9}) (Miller {\it et al.}, 2007).

The critical velocity permits to provide a general definition of the healing length according to $\xi=\hbar/m v_{c}$. 
Apart from an irrelevant numerical factor it coincides with the usual definition $\hbar/\sqrt{2m\mu_d}$ of the healing length in the BEC regime and with the size of Cooper pairs in the opposite BCS limit as already discussed in Sec.~\ref{Sec7.1}. The healing length provides the typical length above which the dynamic description of the system is safely described by the hydrodynamic picture. It is smallest near unitarity (Pistolesi and Strinati, 1996), where it is on the order of the interparticle distance.

\subsection{Dynamic and static structure factor} \label{Sec7.5}

The dynamic structure factor provides an important characterization of quantum many-body systems (see, for example, Pines and Nozi\`eres, 1966; Pitaevskii and Stringari, 2003). At low energy transfer it gives information on the spectrum of collective oscillations, including the propagation of sound, while at higher energies it is sensitive to the behavior of single-particle excitations. In general, the dynamic structure factor is measured through inelastic scattering experiments in which the probe particle is weakly coupled to the many-body system so that the scattering may be described within the Born approximation. 
In dilute gases it can be accessed with stimulated light scattering by using two photon Bragg spectroscopy (Stamper-Kurn {\it et al.}, 1999). The  dynamic structure factor is defined by the expression 
\begin{equation}
S({\bf q},\omega)= Q^{-1}\sum_{m,n}e^{-\beta E_{mn}}|\langle0|\delta\hat{\rho}_{q}|n\rangle|^2\delta(\hbar\omega-\hbar\omega_{mn}) \;,
\label{Sqomega}
\end{equation}
where $\hbar{\bf q}$ and $\hbar\omega$ are, respectively, the momentum and energy transferred by the probe to the sample, $\delta\hat\rho_{\bf q}=\hat{\rho}_{\bf q}-\langle\hat{\rho}_{\bf q}\rangle$ is the fluctuation of the Fourier component  $\hat\rho_q=\sum_j\exp(-i{\bf q}\cdot{\bf r}_j)$ of the density operator, $\omega_{mn}\equiv E_{mn}/\hbar=(E_m-E_n)/\hbar$ are the usual Bohr frequencies and $Q$ is the partition function. The definition of the dynamic structure factor is immediately generalized to other excitation operators like, for example, the spin density operator.

The main features of the dynamic structure factor are best understood in uniform matter, where the excitations are classified in terms of their momentum. From the results of the Sec.~\ref{Sec7.4} one expects that, for sufficiently small momenta, the dynamic structure factor be characterized by a sharp phonon peak and by a continuum of single-particle excitations above the threshold energy $\hbar\omega_{thr}$.  Measurements of the dynamic structure factor in Fermi superfluids can then provide unique information on the gap parameter. Theoretical calculations of the dynamic structure factor in the small $q$ regime were carried out using a proper dynamic generalization of the BCS mean-field approach (Minguzzi, Ferrari and Castin, 2001; B\"uchler, Zoller and Zwerger, 2004). At higher momentum transfer the behavior will depend crucially on the regime considered. In fact, for values of $q$ of the order of the Fermi momentum  the discretized branch no longer survives on the BCS side of the resonance. At even higher momenta the theoretical calculations of Combescot, Giorgini and Stringari (2006) have revealed that on the BEC side of the resonance the response is dominated by a discretized peak correponding to the excitation of free molecules with energy $\hbar^2q^2/4m$. This molecular-like peak has been shown to survive even at unitarity. On the BCS side of the resonance, the molecular signatures are instead completely lost at high momenta and the response is very similar to the one of an ideal Fermi gas.

From the knowledge of the dynamic structure factor one can evaluate the static structure factor, given by 
\begin{equation}
S(q)={\hbar\over N}\int_0^{\infty}d\omega S({\bf q},\omega)= 1+n\int d{\bf r}[g(r)-1]e^{-i{\bf q \cdot r}} 
\label{staticfactor}
\end{equation}
showing that the static structure factor is directly related to the two-body correlation function $g=(g_{\uparrow\uparrow}+g_{\uparrow\downarrow})/2$ discussed in Sec.~\ref{Sec5.2}. The measuremnt of $S(q)$ would then provide valuable information on the correlation effects exhibited by these systems. The behavior of the static structure factor along the BCS-BEC crossover has been calculated by Combescot, Giorgini and Stringari (2006).

\begin{figure}[b]
\begin{center}
\includegraphics[width=8.6cm]{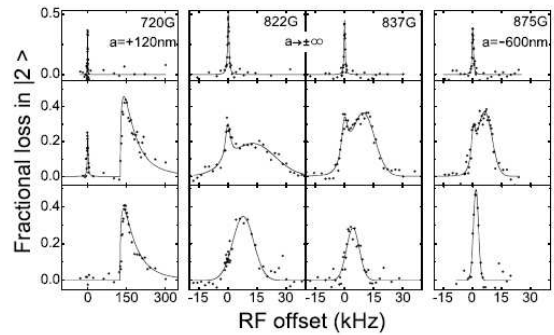}
\caption{RF spectra for various magnetic fields along the BCS-BEC crossover and for different temperatures. The RF 
offset $\delta\nu=\nu-\nu_{23}$ is given relative to the atomic transition $2\to3$. The molecular limit is realized for 
$B$=720 G (first column). The resonance regime is studied for $B$=822 G and 837 G (second and third column). The data 
at 875 G (fourth column) correspond to the BCS side of the crossover. Upper row, signals of unpaired atoms at high 
temperature (larger than $T_F$); middle row, signals for a mixture of unpaired and paired atoms at intermediate 
temperature (fraction of $T_F$); lower row, signals for paired atoms at low temperature (much smaller than $T_F$). From Chin 
{\it et al.} (2004).}
\label{fig7.11}
\end{center}
\end{figure}

\begin{figure}[b]
\begin{center}
\includegraphics[width=8.5cm]{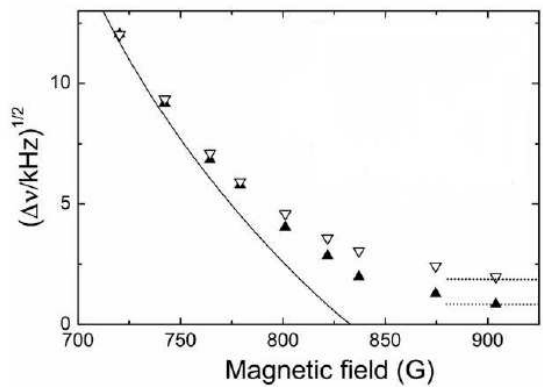}
\caption{Frequency shift $\delta\nu_{max}$ measured in $^6$Li at low temperature as a function of the magnetic field for 
two different configurations corresponding to $T_F$= 1.2 $\mu$K (filled symbols) and 3.6 $\mu$K (open symbols). The solid 
line shows the value $\delta\nu_{max}$ predicted in the free molecular regime where it is essentially given by the molecular binding energy. From Chin {\it et al.} (2004).}
\label{fig7.12}
\end{center}
\end{figure}

\subsection{Radiofrequency transitions} \label{Sec7.6}

In Secs.~\ref{Sec7.4}-\ref{Sec7.5} we have shown that pair-breaking transitions characterize the excitation spectrum associated to the density-density response function and are directly visible in the dynamic structure factor. Information on pair-breaking effects is also provided by transitions which outcouple atoms to a third internal state (T\"orm\"a and Zoller, 2000). Experiments using radiofrequency (RF) excitations have already been performed in Fermi superfluids (Chin {\it et al.}, 2004) in different conditions of temperature and magnetic fields. The basic idea of these experiments is the same as for the measurement of the binding energy of free molecules (see Sec.~\ref{Sec3.1}). 

The structure of the  RF transitions is determined by the Zeeman diagram of the hyperfine states in the presence of an 
external magnetic field.  Starting from a sample where two hyperfine states (hereafter called 1 and 2) are occupied, one 
considers single-particle transitions from the state 2 (with energy $E_2$) to a third, initially unoccupied, state 3 with 
energy $E_3$. The typical excitation operator characterizing these RF transitions has the form $V_{RF}=\lambda\int d{\bf r}[\hat{\psi}_3^\dagger({\bf r})\hat{\psi}_2({\bf r})+ h.c.]$ and does not carry any momentum. The experimental signature of the transition is given by the appearence of atoms in the state 3 or in the reduction of the number of atoms in the state 2. In the absence of interatomic forces  the transition is resonant at the frequency  $\nu_{23}=(E_3-E_2)/h$. If instead the two atoms  interact and  form a molecule the frequency required to induce the transition is higher since part of the energy carried by the radiation is needed to break the molecule. The threshold for the transition is given by the frequency $\nu= \nu_{23}+|\epsilon_b|/h$, where $|\epsilon_b|\sim\hbar^2/ma^2$ is the binding energy of the molecule (we are assuming here that no molecule is formed in the 1-3 channel). The actual dissociation line shape  is determined by the overlap of the molecular state with the continuum and is affected by the final-state interaction between the atom occupying the state 3 and the atom occupying the state 1. In particular, the relationship between the value of the frequency where the signal is maximum and the threshold frequency depends on the value of the scattering length $a_{31}$  characterizing the interaction between the state 3 and the state 1 (Chin and Julienne, 2005). 

For interacting many-body configurations along the BCS-BEC crossover a similar scenario takes place. Also in this case breaking pairs costs energy so that the frequency shifts of the RF transitions provide information on the behavior of the gap, although this information is less direct than in the case of free molecules. The ideal situation would take place if the interaction between the final state 3 and the states 1 and 2 were negligible. In this case the threshold frequency is provided by the simple formula
\begin{equation}
h\delta \nu \equiv h(\nu- \nu_{23})= \epsilon_{k=0} -\mu \;,
\label{threshold}
\end{equation}
where we have considered the most favourable case where the RF transition excites single-particle states with zero momentum. Here $\mu$ is the chemical potential while $\epsilon_k$ is the quasi-particle excitation energy defined in Sec.~\ref{Sec4.2}. Eq.~(\ref{threshold}) reproduces the result $h\delta\nu =|\epsilon_b|$ in the deep BEC limit of free molecules where $\mu$ approaches the value $-|\epsilon_b|/2$. In the opposite BCS regime, where $\mu\to E_F$ and the single-particle gap $\Delta_{\text gap}$ is much smaller than the Fermi energy, one has instead $h\delta\nu=\Delta_{\text gap}^2/2E_F$. The proper inclusion of final-state interactions in the calculation of the RF spectra is a difficult problem that has been the object of several papers (Kinnunen, Rodriguez and T\"orm\"a, 2004; He, Chen and Levin, 2005; Ohashi and Griffin, 2005; Yu and Baym, 2006; Perali, Pieri and Strinati, 2007). 

Typical experimental results on $^6$Li are shown in Fig.~\ref{fig7.11}, where the observed lineshapes are presented for 
different values of the temperature along the BCS-BEC crossover. In $^6$Li the relevant scattering lengths $a_{13}$ ($<0$) 
and $a_{12}$ are both large in modulus and final-state interactions can not be ignored. On the BEC side of the resonance (lowest magnetic field in Fig.~\ref{fig7.11}) one recognizes the emergence of the typical molecular lineshape at low temperature with 
the clear threshold effect for the RF transition.  

In the most interesting region where $k_F|a|\sim1$ many-body effects become important and change the scenario of the RF transitions. While at high temperature (upper row in Fig.~\ref{fig7.11}) the measured spectra still reveal the typical feature 
of the free atom transition, at lower temperatures the lineshapes are modified by interactions in a nontrivial way. This is 
shown in Fig.~\ref{fig7.12} where the shift $\delta\nu_{max}=\nu_{max}-\nu_{23}$, defined in terms of the frequency $\nu_{max}$ where the RF signal is maximum, is reported for two different values of $T_F$. In the deep BEC regime the value of $\delta\nu_{max}$ is independent of the density and directly related to the binding energy of the molecule. Viceversa, at unitarity and on the BCS side it shows a clear density dependence. From these data one extracts the relationship $h\delta\nu_{max}\sim 0.2 E_F$ at unitarity. The dependence on the density is even more dramatic in the BCS regime, due to the exponential decrease of pairing effects as $k_F|a|\to 0$. 

Since pairing effects are density dependent and become weaker and weaker as one approaches the border of the atomic cloud, one can not observe any gap in Fig.~\ref{fig7.11} except on the BEC side of the resonance where the gap is density independent. Spatially resolved RF spectroscopy has been recently become available at unitarity (Shin {\it et al.}, 2007) revealing in a clear way the occurrence of the gap and hence opening new perspectives for a direct comparison with the theoretical predictions in uniform matter.

\section{ROTATIONS AND SUPERFLUIDITY} \label{Chap8}

Superfluidity shows up in spectacular rotational properties. In fact a superfluid can not rotate like a rigid body, due to the irrotationality constraint (\ref{vgrad}) imposed by the existence of the order parameter. At low angular velocity an important macroscopic  consequence of superfluidity is the quenching of the moment of inertia. At higher angular velocities the superfluid can instead carry angular momentum via the formation of vortex lines. The circulation around these vortex lines is quantized. When many vortex lines are created a regular vortex lattice is formed and the angular momentum acquired by the system approaches the classical rigid-body value. Both the quenching of the moment of inertia and the formation of vortex lines have been the object of fundamental investigations in the physics of quantum liquids and have been recently explored in a systematic way also in  dilute Bose-Einstein condensed gases. In this Section we summarize some of the main rotational features exhibited by dilute Fermi gases where first experimental results are already available. In Secs.~\ref{Sec8.1} and \ref{Sec8.2} we discuss the consequences of the irrotationality constraint on the moment of inertia, on the collective oscillations and on the expansion of a rotating gas, while Sec.~\ref{Sec8.3} is devoted to discuss some key properties of quantized vortices and vortex lattices.

\subsection{Moment of inertia and scissors mode} \label{Sec8.1}

The moment of inertia $\Theta$ relative to the $z$ axis is defined as the response of the system to a rotational field $-\Omega\hat{L}_z$ according to $\langle\hat{L}_z\rangle = \Omega \Theta$ 
where $\hat{L}_z$ is the $z$ component of the angular momentum operator and the average is taken on the stationary configuration in the presence of the perturbation.  For a non-interacting gas trapped by a deformed harmonic potential the moment of inertia  can be calculated explicitly. In the small $\Omega$ limit (linear response) one finds the result (Stringari 1996a)  
\begin{eqnarray}
\Theta &=&{mN  \over \omega^2_x-\omega^2_y} \left[(\langle y^2\rangle-\langle x^2\rangle)(\omega^2_x+\omega^2_y) \right. \nonumber \\
 &+& \left. 2(\omega^2_y\langle y^2\rangle-\omega^2_x\langle x^2\rangle) \right] \;.
\label{ThetaS}
\end{eqnarray}
where the expectation values should be evaluated in the absence of rotation.
This result applies both to bosonic and fermionic ideal gases. It assumes $\omega_x\neq \omega_y$, but admits a well defined limit when $\omega_x\to \omega_y $. In the ideal Fermi gas, when the number of particles is large, one can use the Thomas-Fermi relationships  $\langle x^2\rangle\propto 1/\omega^2_x$ and $\langle y^2\rangle \propto 1/\omega_x^2$ for the radii. In this case Eq.~(\ref{ThetaS}) reduces to the rigid value of the moment of inertia:
\begin{equation}
\Theta_{rig}=Nm\langle x^2+y^2\rangle \; .
\label{Thetarig}
\end{equation}
For a non-interacting Bose-Einstein condensed gas at $T=0$, where the radii scale according to $\langle x^2\rangle\propto 1/\omega_x$ and $\langle y^2\rangle\propto 1/\omega_x$, one instead finds that $\Theta \to 0$ as $\omega_x\to \omega_y$. 

Interactions  change the value of the moment of inertia of a Fermi gas in a profound way. To calculate $\Theta$ in the superfluid phase one can use the equations of irrotational hydrodynamics developed in Sec.~\ref{Chap7}, by considering a trap rotating with angular velocity $\Omega$ and looking for the stationary solution in the rotating frame. The equations of motion in the rotating frame are obtained by including the term $-\Omega\hat{L}_z$ in the Hamiltonian. In this frame the trap is described by the time independent harmonic potential $V_{ho}$ of Eq.~(\ref{Vho}) and the hydrodynamic equations admit stationary solutions characterized by the velocity field 
\begin{equation}
{\bf v}=-\delta \Omega \nabla(xy) \;,
\label{valpha}
\end{equation}
where ${\bf v}$ is the irrotational superfluid velocity in the laboratory frame, while $\delta = \langle y^2-x^2\rangle/\langle y^2+x^2\rangle$ is the deformation of the atomic cloud in the $xy$ plane. By evaluating the angular momentum $\langle\hat{L}_z\rangle= m\int d{\bf r} ({\bf r \times v})n$ one  finds that the moment of inertia takes the irrotational form 
\begin{equation}
\Theta = \delta^2 \Theta_{rig} \; ,
\label{Thetairrot}
\end{equation}
which identically vanishes for an axisymmetric configuration, pointing out the crucial role  played  by superfluidity. The result for the moment of inertia holds for both Bose and Fermi superfluids rotating in a harmonic trap. 

The behavior of the moment of inertia at high angular velocity was investigated in the case of BEC's by Recati, Zambelli and Stringari (2001) who found that, for angular velocities larger than $\omega_\perp/\sqrt2$ where $\omega^2_\perp=(\omega^2_x+\omega^2_y)/2$, the adiabatic increase of the rotation can sizably affect the value of the deformation parameter $\delta$ yielding large deformations even if the deformation of the trap is small. Physically this effect is the consequence of the energetic instability of the quadrupole oscillation. At even higher angular velocites a dynamic instability of the rotating configuration was predicted by Sinha and Castin (2001) suggesting a natural route to the nucleation of vortices. These theoretical predictions were confirmed experimentally (Madison {\it et al.}, 2001). Similar predictions have been made also for rotating Fermi gases (Tonini, Werner and Castin, 2006).

\begin{figure}[b]
\begin{center}
\includegraphics[width=8.5cm]{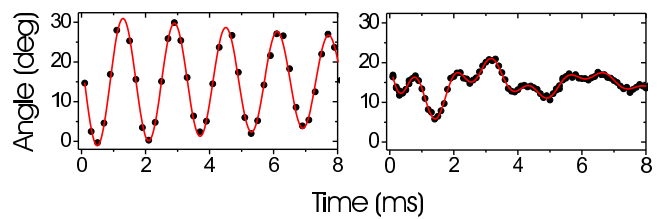}
\caption{Time evolution of the angle characterizing the scissors mode in a Fermi gas at unitarity (left panel) and on the BCS side of the resonance (right panel). The measured frequencies well agree with the theoretical predictions (see text). From Wright {\it et al.} (2007).} 
\label{fig8.1}
\end{center}
\end{figure}

The irrotational nature of the moment of inertia has important consequences on the behavior of the so called {\it scissors mode}. This is an oscillation of the system caused by the sudden rotation of a deformed trap which, in the superfluid case, has frequency (Guery-Odelin and Stringari, 1999)
\begin{equation}
\omega =  \sqrt{\omega^2_x+\omega_y^2} \; .
\label{scissors}
\end{equation}
This result should be compared with the prediction of the normal gas in the collisionless regime where two modes with frequencies   $\omega_{\pm}=|\omega_x\pm\omega_y|$ are found. The occurrence of the low frequency mode $|\omega_x-\omega_y|$ reflects the rigid value of the moment of inertia in the normal phase. The scissors mode, previously observed in a Bose-Einstein condensed gas (Marag\`o {\it et al.}, 2000), has been recently investigated also in ultracold Fermi gases (see Fig.~\ref{fig8.1}) along the BCS-BEC crossover (Wright {\it et al.}, 2007). At unitarity and on the BEC side of the resonance one clearly observes the hydrodynamic oscillation, while when $a$ becomes negative and small the beating between the frequencies $\omega_{\pm}=|\omega_x\pm\omega_y|$ reveals the transition to the normal collisionless regime. If the gas is normal, but deeply collisional as happens at unitarity above the critical temperature, classical hydrodynamics predicts an oscillation with the same frequency (\ref{scissors}) in addition to a low frequency mode of diffusive nature caused by the viscosity of the fluid. This mode, however, is located at too low frequencies to be observable. The persistence of the scissors frequency (\ref{scissors}) has been observed at unitarity in the recent experiment of Wright {\it et al.} (2007) even above $T_c$. This result, together with the findings for the aspect ratio of the expanding gas and for the radial compression mode (see discussion in Secs.~\ref{Sec7.2}-\ref{Sec7.3}), confirms that near resonance the gas behaves hydrodynamically in a wide range of temperatures below and above the critical temperature. This makes the distinction between the superfluid and the normal phase based on the study of the collective oscillations, a difficult task.

Promising perspectives to distinguish between superfluid and classical (collisional) hydrodynamics are provided by the study of the collective oscillations excited in a rotating gas.  In fact, in the presence of vorticity $\nabla \times {\bf v} \neq 0$, the equations of collisional hydrodynamics contain an additional term depending on the curl of the velocity field, which is absent in the equations of superfluid hydrodynamics. The resulting consequences on the behavior of the scissors mode have been discussed by Cozzini and Stringari (2003).

\begin{figure}[b]
\begin{center}
\includegraphics[width=8.5cm]{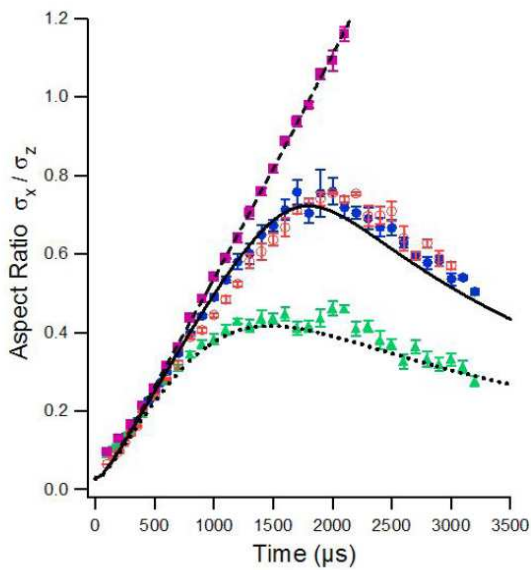}
\caption{(color online). Aspect ratio versus expansion time of a unitary gas for different values of the initial angular velocity $\Omega_0$ (in units of the frequency $\omega_z$ of the long axis of the trap) and of the temperature of the gas (parametrized by the energy per particle $E/N$). Squares (purple): no angular velocity; solid circles (blue): $\Omega_0/\omega_z=0.40$, $E/(NE_F)=0.56$; open circles (red): $\Omega_0/\omega_z=0.40$, $E/(NE_F)=2.1$; triangles (green): $\Omega_0/\omega_z=1.12$, $E/(NE_F)=0.56$. The dashed, solid and dotted lines are the results of calculations based on irrotational hydrodynamics. From Clancy, Luo and Thomas (2007).} 
\label{fig8.3}
\end{center}
\end{figure}

\subsection{Expansion of a rotating Fermi gas} \label{Sec8.2}

Another interesting consequence of the irrotational nature of the superfluid motion concerns the expansion of a rotating gas. Let us suppose that a trapped superfluid Fermi gas is initially put in rotation with a given value of angular velocity (in practice this can be realized through a sudden rotation of a deformed trap which excites the scissors mode). The gas is later released from the trap. In the plane of rotation the expansion along the short axis of the cloud will initially be faster than the one along the long axis, due to larger gradients in the density distribution. However, differently from the hydrodynamic expansion of a non rotating gas where the cloud takes at some time a symmetric shape, in the rotating case the deformation of the cloud can not vanish. In fact, due to the irrotational constraint, this would result in a vanishing value of the angular momentum  and hence in a violation of angular momentum conservation. The consequence is that the angular velocity of the expanding cloud will increase as the value of the deformation is reduced, but can not become too large because of energy conservation. The result is that the deformation parameter will acquire a minimum value during the expansion but will never vanish. This non trivial consequence of irrotationality was first predicted (Edwards {\it et al.}, 2002) and observed (Hechenblaikner {\it et al.}, 2002) in Bose-Einstein condensed atomic gases. Very recently the experiment has been repeated in a cold Fermi gas at unitarity (Clancy, Luo and Thomas, 2007). Fig.~\ref{fig8.3} reports the measured aspect ratio as a function of the expansion time for different initial values of the angular velocity. It shows that, if the gas is initially rotating (lower curves), the aspect ratio never reaches the value 1 corresponding to a vanishing deformation. The experimental data are very well reproduced by the solutions of the equations of irrotational hydrodynamics (solid and dotted line). A remarkable feature is that the same behavior for the aspect ratio is found not only in the superfluid regime but also above the critical temperature, revealing that even in the normal phase the dynamics of the expansion is described by the equations of irrotational hydrodynamics. The reason is that viscosity effects are very small in the normal phase and that the relevant time scales in this experiment are too short to generate a rigid component in the velocity field.

\begin{figure}[b]
\begin{center}
\includegraphics[width=8.5cm]{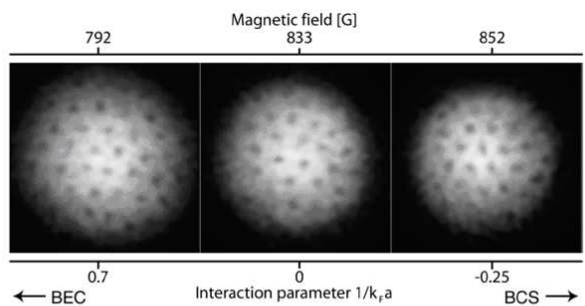}
\caption{Experimental observation of quantized vortices in a superfluid Fermi gas along the BCS-BEC crossover (Zwierlein {\it et al.}, 2005b).} 
\label{fig8.4}
\end{center}
\end{figure}

\subsection{Quantized vortices} \label{Sec8.3}

The existence of quantized vortices is one of the most important predictions of superfluidity. Recent experiments have confirmed  their existence also in ultracold Fermi gases along the BCS-BEC crossover (see Fig.~\ref{fig8.4}). In these experiments vortices are produced by spinning the atomic cloud with a laser beam and are observed by imaging the released cloud of molecules, which are stabilized through a rapid sweep of the scattering length to small and positive values during the first ms of the expansion. This technique, which is similar to the one employed to measure the condensate fraction of pairs (see Sec.~\ref{Sec5.2}), increases the contrast of the vortex cores and therefore their visibility. 

A quantized vortex along the $z$ axis is associated with the appearence of a phase in the order parameter [see Eqs.~(\ref{F})-(\ref{Fshortr})] of the form $\exp({i\phi})$ where $\phi$ is the azimuthal angle. This yields the complex form 
\begin{equation}
\Delta({\bf r})= \Delta(r_\perp,z)\exp(i\phi) \;,
\label{deltaphi}
\end{equation}
for the order parameter $\Delta$ where, for simplicity, we have assumed that the system exhibits axial symmetry and we have used cylindrical coordinates. The velocity field ${\bf v}=(\hbar /2m)\nabla \phi$ of the vortex configuration has a tangential form (${\bf v\cdot r}_\perp=0$) with modulus $v= \hbar /2m r_\perp$,
which increases as one approaches the vortex line, in contrast to the rigid body dependence ${\bf v}={\bf \Omega \times r}$ characterizing the rotation of a normal fluid. 
The circulation is quantized according to the rule
\begin{equation}
\oint {\bf v \cdot} d{\bf \ell} = {\pi \hbar \over m} \;,
\label{circulation}
\end{equation}
which is smaller by a factor $2$ with respect to the case of a Bose superfluid with the same atomic mass $m$. Vortices with higher quanta of circulation can also be considered. The value of the circulation is independent of the radius of the contour. This is a consequence of the fact that the vorticity is concentrated on a single line and hence deeply differs from the  vorticity  $\nabla {\bf \times v}=2{\bf \Omega}$ of the rigid body rotation.

The angular momentum carried by the vortex is given by the expression
\begin{equation}
\langle\hat{L}_z\rangle = m\int d{\bf r} ({\bf r \times v}) n({\bf r})= N{\hbar \over 2}  \;,
\label{Lv}
\end{equation}
holding if the vortex line coincides with the symmetry axis of the density profile. If the vortex is displaced towards the periphery of a trapped gas the angular momentum takes a smaller value. In this case the axial symmetry of the problem is lost and the order parameter can not be written in the form (\ref{deltaphi}).

A first estimate of the  energy of the vortex line is obtained using macroscopic arguments based on hydrodynamics and considering, for simplicity, a gas confined in a cylinder of radial size $R$. The energy $E_v$ acquired by the vortex is mainly determined by the hydrodynamic kinetic energy $(m/2)n\int d{\bf r}v^2$ which yields the following estimate for the vortex energy:
\begin{equation}
E_v = {N\hbar \over 4m R^2}\ln{R\over \xi} \;,
\label{EvM}
\end{equation}
where we have introduced the radius $\xi$ of the core of the vortex of the order of the healing length (see Sec.~\ref{Sec7.4}). The need for the inclusion of the core size $\xi$ in Eq.~(\ref{EvM}) follows from the logarithmic divergence of the integral $n\int d{\bf r}v^2$  at short radial distances. Eq.~(\ref{EvM}) can be used to evaluate the critical angular velocity $\Omega_c$ for the existence of an energetically stable vortex line. This value is obtained by imposing that the change in the energy $E-\Omega_c\langle\hat{L}_z\rangle$ of the system in the frame rotating with angular velocity $\Omega_c$ be equal to $E_v$. One finds $\Omega_c = (\hbar / 2m R^2)\ln(R/\xi)$. By applying this estimate to a harmonically trapped configuration with $R_{TF}\sim R$ and by neglecting the logarithmic term which provides a correction of order of unity, we find  $\Omega_c/\omega_\perp \simeq \hbar\omega_\perp /E_{ho}$ where $\omega_\perp$ is the radial frequency of the harmonic potential and $E_{ho} \sim m\omega^2_\perp R^2_{TF}$ is the harmonic oscillator energy. The above estimate shows that in the Thomas-Fermi regime, $E_{ho}\gg \hbar\omega_{\perp}$, the critical frequency is much smaller than the radial trapping frequency, thereby suggesting that vortices should be easily produced in slowly rotating traps. This conclusion, however, does not take into account the fact that the nucleation of vortices is strongly inhibited at low angular velocities by the occurrence of a barrier. For example in rotating Bose-Einstein condensates it has been experimentally shown that it is possible to increase the angular velocity of the trap up to values higher than $\Omega_c$ without generating vortical states. Under these conditions the response of the superfluid is governed by the equations of irrotational hydrodynamics (see Sec.~\ref{Sec8.1}). 

A challenging problem concerns the visibility of the vortex lines. Due to the smallness of the healing length, especially at unitarity, they can not be observed {\it in situ}, but only after expansion. Another difficulty is the reduced contrast in the density with respect to the case of Bose-Einstein condensed gases. Actually, while the order parameter vanishes on the vortex line the density does not, unless one works in the deep BEC regime. In the opposite BCS regime the order parameter is exponentially small and  the density profile is practically unaffected by the presence of the vortex. 

The explicit behavior of the density near the vortex line as well as the precise  calculation of the vortex energy requires the implementation of a microscopic calculation. This can be carried out following the lines of the mean-field BCS theory developed in Sec.~\ref{Sec5.1}. While this approach is approximate, it nevertheless provides a useful consistent description of the vortical structure along the whole crossover. 
The vortex is described by the solution of the Bogoliubov- de Gennes Eq.~(\ref{BdGnonuniform}) corresponding to the ansatz
\begin{eqnarray}
u_i({\bf r})&=& u_n(r_\perp)e^{-im\phi}e^{ik_zz}/\sqrt{2\pi L} \nonumber \\
v_i({\bf r})&=& v_n(r_\perp)e^{-i(m+1)\phi}e^{ik_zz}/\sqrt{2\pi L} \;,
\label{uvvortex}
\end{eqnarray}
for the normalized functions $u_i$ and $v_i$ where ($n,m,k_z$) are the usual quantum numbers of cylindrical symmetry and $L$ is the length of the box in the $z$ direction. The ansatz (\ref{uvvortex}) is consistent with the dependence (\ref{deltaphi}) of the order parameter $\Delta$ on the phase $\phi$. Calculations of the vortex structure based on the above approach, have been carried out by several authors (Nygaard {\it et al.}, 2003; Machida and Koyama, 2005; Chien {\it et al.}, 2006; Sensarma, Randeria and Ho, 2006). A generalized version of the Bogoliubov-de Gennes equations based on the density functional theory was used instead by Bulgac and Yu (2003). An important feature emerging from these calculations is that, near the vortex line, the density contrast is reduced at unitarity with respect to the BEC limit and is absent in the BCS regime.   

At higher angular velocities more vortices can be formed giving rise to a regular vortex lattice. In this limit  the angular momentum acquired by the system approaches the classical rigid-body value and the rotation is similar to the one of a rigid body, characterized by the law $\nabla\times {\bf v}= 2{\bf \Omega}$. Using result (\ref{circulation}) and averaging the vorticity over several vortex lines one finds $\nabla\times{\bf v}= (\pi\hbar/m) n_v \hat{{\bf z}}$, where $n_v$ is the number of vortices per unit area, so that the density of vortices $n_v$ is related to the angular velocity $\Omega$ by the relation $n_v=2m \Omega/ \pi \hbar$ showing that the distance between vortices (proportional to $1/\sqrt{n_v}$) depends on the angular velocity but not on the density of the gas. The vortices form thus a regular lattice even if the average density is not uniform as happens in the presence of harmonic trapping. This feature, already pointed out in the case of Bose-Einstein condensed gases, has been confirmed in the recent experiments on Fermi gases (see Fig.~\ref{fig8.4}). It is worth noticing that, due to the repulsive quantum pressure effect characterizing Fermi gases, one can realize trapped configurations with a large size $R_\perp$ hosting a large number of vortices $N_v=\pi R^2_\perp n_v$, even with relatively small values of $\Omega$. For example choosing $\Omega=\omega_\perp/3$, $\omega_z\simeq \omega_\perp$ and $N=10^6$, one predicts $N_v\sim 130$ at unitarity which is significantly larger than the number of vortices that one can produce in a dilute Bose gas with the same angular velocity.

At large angular velocity the vortex lattice is responsible for a bulge effect associated with the increase of the radial size of the cloud and consequently with a modification of the aspect ratio. In fact, in the presence of an average rigid rotation, the effective potential felt by the atoms is given by $V_{ho}-(m/2) \Omega^2 r^2_\perp$. The new Thomas-Fermi radii satisfy the  relationship
\begin{equation}
{R^2_z \over R^2_\perp} = {\omega^2_\perp-\Omega^2 \over \omega_z^2} \;,
\label{bulge}
\end{equation}
showing that at equilibrium the angular velocity can not overcome the radial trapping frequency. This formula can be used to determine directly the value of $\Omega$ by just measuring the {\it in situ} aspect ratio of the atomic cloud. 

Important consequences of the presence of vortex lines concern the frequency of the collective oscillations. For example, using a sum-rule approach (Zambelli and Stringari, 1998) it is possible to show that the splitting $\Delta \omega$ between the $m=\pm 2$
quadrupole frequencies is given by the formula 
\begin{equation}
\Delta \omega= \omega(m=+2)-\omega(m=-2)=2{\frac{\ell_{z}}{m\langle r_{\perp}^{2}\rangle }} \;,
\label{33F}
\end{equation}
where $\ell_z=\langle L_{z}\rangle /N$ is the angular momentum per particle carried by the vortical configuration. For a single vortex line $\ell_z$ is equal to $\hbar/2$, while for a vortex lattice $\ell_z$ is given by the rigid-body value $\Omega m \langle r^2_{\perp}\rangle$. The splitting (\ref{33F}), and hence the angular momentum $\ell_z$ can be directly measured by producing a sudden quadrupole deformation in the $x-y$ plane and imaging the corresponding precession $\dot{\phi}= \Delta\omega/4$ of the angle $\phi$ of the symmetry axis of the deformation during the quadrupole oscillation. This precession phenomenon has been already observed in the case of Bose-Einstein condensates containing quantized vortex lines (Chevy, Madison and Dalibard, 2000). For a single vortex line this experiment gives direct access to the quantization of the angular momentum (\ref{Lv}) carried by the vortex, in analogy with the famous Vinen experiment of superfluid helium (Vinen, 1961).

In the presence of many vortex lines the collective oscillations of the system can be  calculated using the equations of rotational hydrodynamics (Cozzini and Stringari, 2003). In fact, from a macroscopic point of view, the vortex lattice behaves like a classical body rotating in a rigid way. For example, in the case of axisymmetric configurations one finds the result 
\begin{equation}
\omega(m=\pm2)=\sqrt{2\omega^2_{\perp}-\Omega^2}\pm\Omega
\label{m=2R}
\end{equation}
for the frequencies of the two $m=\pm 2$ quadrupole modes, which is consistent with the sum-rule result (\ref{33F}) for the splitting in the case of a rigid rotation. Other important modes affected by the presence of the vortex lattice are the compressional $m=0$ oscillations resulting from the coupling of the radial and axial degrees of freedom. These modes were discussed in Sec.~\ref{Sec7.3} in the absence of rotation. The effect of the rotation in a Fermi gas has been recently discussed by Antezza, Cozzini and Stringari (2007). In the centrifugal limit $\Omega \to \omega_{\perp}$ the frequency of the radial mode  approaches the value $\omega=2\omega_\perp$ independent of the equation of state, while the frequency of the  axial mode approaches the value
\begin{equation}
\omega = \sqrt{\gamma +2}\;\omega_z \; ,
\label{m=0-R}
\end{equation}
where $\gamma$ is the coefficient of the polytropic equation of state (see Sec.~\ref{Sec7.3}).  

It is finally worth noticing that the achievement of the centrifugal limit for a superfluid containing a vortex lattice can not be ensured  on the BCS side of the resonance. In fact, due  to the bulge effect (\ref{bulge}), the centrifugal limit is associated with a strong decrease of the density and hence, for $a<0$, with an exponential decrease of the order parameter $\Delta$. It follows that the superfluid  can not support rotations with values of $\Omega$ too close to $\omega_\perp$  and that the system will exhibit a transition to the normal phase (Zhai and Ho, 2006; Veillette {\it et al.}, 2006) [see also Moller and Cooper (2007) for a recent discussion of the new features exhibited by the rotating BCS Fermi gas]. If $\Omega$ becomes too close to $\omega_\perp$, superfluidity will be eventually lost also at resonance and on the BEC side of the resonance because the system enters the Quantum Hall regime (for the case of bosons see, {\it e.g.}, Cooper, Wilkin and Gunn, 2001; Regnault and Jolicoeur, 2003) 

\section{SPIN POLARIZED FERMI GASES AND FERMI MIXTURES} \label{Chap9} 

The description of Fermi superfluidity presented in the previous Sections was based on the assumption that the gas has an equal number of atoms occupying two different spin states. One can also consider more complex configurations of spin imbalance where the number of atoms  in the two spin states is different ($N_\uparrow \neq N_\downarrow$) as well as mixtures of atomic species with different masses ($m_\uparrow \neq m_\downarrow$), including Fermi-Fermi and Bose-Fermi mixtures. Recent realizations of these novel configurations are opening new perspectives for both experimental and theoretical research.

\begin{figure}[b]
\begin{center}
\includegraphics[width=7.5cm]{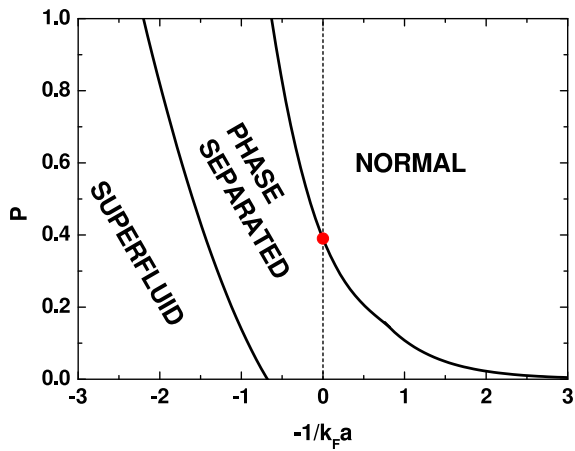}
\caption{Qualitative phase diagram as a function of the interaction strength $-1/k_Fa$ and of the polarization $P$. The circle at unitarity corresponds to the critical value $P_c=0.39$ discussed in Sec.~\ref{Sec9.2}. The Fermi wavevector corresponds here to the total average density: $k_F=[3\pi^2(n_\uparrow+n_\downarrow)]^{1/3}$. Notice that the possible occurrence of the FFLO phase is not considered on this diagram.} 
\label{fig9.1}
\end{center}
\end{figure}

\subsection{Equation of state of a spin polarized Fermi gas} \label{Sec9.1}

The problem of spin imbalance has an old story in the context of BCS theory of superconductivity. In superconductors, due to the fast relaxation between different spin states leading to balanced spin populations, the only possibility to create the asymmetry is to add an external magnetic field. However, in bulk superconductors this field is screened by the orbital motion of electrons (Meissner effect). The situation in superfluid Fermi gases is more favourable. In fact, in this case, the relaxation time is very long and the numbers of atoms occupying different spin states can be considered as independent variables. 

Let us recall that the mechanism of BCS superfluidity, in the regime of small and negative scattering lengths, arises from the pairing of particles of different spin occupying states with opposite momenta, close to the Fermi surface. This mechanism is  inhibited by the presence of spin imbalance, since the Fermi surfaces of the two components do not coincide and pairs with zero total momentum are difficult to form. Eventually, if the gap between the two Fermi surfaces is too large, superfluidity is broken and the system undergoes a quantum phase transition towards a normal state. The existence of a critical value for the polarization is easily understood by noticing that, at zero temperature, the unpolarized gas is superfluid, while a fully polarized gas is normal due to the absence of interactions. The occurrence of such a transition was first suggested in the papers by Clogston (1962) and Chandrasekhar (1962) who predicted the occurrence of a first order transition from the normal to the superfluid state. This transition takes place when the gain in the grand-canonical energy associated with the finite polarization of the normal phase is equal to the energy difference between the normal and the superfluid unpolarized states. In the BCS regime one finds the critical condition [see Eq.~(\ref{E0sup})]
\begin{equation}
h\equiv{\mu_\uparrow -\mu_\downarrow \over 2}={\Delta_{\text gap} \over \sqrt2} \;,
\label{clogston}
\end{equation}
where $\Delta_{\text gap}=(2/e)^{7/3}E_F e^{\pi/2k_Fa}$ is the BCS gap and we have assumed the spin-down particles as the minority component. The chemical potential difference $h$ in the above equation plays the role of an effective magnetic field. In terms of the polarization 
\begin{equation}
P=\frac{N_\uparrow-N_\downarrow}{N_\uparrow+N_\downarrow} \;, 
\label{equation}
\end{equation}
it can be expressed through the relationship $h=2E_FP/3$, which holds if $P\ll 1$. The condition (\ref{clogston}) then immediately yields the critical value of $P$ at which the system phase separates:
\begin{equation}
P_c=\frac{3}{\sqrt{8}}\left(\frac{2}{e}\right)^{7/3} e^{\pi/2k_Fa} \;.
\label{Pc-BCS}
\end{equation}
For $P>P_c$ the system is normal and corresponds to a uniform mixture of the two spin components well described by the non-interacting model. For $P<P_c$ the system is instead in a mixed state, where the unpolarized BCS superfluid coexists with the normal phase which accommodates the excess polarization. In this mixed state, the chemical potential difference of the normal phase retains the critical value (\ref{clogston}) irrespective of polarization, a decrease in $P$ being accounted for by an increase in the volume fraction of the superfluid phase which eventually occupies the entire volume for $P=0$. An important remark concerning the Clogston-Chandrasekhar condition (\ref{clogston}) is that the critical effective magnetic field $h$ is smaller than the superfluid gap $\Delta_{\text gap}$. If one had $h>\Delta_{\text gap}$ the above scenario would not apply because the system would prefer to accommodate the excess polarization by breaking pairs and creating quasi-particles. The gapless superfluid realized in this way would be homogeneous and the transition to the normal state would be continous. Such a uniform phase is indeed expected to occur in the deep BEC regime (see below).

The physical understanding of polarized Fermi gases became more complicated when exotic superfluid phases were proposed such as the inhomogeneous Fulde-Ferrell-Larkin-Ovchinnikov (FFLO) state (Fulde and Ferrell, 1964; Larkin and Ovchinnikov, 1964). Other alternative states include the breached pair or Sarma state (Sarma, 1963; Liu and Wilczek, 2003) and states with a deformed Fermi surface (Sedrakian {\it et al.}, 2005). In the FFLO state Cooper pairs carry a finite momentum resulting in a spontaneous breaking of translational symmetry with a periodic structure of the order parameter of the form $\Delta(x)\propto\cos(q x)$, where the direction of the $x$ axis is arbitrary. The wavevector $q$ is proportional to the difference of the two Fermi wavevectors $q\propto k_{F\uparrow}-k_{F\downarrow}$ with a proportionality coefficient of order unity. The excess spin-up atoms are concentrated near the zeros of the order parameter $\Delta(x)$. One can show that in the BCS limit the FFLO phase exists for $h<0.754\Delta_{\text gap}$, corresponding to the tiny interval of polarization $0<P<1.13\Delta_{\text gap}/E_F$, and at $P>1.13\Delta_{\text gap}/E_F$ the system is normal (see, {\it e.g.}, Takada and Izuyama, 1969). In the deep BCS regime this scenario is more energetically favourable compared to the Clogston-Chandrasekhar transition which would take place at $P_c=1.06\Delta_{\text gap}/E_F$. The FFLO state is of interest both in condensed matter physics and in elementary particle physics, even though direct experimental evidences of this phase are still lacking [for recent reviews, see Casalbuoni and Nardulli (2004) and Combescot (2007)].

In ultracold gases the BCS regime is not easily achieved due to the smallness of the gap parameter and one is naturally led to explore configurations with larger values of $k_F|a|$, where the concept of Fermi surface looses its meaning due to the broadening produced by pairing. A major question is whether the FFLO phase survives when correlations are stronger and if it can be realized in trapped configurations (Mizushima, Machida and Ichioka, 2005; Sheehy and Radzihovsky 2007; Yoshida and Yip, 2007). As a function of the interaction strength, parametrized by $1/k_{F}a$ where $k_F=[3\pi^2(n_\uparrow+n_\downarrow)]^{1/3}$ is fixed by the total density, different scenarios can take place as schematically shown in Fig.~\ref{fig9.1} (Sheehy and Radzihovsky, 2006 and 2007; Son and Stephanov, 2006; Pao, wu and Yip, 2006; Hu and Liu, 2006; Iskin and S\'a de Melo, 2006b; Parish {\it et al.}, 2007a). An important region of the phase diagram is the deep BEC regime of small and positive scattering lengths, where the energetically favourable phase consists of a uniform mixture of a superfluid gas of bosonic dimers and of a normal gas of spin polarized fermions. In this regime one expects that the normal uniform gas exists only for $P=1$, corresponding to the fully polarized ideal Fermi gas. The general problem of an interacting mixture of bosons and fermions was investigated by Viverit, Pethick and Smith (2000) who derived the conditions of miscibility in terms of the values of the densities and masses of the two components and of the boson-boson and boson-fermion scattering lengths (see Sec.~\ref{Sec9.5}). In particular, for $P\simeq1$ corresponding to a small number of bosonic dimers in a fully polarized Fermi sea, the relevant condition for the solubility of the mixture reads [see Eq.~(\ref{BFstability}) below]
\begin{equation}
k_{F}\leq \frac{4\pi}{2^{1/3}9}\frac{a_{dd}}{a_{ad}^2} \;.
\label{stabilityBF}
\end{equation}
By using the values $a_{dd}=0.60a$ and $a_{ad}=1.18a$ for the dimer-dimer and atom-dimer scattering length respectively, one finds that the uniform phase exists for $1/k_Fa>2.1$. This Bose-Fermi picture, however, looses its validity as one approaches the resonance region and more detailed analyses are needed to understand the phase diagram of the system close to unitarity (Pilati and Giorgini, 2007). In the presence of harmonic trapping the conditions of phase separation in the BEC regime change significantly due to the non-uniform effective potentials felt by the two components. The density profiles of the bosonic dimers and of the unpaired fermions have been investigated by Pieri and Strinati (2006) within the local density approximation. 

The determination of the energetically favourable configuration in the unitary regime of infinite scattering length is a difficult problem (Carlson and Reddy, 2005). Will the unpolarized superfluid and the polarized normal gas co-exist like in the BEC regime or will they separate? Will the FFLO phase play any role? First experiments carried out with spin imbalanced trapped Fermi gases close to unitarity suggest the occurrence of a phase separation between an unpolarized superfluid and a polarized normal phase (Partridge {\it et al.}, 2006a; Shin {\it et al.}, 2006). 

Other important questions that will not be addressed in this review concern the phases of these spin polarized Fermi gases at finite temperature (Gubbels, Romans and Stoof, 2006; Parish {\it et al.}, 2007a; Chien {\it et al.}, 2007) and the occurrence of $p$-wave superfluid phases at very low temperatures (Bulgac, Forbes and Schwenk, 2006).

\begin{figure}[b]
\begin{center}
\includegraphics[width=8.5cm]{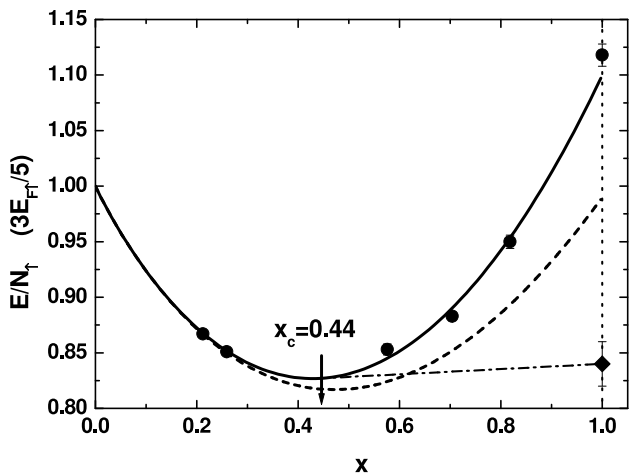}
\caption{Equation of state of a normal Fermi gas as a function of the concentration $x$. The solid line is a polynomial best fit to the QMC results (circles). The dashed line corresponds to the expansion (\ref{Ex}). The dot-dashed line is the coexistence line
between the normal and the unpolarized superfluid states and the arrow indicates the critical concentration $x_c$ above which the system phase separates. For $x=1$, both the energy of the normal and of the superfluid (diamond) states are shown.} 
\label{fig9.2}
\end{center}
\end{figure}

\subsection{Phase separation at unitarity} \label{Sec9.2}

A possible scenario for the equation of state of spin imbalanced configurations at unitarity is based on the phase separation between an unpolarized superfluid and a polarized normal gas similar to the Clogston-Chandrasekhar transition discussed in the Sec.~\ref{Sec9.1} (De Silva and Mueller, 2006a; Haque and Stoof 2006; Chevy, 2006a). An important ingredient of this scenario, that is not accounted for by the mean-field description, is the proper inclusion of interaction effects in the normal phase (Chevy 2006b; Lobo {\it et al.}, 2006b; Bulgac and Forbes, 2007). While there is not at present a formal proof that the phase separated state is the most energetically favourable, the resulting predictions well agree with recent experimental findings (see Sec.~\ref{Sec9.3}). 

The unpolarized superfluid phase was described in details in Secs.~\ref{Chap4} and \ref{Chap5} and is characterized, at unitarity, by the equation of state 
\begin{equation}
{E_S\over N}= {3\over 5}E_F(1+\beta) \;,
\label{Esuperfluid}  
\end{equation}
where $E_S$ is the energy of the system, $N$ is the total number of atoms and $E_F=(\hbar^2/2m)(3\pi^2 n_S)^{2/3}$ is the Fermi energy, where $n_S=2n_\uparrow=2n_\downarrow$ is the total density of the gas. Starting from (\ref{Esuperfluid}) one can derive the pressure $P_S=-\partial E/\partial V $ and the chemical potential $\mu_S=\partial E_S/\partial N$. 

Differently from the superfluid the normal phase is polarized and consequently its equation of state will depend also on the concentration
\begin{equation}
x= n_\downarrow / n_\uparrow \;,
\label{x}
\end{equation}
which, in the following, will be assumed to be smaller or equal to $1$, corresponding to $N_\uparrow \ge N_\downarrow$. A convenient way to build the $x$-dependence of the equation of state is to take the point of view of a dilute mixture where a few spin-down atoms are added to a non-interacting gas of spin-up particles. When $x\ll 1$, the energy of the system can be written in the form (Lobo {\it et al.}, 2006b)
\begin{equation}
{E_N(n_{\uparrow},x)\over N_\uparrow}=  {3\over 5}E_{F^\uparrow}[1-Ax+{m\over m^*}x^{5/3} + ...] \;,
\label{Ex}
\end{equation}
where $E_{F\uparrow}=(\hbar^2/2m)(6\pi^2 n_\uparrow)^{2/3}$ is the Fermi energy of the spin-up particles. 
The first term in Eq.~(\ref{Ex}) corresponds to the energy per particle of the non-interacting gas, while the term linear in $x$ gives the binding energy of the spin-down particles. Eq.~(\ref{Ex}) assumes that, when we add spin-down particles creating a small finite density $n_\downarrow$, they form a Fermi gas of quasi-particles with effective mass $m^\ast$ occupying, at zero temperature, all the states with wavevector $k$ up to $k_{F_\downarrow}=(6\pi^2n_\downarrow)^{1/3}$ and contributing to the total energy (\ref{Ex}) with the quantum pressure term proportional to $x^{5/3}$. The interaction between the spin-down and spin-up particles is accounted for by the dimensionless parameters $A$ and $m/m^\ast$. The expansion (\ref{Ex}) should in principle include  additional terms originating from the interaction between quasi-particles and exhibiting a higher order dependence on $x$. 

The values of the coefficients entering (\ref{Ex}) can be calculated using fixed-node diffusion Monte Carlo simulations where one spin-down atom is added to a non-interacting Fermi gas of spin-up particles. The results of these calculations are $A=0.97(2)$ and $m/m^*=1.04(3)$ (Lobo {\it et al.}, 2006b). The same value of $A$ has been obtained from an exact diagrammatic Monte Carlo calculation (Prokof'ev and Svistunov, 2007) and, quite remarkably, also employing a simple variational approach based on a single particle-hole wavefunction (Chevy, 2006b, Combescot {\it et al.}, 2007). The prediction of Eq.~(\ref{Ex}) for the equation of state is reported in Fig.~\ref{fig9.2}, where we also show the FN-DMC results obtained for finite values of the concentration $x$. It is remarkable to see that the expansion (\ref{Ex}) reproduces quite well the best fit to the FN-DMC results up to the large values of $x$ where the transition to the superfluid phase takes place (see discussion below). In particular, the repulsive term in $x^{5/3}$, associated with the Fermi quantum pressure of the minority species, plays a crucial role in determining the $x$ dependence of the equation of state. Notice that when $x=1$ ($N_\uparrow=N_\downarrow$) the energy per particle of the normal phase is smaller than the ideal gas value $6E_{F\uparrow}/5$, reflecting the attractive nature of the force, but larger than the value in the superfluid phase $(0.42)6E_{F\uparrow}/5$ (see Sec.~\ref{Sec5.2}).

We can now determine the conditions of equilibrium between the normal and the superfluid phase. A first condition is obtained by imposing that the pressures of the two phases be equal
\begin{equation}
{\partial E_S \over \partial V_S} = {\partial E_N \over \partial V_N} \;, 
\label{PN=PS} 
\end{equation}
where $V_S$ and $V_N$ are the volumes occupied by the two phases respectively. A second condition is obtained by requiring that
the chemical potential of each pair of spin up-spin down particles be the same in the two phases. In order to exploit this latter condition one takes advantage of the thermodynamic identity $\mu_S=(\mu_\uparrow +\mu_\downarrow)/2$ for the chemical potential in the superfluid phase yielding the additional relation
\begin{equation}
{\partial E_S \over \partial N}= {1\over 2} \left({\partial E_N \over \partial N_\uparrow} + {\partial E_N \over \partial N_\downarrow} \right) \;,
\label{musupdown}
\end{equation}
where we have used the expression $\mu_{\uparrow(\downarrow)}=\partial E_N /\partial N_{\uparrow(\downarrow)}$ for the chemical potentials of the spin-up and spin-down particles calculated in the normal phase. Eq.~(\ref{musupdown}), combined with (\ref{PN=PS}), permits to determine the critical values of the thermodynamic parameters characterizing the equilibrium between the two phases.
For example, if applied to the BCS regime where $E_N=3/5(N_\uparrow E_{F\uparrow}+N_\downarrow E_{F\downarrow})$ and $E_S=E_N-3N\Delta_{\text gap}^2/8E_F$ [see Eq.~(\ref{E0sup})], this approach reproduces the Clogston-Chandrasekhar condition (\ref{clogston}) (Bedaque, Caldas and Rupak, 2003). At unitarity instead, a calculation based on the QMC values of $E_N$ and $E_S$ yields the values $x_{c}=0.44$, corresponding to $P_c=(1-x_c)/(1+x_c)=0.39$, and $(n_N/n_S)_{c}= 0.73$, where $n_N=n_\uparrow+n_\downarrow$ is the density of the normal phase in equilibrium with the superfluid (Lobo {\it et al.}, 2006b). For values of the polarization larger than $P_{c}=0.39$ the stable configuration is the uniform normal phase, while if we increase the number of the spin down particles (corresponding to a reduction of $P$) there will be a phase separation between a normal phase with the concentration $x_{c}=0.44$ and a superfluid unpolarized phase. The phase transition has first order character consistently with the critical value $h=0.81\Delta_{\text gap}$ being smaller than the superfluid gap. In particular, while the spin-up density is practically continuous, the density of the spin-down particles exhibits a significant jump at the transition. Let us finally point out that the parameters characterizing the transition between the superfluid and the normal phases depend in a crucial way on the many-body scheme employed for the calculation. For example, if instead of the Monte Carlo results we use the BCS mean-field theory of Sec.~\ref{Sec5.1} and the non-interacting expression $E_N=3/5(N_\uparrow E_{F\uparrow}+N_\downarrow E_{F\downarrow})$ for the energy of the normal phase, we would find the very different value $x_{c}= 0.04$ for the critical concentration.

\begin{figure}[b]
\begin{center}
\includegraphics[width=8.5cm]{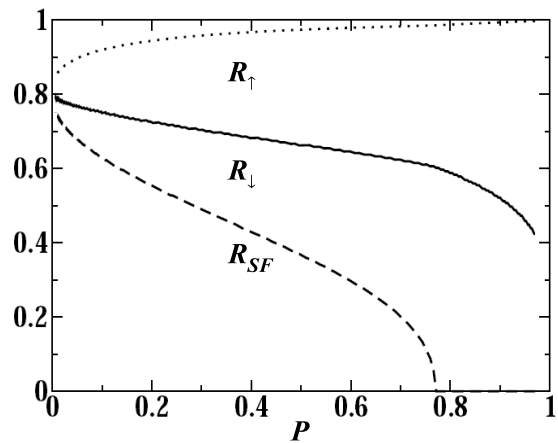}
\caption{Radii of the three phases in the trap in units of the radius $R_\uparrow^0=a_{ho}(48 N_\uparrow)^{1/6}$ of a non-interacting fully polarized gas.} 
\label{fig9.3}
\end{center}
\end{figure}

\begin{figure}[b]
\begin{center}
\includegraphics[width=8.5cm]{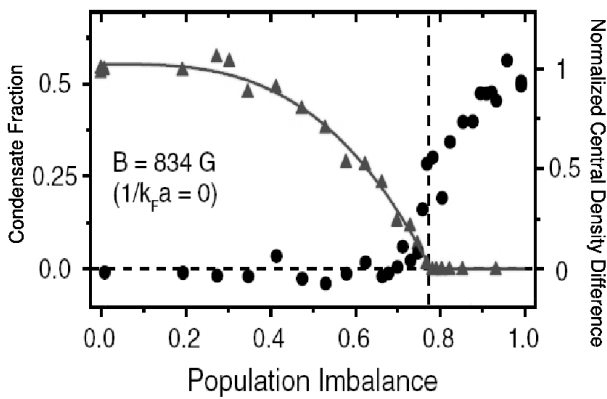}
\caption{Phase separation in an interacting Fermi gas at unitarity. Normalized central density difference (circles) and condensate fraction (triangles). From both sets of data one extract the same value $\sim 0.75$ for the critical polarization. From Shin {\it et al.} (2006).} 
\label{fig9.4}
\end{center}
\end{figure}

\begin{figure}
\begin{center}
\includegraphics[width=8.5cm]{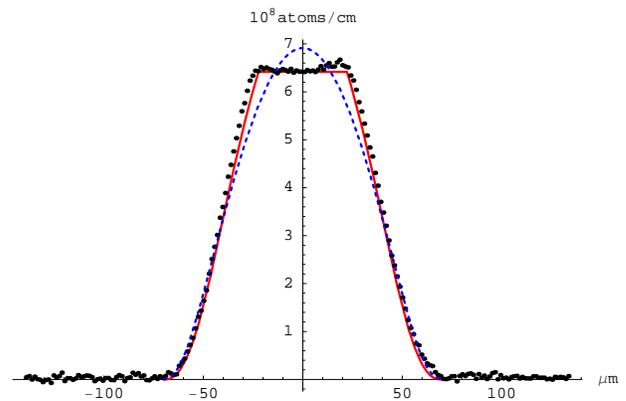}
\caption{Double integrated density difference measured at unitarity in an ultracold trapped Fermi gas of $^6$Li with polarization $P=0.58$  (from Shin {\it et al.}, 2006). The theoretical curves correspond to the theory of Secs.~\ref{Sec9.2}-\ref{Sec9.3} based on the local density approximation (red solid line) and to the predictions of a non-interacting gas with the same value of $P$ (dashed line). (From Lobo, Recati and Stringari, 2007).} 
\label{fig9.6}
\end{center}
\end{figure}

\subsection{Phase separation in harmonic traps at unitarity} \label{Sec9.3}

The results presented in Sec.~\ref{Sec9.2} can be used to calculate the density profiles in the presence of harmonic trapping. We make use of the local density approximation which permits to express the local value of the chemical potential of each spin species as
\begin{equation}
\mu_{\uparrow(\downarrow)}({\bf r})= \mu^0_{\uparrow(\downarrow)} - V_{ho}({\bf r}) \; .
\label{LDAspin}
\end{equation}
The chemical potentials $\mu^0_{\uparrow(\downarrow)}$ are fixed by imposing the proper normalization to the spin-up and spin-down densities and the condition $\mu_{\uparrow(\downarrow)}^0-V_{ho}(r=R_{\uparrow(\downarrow)})=0$ defines the Thomas-Fermi radii $R_{\uparrow(\downarrow)}$ of the two species. In our discussion we assume isotropic trapping in order to simplify the formalism. A simple scaling transformation permits to apply the results also to the case of anisotropic trapping. 

For small concentrations of the spin-down particles ($N_{\downarrow}\ll N_{\uparrow}$) only the normal state is present and one can ignore the change in $\mu_\uparrow$ due to the attraction of the spin-down atoms. In this case $n_\uparrow$ reduces to the Thomas-Fermi profile (\ref{n0r}) of an ideal gas, whereas $n_\downarrow$ is a Thomas-Fermi profile with a modified harmonic potential $V_{ho}\to (1+3A/5)V_{ho}$. The confining potential felt by the spin-down atoms is stronger due to the attraction produced by the spin-up atoms. One can also calculate the frequency of the dipole oscillation of the spin-down particles. This is given by ($i=x,y,z$)
\begin{equation}
\omega^{dipole}_i=\omega_i\sqrt{(1+{3\over 5}A){m\over m^*}} \simeq 1.23 \omega_i \;,
\label{omegaspin}
\end{equation} 
and is affected by both the interaction parameter $A$ and the effective mass parameter $m/m^*$ entering the expansion (\ref{Ex}).

When in the center of the trap the local concentration of spin-down particles reaches the critical value $x_c=0.44$ (see Sec.~\ref{Sec9.2}) a superfluid core starts to nucleate in equilibrium with a polarized normal gas outside the superfluid. The radius $R_{SF}$ of the superfluid is determined by the equilibrium conditions between the two phases discussed in the previous Section. Towards the periphery of the polarized normal phase the density of the spin-down particles will eventually approach a vanishing value corresponding to $R_\downarrow$. For even larger values of the radial coordinate only the density of the spin-up particles will be different from zero up to the radius $R_\uparrow$. In this peripherical region the normal phase corresponds to a fully polarized,  non-interacting Fermi gas.  

By using the Monte Carlo equation of state discussed in Sec.~\ref{Sec9.2} one finds that the superfluid core disappears for polarizations $P>P_c^{trap}=0.77$ (Lobo {\it et al.}, 2006b). The difference between the critical value $P_c^{trap}$ and the value $P_c=0.39$ obtained for uniform gases reflects the inhomogeneity of the trapping potential. In Fig.~\ref{fig9.3} we show the calculated radii of the superfluid and of the spin-down and spin-up components as a function of $P$. The radii are given in units of the non-interacting radius of the majority component $R^0_\uparrow = a_{ho}(48N_\uparrow)^{1/6}$. The figure explicitly points out the relevant features of the problem. When $P\to 0$ one approaches the standard unpolarized superfluid phase where the three radii ($R_{SF}$, $R_\uparrow$ and $R_\downarrow$) coincide with the value $(1+\beta)^{1/4}R^0_\uparrow \simeq 0.81 R^0_\uparrow$ [see Eq.~(\ref{Runitarity})]. By increasing $P$ one observes the typical shell structure with $R_{SF}< R_\downarrow < R_\uparrow$. The spin-up radius increases and eventually approaches the non-interacting value when $P\to 1$. The spin-down radius instead decreases and eventually vanishes as $P\to 1$. Finally the radius $R_{SF}$ of the superfluid component decreases and vanishes for $P_{c}^{trap}=0.77$, corresponding to the disappearence of the superfluid phase. 

The predicted value for the critical polarization is in good agreement with the findings of the MIT experiments (Zwierlein {\it et al.}, 2006a; Shin {\it et al.}, 2006), where the interplay between the superfluid and the normal phase was investigated by varying the polarization $P$ of the gas. The experimental evidence for superfluidity in a spin imbalanced gas emerges from measurements of the condensate fraction and of the vortex structure in fast rotating configurations (Zwierlein {\it et al.}, 2006a). These time-of-flight measurements were performed by rapidly ramping the scattering length to small and positive values after opening the trap, in order to stabilize fermion pairs and increase the visibility of the bimodal distribution and of vortices (see discussion in Sec.~\ref{Sec5.2.2} and in Sec.~\ref{Sec8.3}). In another paper (Shin {\it et al.}, 2006) the {\it in situ} density difference $n_{\uparrow}({\bf r})-n_{\downarrow}({\bf r})$ was directly measured with phase contrast techniques. Phase separation was observed by correlating the presence of a core region with $n_{\uparrow}-n_{\downarrow}=0$ with the presence of a condensate of pairs (see Fig.~\ref{fig9.4}). At unitarity these results reveal that the superfluid core appears for $P\le 0.75$. An interesting quantity that can be directly extracted from these measurements is the double integrated density difference 
\begin{equation}
n^{(1)}_d(z)=\int dx dy [n^\uparrow({\bf r})-n^\downarrow({\bf r})] \;,
\label{axialdensity}
\end{equation}
which is reported in Fig.~\ref{fig9.6} for a unitary gas with $P=0.58$. The figure reveals the occurrence of a characteristic central region  where $n_d^{(1)}(z)$ is constant. The physical origin of this plateau can be understood using the local density approximation (De Silva and Mueller, 2006a; Haque and Stoof, 2006): it is a consequence of the existence of a core region with $n^\uparrow({\bf r})=n^\downarrow({\bf r})$ which is naturally interpreted as the superfluid core. Furthermore, the value of $z$ where the density exhibits the cusp corresponds to the Thomas-Fermi radius $R_{SF}$ of the superfluid. In the same figure we show the theoretical predictions, based on the Monte Carlo results for the equation of state of the superfluid and normal phases, which well agree with the experimental data (Lobo, Recati and Stringari, 2007).  

The theoretical predictions discussed above are based on a zero temperature assumption and on the local density approximation applied to the various phases of the trapped Fermi gas. While the applicability of the LDA seems to be adequate to describe the MIT data, the Rice experiments (Partridge {\it et al.}, 2006a), carried out with a very elongated trap, show that in this case surface tension effects, not accounted for by LDA, play a major role. It is possible to prove (De Silva and Mueller, 2006a; Haque and Stoof, 2006) that the double integrated density difference, when evaluated within the LDA, should exhibit a monotonic non increasing behavior as one moves from the trap center. The  non monotonic  structure of the density observed in the Rice experiment can be explained through the inclusion of surface tension effects. For a recent discussion of surface tension and thermal effects in spin imbalanced configurations see De Silva and Mueller (2006b), Partridge {\it et al.} (2006b) and Haque and Stoof (2007).

Other important questions concern the dynamic behavior of these spin polarized configurations. The problem is interesting since the dynamic behavior of the superfluid is quite different from the one of the normal gas, the latter being likely governed, at very low temperature, by collisionless kinetic equations rather than by the equations of hydrodynamics. As a consequence the role of the boundary separating the two phases should be carefully taken into account for a reliable prediction of the dynamic properties, such as the collective oscillations and the expansion.

\subsection{Fermi superfluids with unequal masses} \label{Sec9.4}

When one considers mixtures of Fermi gases belonging to different atomic species and hence having different masses new interesting issues emerge. One should point out that even if the masses are different, configurations where the atomic densities of the two components are equal result in equal Fermi momenta: $k_{F\uparrow}=k_{F\downarrow}=(3\pi^2n)^{1/3}$, where $n=n_\uparrow+n_\downarrow$ is the total density. This means that the mechanism of Cooper pairing, where two atoms of different spin can couple to form a pair of zero momentum, is still valid. At $T=0$ the BCS mean-field theory predicts a simple scaling behavior of the equation of state in terms of the reduced mass of the two atoms, which holds in the whole BCS-BEC crossover. However, in the BCS regime, correlations beyond the mean-field approximation introduce a non trivial dependence on the mass ratio in the superfluid gap (Baranov, Lobo and Shlyapnikov, 2007). Furthermore, in the BEC regime, theoretical studies of four-fermion systems (Petrov, 2003; Petrov, Salomon and Shlyapnikov, 2005) emphasize the crucial role of the mass ratio on the interaction between dimers, resulting in instabilities if the mass ratio exceeds a critical value (see Sec.~\ref{Sec3.3}). First quantum Monte Carlo results have also become available in the crossover, exploring the dependence of the equation of state and of the superfluid gap on the mass ratio (von Stecher, Greene and Blume 2007; Astrakharchik, Blume and Giorgini, 2007). These studies point out significant deviations from the predictions of the BCS mean-field theory. Important questions emerge also at finite temperature, where one can immediately conclude that the simple scaling in terms of the reduced mass can not hold since, for example, the BEC transition temperature of composite bosons (see Sec.~\ref{Sec4.5}) should depend on the molecular mass, $M=m_\uparrow+m_\downarrow$, rather than on the reduced mass. Other interesting scenarios, not addressed in this review, refer to the interplay between unequal masses and unequal populations of the two spin components (see, {\it e.g.}, Wu, Pao and Yip, 2006; Iskin and S\'a de Melo, 2006b; Parish {\it et al.}, 2007b; Iskin and S\'a de Melo, 2007a). The experimental realization of fermionic heteronuclear mixtures in the strongly correlated regime is at present actively pursued in different laboratories (see, {\it e.g.}, Wille {\it et al.}, 2007).

\subsubsection{Equation of state along the crossover}

The BCS mean-field approach developed in Sec.~\ref{Sec5.1} can be generalized in a straightforward manner to the case of unequal masses. Starting from the BCS Hamiltonian (\ref{HBCS}), where the single-particle energy term contains now the mass $m_{\uparrow(\downarrow)}$ and the chemical potential $\mu_{\uparrow(\downarrow)}$ of the two spin components, one follows the same steps leading to the gap and number equations (\ref{BCS1})-(\ref{BCS2}). These equations, as well as Eqs.~(\ref{uv})-(\ref{elemexcit}) for the quasi-particle amplitudes and excitation energies, read exactly the same as in the equal mass case, with the only difference that the atomic mass $m$ should be replaced by twice the reduced mass 
\begin{equation}
m\to 2m_r=2{m_\uparrow m_\downarrow\over m_\uparrow + m_\downarrow} \;,
\label{mr}
\end{equation} 
and that the single-particle energy should be replaced by $\eta_k\to \hbar^2k^2/4m_r-\mu$, where $\mu=(\mu_\uparrow+\mu_\downarrow)/2$ is the average chemical potential. In units of the reduced Fermi energy
\begin{equation}
E_F^r= {1\over 2} \left(E_{F_\uparrow}+E_{F_\downarrow} \right)={\hbar^2k_F^2 \over 4m_r} \;,
\label{EFr}
\end{equation}
the resulting values of the chemical potential $\mu$ and of the order parameter $\Delta$, for a given value $1/k_Fa$ of the interaction strength, are then independent of the mass ratio $m_\uparrow/m_\downarrow$.

At unitarity dimensionality arguments permit to write the energy of the system in the general form $E_S/N=3/5[1+\beta(m_\uparrow/m_\downarrow)]E_F^r$,
where $N=N_\uparrow+N_\downarrow$ is the total number of particles. In the BCS mean-field approach the value of $\beta$ is given by $\beta=-0.41$ (see Table~\ref{Tab2} in Sec.\ref{Chap6}). First results based on QMC simulations, suggest that the dependence of $\beta$ on the mass ratio is very weak (Astrakharchik, Blume and Giorgini, 2007). On the contrary, the corresponding value of the superfluid gap $\Delta_{\text gap}$ is strongly reduced by increasing the mass ratio. 

In the BEC regime, the mean-field approach predicts that the binding energy of the dimers is given by the formula $\epsilon_b=- \hbar^2/2m_ra^2$ and that these molecules interact with the same scattering length as in the symmetric mass case ($a_{dd}=2a$). While the result for the binding energy is correct, the actual relationship between $a_{dd}$ and $a$ is sensitive to the value of the mass ratio, approaching the value $0.60$ when $m_\uparrow=m_\downarrow$ (Petrov, 2003; Petrov, Salomon and Shlyapnikov, 2005).

\subsubsection{Density profiles and collective oscillations}

Also in the case of unequal masses the equation of state of uniform systems can be used to evaluate the density profiles of the harmonically trapped configurations. In the local density approximation the chemical potential of each atomic species varies in space according to the law (\ref{LDAspin}) where, however, the confining potential is now spin dependent being related to different magnetic and optical properties of the two atomic species and should be replaced by $V_{ho}({\bf r})\to V_{ho}^{\uparrow(\downarrow)}({\bf r})$. Since the chemical potential of the superfluid phase is given by the average $\mu_S=(\mu_\uparrow+\mu_\downarrow)/2$, the corresponding density is determined at equilibrium by the Thomas-Fermi relation
\begin{equation} 
\mu_S(n)=\mu_0 - \tilde{V}_{ho}({\bf r}) \;,
\label{densityeq}
\end{equation}
where
\begin{equation}
\tilde{V}_{ho}({\bf r}) = \frac{M}{4}
\left( \tilde{\omega}_x^2x^2 +\tilde{\omega}_y^2y^2 +\tilde{\omega}_z^2z^2\right) \;,
\label{Vbar}
\end{equation}
is the average of the trapping potentials. The effective frequencies $\tilde{\omega}_i$ are given by
\begin{equation}
\tilde{\omega}_i^2 ={m_\uparrow(\omega^\uparrow_i)^2 + m_\downarrow(\omega^\downarrow_i)^2 \over m_\uparrow  + m_\downarrow} \;,
\label{omegatilde}
\end{equation}
where $\omega^{\uparrow(\downarrow)}_i$ are the oscillator frequencies relative to $V_{ho}^{\uparrow(\downarrow)}$, whereas $M=m_\uparrow+m_\downarrow$ denotes the mass of the pair. If the oscillator lengths of the two atomic components coincide, {\it i.e.} if ${m_\uparrow}\omega^{\uparrow}_i=m_{\downarrow}\omega^{\downarrow}_i$, the effective frequencies (\ref{omegatilde}) take the simplified form $\tilde{\omega}_i=\sqrt{\omega^\uparrow_i\omega^\downarrow_i}$.

In the superfluid phase the densities of the spin-up and spin-down atoms are equal, even if the trapping potentials, in the absence of interatomic forces, would give rise to different profiles. In principle, even if one of the two atomic species does not feel directly any external potential (for example $\omega_\downarrow=0$), it can be trapped due to the interaction with the other species. At unitarity, one can show (Orso, Pitaevskii and Stringari, 2007) that the trapped superfluid configuration is energetically favourable if the condition $(1+\beta)M/m_\downarrow<1$ is satisfied. This condition is easily fulfilled if $m_\uparrow\le m_\downarrow$. 

The dynamic behavior of Fermi superfluids with unequal masses can be studied by properly generalizing the equations of hydrodynamics which take the form (\ref{continuity})-(\ref{euler}) with $V_{ho}$ replaced by the effective trapping potential (\ref{Vbar}) and where $m$ is replaced by half of the pair mass $m\to M/2$. In uniform systems ($\tilde{V}_{ho}=0$) the equations give rise to the propagation of sound with the sound velocity fixed by the thermodynamic relation $Mc^2=2n_S\partial \mu_S/\partial n_S$.
In the BCS limit, where the equation of state approaches the ideal gas expression the sound velocity takes the value $c=\hbar k_F\sqrt{1/(3m_\uparrow m_\downarrow)}$  where $k_F=(3\pi^2 n)^{1/3}$. At unitarity the sound velocity is given by the above ideal gas value multiplied by the factor $\sqrt{1+\beta}$. 

For harmonically trapped configurations the hydrodynamic equations can be used to study the expansion of the gas after opening the trap as well as the collective oscillations (Orso, Pitaevskii and Stringari, 2007). The effect of the mass asymmetry is accounted for through the effective frequencies (\ref{omegatilde}) as well as through the changes in the equation of state $\mu_S$. At unitarity, where the density dependence of the chemical potential keeps the $n^{2/3}$ power law, the same results discussed in Sec.~\ref{Sec7.3} hold with the oscillator frequencies replaced by $\tilde{\omega}_i$. Notice that the effective frequencies $\tilde{\omega}_i$ differ in general from either $\omega^\uparrow_i$ and $\omega^\downarrow_i$. Actually in the superfluid phase the two atomic species can not oscillate independently, but always move in phase as a consequence of the pairing mechanism produced by the interaction.

\subsection{Fermi-Bose mixtures} \label{Sec9.5}

The problem of quantum degenerate mixtures of a spin polarized Fermi gas and a Bose gas has been the object of considerable experimental and theoretical work. Particularly interesting are the mixtures where the interactions between the fermions and the bosons are tunable by means of a Feshbach resonance. This is the case, for example, of the $^{40}$K-$^{87}$Rb system which has been extensively investigated by the groups in Florence (Modugno {\it et al.}, 2002; Ferlaino {\it et al.}, 2006; Zaccanti {\it et al.}, 2006; Modugno, 2007) and in Hamburg (Ospelkaus {\it et al.}, 2006a; Ospelkaus {\it et al.}, 2006b). In these configurations the mixture is characterized by a repulsive boson-boson interaction with scattering length $a_{BB}\simeq 5.3$nm and a magnetically tunable boson-fermion interaction parametrized by the scattering length $a_{BF}$. Both the observation of an induced collapse of the mixture for large negative values of $a_{BF}$ and of phase separation for large positive values of $a_{BF}$ have been reported. 

From the theoretical point of view these experimental findings can be understood using a mean-field approach. The conditions of stability of a Bose-Fermi mixture in uniform systems at $T=0$ have been investigated by Viverit, Pethick and Smith (2000). Depending on the sign of $a_{BF}$ different scenarios can apply. 

If $a_{BF}<0$, the only relevant condition is determined by mechanical stability, {\it i.e.} by the requirement that the energy of the mixture must increase for small fluctuations in the density of the two components. Starting from a mean-field energy functional, which includes to lowest order the interaction effects between bosons and between fermions and bosons, the linear stability requirement fixes an upper limit on the fermionic density $n_F$ irrespective of the value $n_B$ of the bosonic density
\begin{equation}
n_F^{1/3}|a_{BF}|\le \left(\frac{4\pi}{3}\right)^{1/3}\frac{a_{BB}}{|a_{BF}|}\frac{m_B/m_F}{(1+m_B/m_F)^2} \;,
\label{BFstability}
\end{equation}
where $m_{B}$ and $m_F$ denote, respectively, the masses of bosons and fermions and $a_{BB}$ is assumed to be positive to ensure the stability of the configurations where only bosons are present. If $n_F$ exceeds the upper bound (\ref{BFstability}) the system collapses. 

If instead $a_{BF}>0$, the uniform mixture can become unstable against phase separation into a mixed phase and a purely fermionic one. While the fermionic density can not in any case violate the condition (\ref{BFstability}), yet for each value of $n_F$ there exists a critical bosonic density $n_B^c(n_F)$ above which the system phase separates. The function $n_B^c(n_F)$ is nontrivial and for a given pair of densities ($n_F$ and $n_B$) one has to check whether the phase separated configuration is in equilibrium and is energetically favourable compared to the uniform mixture. For vanishingly small bosonic densities $n_B$, the relevant condition for $n_F$ coincides with (\ref{BFstability}). It is worth noticing that the repulsive $a_{BF}$ scenario describes the mixtures of fermions and composite bosons in the deep BEC regime considered in Sec.~\ref{Sec9.1}. 

In the presence of harmonic trapping the conditions for the collapse and for the phase separation change. The density profiles of the two components have been investigated by M{\o}lmer (1998) using the same mean-field energy functional described above in the local density approximation. The results are consistent with the scenario of a collapsed state if $a_{BF}$ is large and negative and of a phase separated state (a core of bosons surrounded by a shell of fermions) in the opposite regime of large and positive $a_{BF}$. These two scenarios are in agreement with the features observed in $^{40}$K-$^{87}$Rb mixtures (Ospelkaus {\it et al.}, 2006b; Zaccanti {\it et al.}, 2006; Modugno, 2007). 

The collective oscillations in harmonically trapped Bose-Fermi mixtures have also been the object of a considerable number of theoretical investigations (see, {\it e.g.}, the recent work by Maruyama and Bertsch, 2007, and references therein). In particular, the monopole (Maruyama, Yabu and Suzuki, 2005) and the dipole mode (Maruyama and Bertsch, 2007) have been studied at zero temperature using a dynamic approach based on the solution of a coupled system of time-dependent equations: the Gross-Pitaevskii equation for the bosons and the collisionless Vlasov equation for the fermions. These studies point out the existence of a characteristic damping in the motion of the fermionic component affecting both types of oscillations. 

An important aspect of Bose-Fermi mixtures concerns the boson-induced interactions experienced by the otherwise non-interacting fermionic atoms. The physical origin of the induced interactions is the polarization of the bosonic medium which acts as an effective potential between the fermions. The density-density response function of the bosons is the relevant quantity to describe this effect and the induced interaction is thus frequency and wavevector dependent. At low frequencies and long wavelenghts it is always attractive, irrespective of the sign of $a_{BF}$, is independent of the density of bosons and reproduces the mechanism of instability discussed in the paragraph after Eq.~(\ref{BFstability}) (Bijlsma, Heringa and Stoof, 2000; Viverit, Pethick and Smith, 2000). The physical picture is similar to the effective attraction between $^3$He atoms in solution in superfluid $^4$He (Edwards {\it et al.}, 1965) and to the most famous phonon-induced attraction between electrons in ordinary superconductors (see, for example, de Gennes, 1989). 

An additional major interest of exploiting heteronuclear Feshbach resonances (including the Fermi-Fermi mixtures considered in Sec.~\ref{Sec9.4}) is the possibility of creating polar molecules characterized by long-range anisotropic interactions which are expected to have a profound impact on the many-body physics (Baranov {\it et al.}, 2002). Furthermore, the creation of such heteronuclear molecules in optical lattices (Ospelkaus {\it et al.}, 2006c) might lead to important applications in quantum information processing (Micheli, Brennen and Zoller, 2006).

\section{FERMI GASES IN OPTICAL LATTICES} \label{Chap10} 

The availability of optical lattices has opened new frontiers of research in the physics of ultracold atomic gases (for a recent review, see Morsch and Oberthaler, 2006, and Bloch, Dalibard and Zwerger, 2007). In the case of Fermi gases the problem is closely related to the physics of electrons in metals or semiconductors. However, optical lattices differ favorably from traditional crystals in many important aspects. The period of the optical lattice is macroscopically large, which simplifies the experimental observation. The lattice can be switched off, and its intensity can be tuned at wish. Atoms, differently from  electrons, are neutral and furthermore their interaction is tunable thanks to the existence of Feshbach resonances. The lattices are static and practically perfect, being free of defects. Disorder can be added in a controllable way by including random components in the optical field. Lastly, it is easy to produce one- and two-dimensional structures.

An atom in a monochromatic electric field feels a time averaged potential proportional to the square of the field amplitude
(see, {\it e.g.}, Pethick and Smith, 2002; Pitaevskii and Stringari 2003). It is useful to work sufficiently close to the frequency $\omega_0$ of the absorption line of an atom, where the force on the atom becomes strong. The atoms are pulled into the strong-field region for $\omega<\omega_0$ (``red detuning'') and pushed out of it for $\omega>\omega_0$ (``blue detuning'').

One-dimensional periodic potentials can be produced by a standing light wave. In this case, the potential energy is conveniently written as $V_{opt}\left(z\right) =sE_{R}\sin^2\left( Kz\right)$, where $K$ is the wavevector of the laser, $E_{R}=\hbar^{2}K^{2}/2m$ is the ``recoil energy''  and $s$ is the dimensionless parameter proportional to the laser field intensity. Typical values of $s$ in experiments range from $1$ to $20$. The potential has a period $d=\lambda/2=\pi/K$, where $\lambda$ is the wavelength of the laser. If the two counterpropagating laser beams interfere under an angle $\theta$ less than $180^o$ the period is increased by the factor $\sin(\theta/2)^{-1}$. By using three mutually orthogonal laser beams one can  generate a potential of the form
\begin{equation}
V_{opt}\left( \mathbf{r}\right) =sE_{R}\left[ \sin ^{2}\left( Kx\right)
+\sin ^{2}\left( Ky\right) +\sin ^{2}\left( Kz\right) \right] \;.
\label{3D}
\end{equation}
Notice that in  experiments atoms are trapped by additional confining potentials, in most cases of harmonic form. To preserve the effective periodicity of the problem the harmonic potential should vary slowly with respect to the period of the lattice. This demands the condition $E_R\gg \hbar \omega_{ho}$, which is easily satisfied in experiments.

\subsection{Ideal Fermi gases in optical lattices} \label{Sec10.1}

\begin{figure}[b]
\begin{center}
\includegraphics[width=8.5cm]{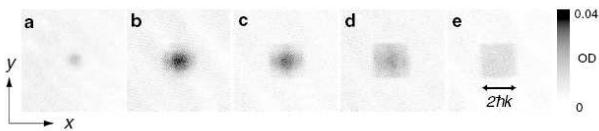}
\caption{Time-of-flight images obtained after adiabatically ramping down the optical lattice. Image (a) is obtained with $N_\sigma=3500$ and $s=5E_R$. Images (b)-(e) are obtained with $N_\sigma=15000$ and correspond to $s=5E_R$ (b), $s=6E_R$ (c), $s=8E_R$ (d) and $s=12E_R$ (e). The images show the optical density (OD) integrated along the vertically oriented $z$ axis after 
9 ms of ballistic expansion. From K\"ohl {\it et al.} (2005).}
\label{fig10.1}
\end{center}
\end{figure}

\subsubsection{Fermi surface and momentum distribution} \label{Sec10.1.1}

An advantage of experiments with cold fermions is the possibility of realizing a non-interacting gas by creating a spin-polarized configuration. As explained in Sec.~\ref{Sec3.1}, the interaction between fermionic atoms with parallel spins is in fact negligible. 

The quantum mechanical description of the motion of a particle in a periodic external field was developed by Bloch (Bloch, 1928). In 1D the wavefunctions have the Bloch form  $\psi_{q_zn}(z)=\exp(iq_{z}z)u_{q_zn}(z)$ and are classified in terms of the quasi-momentum $p_{z}\equiv \hbar q_{z}$, while $n$ is a discrete number labelling the Bloch band. The values of the wavevector $q_{z}$ differing by the reciprocal lattice vector $2K$ are physically equivalent and it is therefore enough to restrict the values of $q_z$ to the first Brillouin zone: $-K<q_{z}<K$. The function $u_{q_zn}(z)$ is a periodic function of $z$ with period $d$. The formalism is straightforwardly generalized to 3D lattices where the eigenstates of the Hamiltonian with the potential (\ref{3D}) are products of  wavefunctions of the Bloch form  along the three directions and are hence classified in terms of the 3D quasi-momentum ${\bf p}$. In a Fermi gas at zero temperaure all the states with excitation energy $\epsilon(\mathbf{p})=E(\mathbf{p})-E(0)$ are occupied up to values such that $\epsilon(\mathbf{p})=E_F$, where $E_F$ is the Fermi energy. The corresponding values of ${\bf p}$ characterize the Fermi surface. 

Experimentally one can measure the momentum distribution of these non-interacting configurations by imaging the atomic cloud after  release from the trapping potential. In fact, the spatial distribution $n({\bf r})$ of a non-interacting expanding gas reproduces asymptotically the initial momentum distribution $n({\bf p})$ according to the law $n(\mathbf{r},t)\to (m/t)^3 n(\mathbf{p}=m\mathbf{r}/t)$.

In order to have access to the quasi-momentum distribution  a practical procedure consists of switching off the lattice potential in an adiabatic way so that each state in the lowest energy band with quasi-momentum ${\bf p}$ is adiabatically transformed into a state with momentum ${\bf p}$. The condition of adiabaticity requires that the lattice potential be switched off in times longer than the inverse of the energy gap between the first and second band. 

The Bloch problem can be solved analytically in the tight binding approximation, holding for large lattice heights. For a 3D cubic lattice one finds the result ($\mathbf{p}=\hbar\mathbf{q}$)
\begin{eqnarray}
\epsilon_{\bf p} = 2\delta \left[ \sin^{2}(q_{x}d/2) + \sin^{2}(q_{y}d/2) 
+ \sin^{2}(q_{z}d/2) \right]   
\label{tight3D}
\end{eqnarray}
for the dispersion law of the single-particle excitations in the lowest band. The band width $2\delta$ decreases exponentially for large $s$ according to $\delta= 8E_Rs^{3/4}\exp(-2\sqrt{s})/ \sqrt{\pi}$ (Zwerger, 2003), and is inversely proportional to the tunneling rate through the barriers. The energy gap between the first and second band  coincides with the energy splitting $\hbar\omega_{opt}= 2 \sqrt{s}E_R $ between the states in the harmonic potential produced by the optical potential (\ref{3D}) around each local minimum. When ${\bf p}\to 0$, Eq.~(\ref{tight3D}) takes the simple form $\epsilon_{\bf p}=p^2/2m^*$, the effective mass being related to the band width by the relation $m^*=\hbar^2/\delta d^2$. The explicit dependence of $m^*$ on $s$, including the case of small laser intensities, was calculated by Kr\"{a}mer {\it et al.} (2003). 

The dispersion law (\ref{tight3D}) results from the removal of the degeneracy between the lowest energy states of each well produced by the interwell tunneling. This means that the number of levels in the band is equal to the number of the lattice cells. Then, if one works within the first Bloch band, the maximum achievable density is $n^{max}_\sigma=1/d^3$.

The inclusion of harmonic trapping can be accounted for by introducing the semiclassical distribution function which, at zero temperature, takes the form $f\left(\mathbf{p},\mathbf{r}\right)=\Theta\left[E_{F}-H\left(\mathbf{p,r}\right)\right]$ where $E_F$ is the Fermi energy. In the above equation ${\bf p}$ is the quasi-momentum variable and $H=\epsilon_{\bf p}+V_{ho}({\bf r})$, with $\epsilon_{\bf p}$ given by (\ref{tight3D}). Starting from the distribution function, one can evaluate the quasi-momentum ($qm$) distribution by integrating over  $\bf r$:
\begin{equation}
n^{(qm)}({\bf p})= {\sqrt 2 \over 3 \pi^2 \hbar^3} \left({E_F-\epsilon_{\bf p} \over m\omega^2_{ho}}\right)^{3/2}\Theta\left(E_F-\epsilon_{\bf p}\right) \;,
\label{QMtrap}
\end{equation}
where $\omega_{ho}\ll \omega_{opt}$ is the usual geometrical average of the harmonic frequencies of the potential $V_{ho}({\bf r})$ and ${\bf p}$ is restricted to the first Brillouin zone.

If the Fermi energy is much smaller than the band width $2\delta$, one can expand the dispersion law up to terms quadratic in ${\bf p}$. In this case one recovers the same Thomas-Fermi form (\ref{n0k}) holding for the momentum distribution in the absence of the optical potential, the only difference being the presence of an effective-mass term which renormalizes  the trapping frequencies. 
In the opposite regime $E_F\gg 2\delta$, but still $E_F<\hbar\omega_{opt}$, the quasi-momentum distribution becomes flat within the first Brillouin zone, giving rise to a characteristic cubic shape for the Fermi surface. In this limit 
we find 
\begin{equation}
E_F=\left({3\pi^2 \over 32}\right)^{2/3}{(\hbar\omega_{ho})^2 \over E_R}N_\sigma^{2/3} \; .
\label{EF2/3}
\end{equation}
The experimental investigation of the Fermi surface in a 3D optical lattice was carried out by K\"{o}hl {\it et al.} (2005), who observed the transition from the spherical to the cubic shape by increasing the intensity of the laser generating the optical lattice (see  Fig.~\ref{fig10.1}). For $s=12$ [panel (e) in the figure], corresponding to $\delta\sim 10$ nK, the values of the relevant  parameters were $E_R=348$ nK, $\hbar\omega_{ho}=2\pi\hbar$ 191 Hz=9.2 nK and $N_\sigma=15000$, so that the conditions $\hbar\omega_{opt} \gg E_F\gg 2\delta$, needed to reach the cubic shape for the Fermi surface, were well satisfied in this experiment. 

In the same limit $E_F\gg 2\delta$ the coarse-grained density distribution takes the constant value $n^{max}_\sigma=1/d^3$ within the ellipsoid fixed by the radii ($i=x,y,z$)
\begin{equation}
R_i= \left({3\over 4\pi}\right)^{1/3} N_\sigma^{1/3} d {\omega_{ho} \over \omega_i} \;.
\label{R1/3}
\end{equation} 
where $d$ is the periodicity of the laser field. The constant value of the density reflects the insulating nature of the system.

\subsubsection{Bloch oscillations} \label{Sec10.1.2}

Atomic gases confined by optical lattices are well suited to study Bloch oscillations (these oscillations are difficult to observe in natural crystals because of the scattering of electrons by the lattice defects). Let us consider a 1D optical lattice in the presence of a gravitational field $V_{ext}=-mgz$ oriented along the direction of the lattice. The dynamics of atoms in the lowest band can be described in the semiclassical approximation using the effective single-particle Hamiltonian
\begin{equation}
H\left( \mathbf{p,r}\right) =\frac{\mathbf{p}_{\perp }^{2}}{2m}+\epsilon_{p_z}+V_{ext}\left( \mathbf{r}\right) \;.  
\label{Heff}
\end{equation}
The Hamilton equation $dp_{z}/dt=-\partial H /\partial z $ then yields the obvious solution $p_{z}=mgt.$ In this problem it is convenient not to
restrict $p_{z}$ to the first Brillouin zone, but to allow $p_{z}$ to take also larger values. Then all the observable physical quantities must be periodic functions of $p_{z}$ with period $2K\hbar$. From the time dependence of  $p_{z}$ it follows that these quantities oscillate in time with the frequency $\omega _{B}=mgd/\hbar $. The periodicity of these Bloch oscillations is ensured by the periodicity of the optical lattice and the theory is applicable if $\omega_{B} \ll \omega_{opt}$.

Bloch oscillations have been observed in Bose gases both above (Ben Dahan {\it et al.}, 1996; Clad\'e {\it et al.}, 2006; Ferrari {\it et al.}, 2006) and below (Anderson and Kasevich, 1998; Morsch {\it et al.}, 2001) the critical temperature for Bose-Einstein condensation. A general limitation in the study of Bloch oscillations with bosons is due to instabilities and damping effects produced by the interactions. Precise measurements have been recently achieved in a dense BEC gas of $^{39}$K, near a Feshbach resonance which permits to tune the scattering length to vanishing values (Roati {\it et al.}, 2007). 

The use of polarized fermions permits to work with relatively dense gases because of the absence of $s$-wave collisions. In the experiment of Roati {\it et al.} (2004) a Fermi gas of $^{40}$K atoms was initially confined in a harmonic trap so that the external potential takes the form $V_{ext}({\bf r}) =V_{ho}({\bf r})-mgz$.  The quasi-momentum distribution is obtained by integrating the $T=0$  distribution function  in all the variables except $p_{z}$:  
\begin{equation}
n^{(qm)}( p_{z}) \propto 
\left(E_{F}-\epsilon_{p_z}\right)^{5/2} \Theta(E_F-\epsilon_{p_z}) \; .  
\label{npz}
\end{equation}
As long as $E_F$ is smaller than $2\delta$ the quasi-momentum distribution is localized in a narrow region around $p_z=0$, while the contrast deteriorates for larger values of $E_F$. 

At $t=0$ the vertical harmonic confinement is suddenly switched off and the atoms evolve in the presence of the lattice and gravitational potentials. At the initial time the quasi-momentum distribution is centered at $p_{z}=0$. It will later move  according to the law $n^{(qm)}\left(p_{z},t\right)=n^{(qm)}\left( p_{z}-mgt\right)$. When the cloud reaches the edge of the Brillouin zone $K$, it reappears on the opposite side and the quasi-momentum distribution acquires a two-peak character. At $t=2\pi /\omega_{B}$ it regains its initial shape. 

After a given evolution time the lattice potential is adiabatically switched off in order to transfer the quasi-momentum distribution into the momentum one. The cloud is then imaged after a given time of free expansion. In this experiment it was possible to observe about 100 Bloch periods. The high precision achievable in the measurement of Bloch frequencies opens new perspectives in sensitive measurements of weak forces, like the Casimir-Polder force between atoms and a solid substrate (Carusotto {\it et al.}, 2005).

\begin{figure}[b]
\begin{center}
\includegraphics[width=8.5cm]{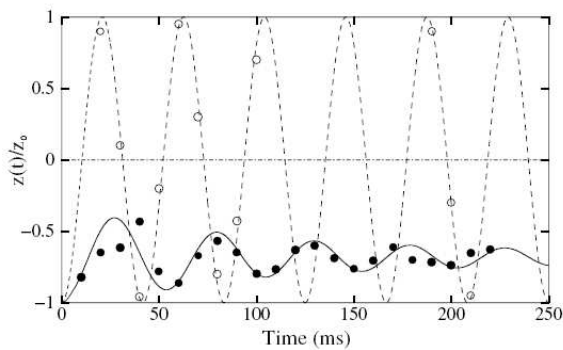}
\caption{Dipole oscillations of a Fermi gas of $^{40}$K atoms at $T=0.3\;T_F$ in the presence (solid symbols and solid line) and in
the absence (open symbols and dashed line) of a lattice with height $s=3$. The lines correspond to the theoretical predictions and the symbols to the experimental results. The horizontal dot-dashed line represents the trap minimum. From Pezz\'e {\it et al.} (2004).}
\label{fig10.3}
\end{center}
\end{figure}

\subsubsection{Center of mass oscillation} \label{Sec10.1.3}

In addition to Bloch oscillations it is also interesting to study the consequence of the lattice on the oscillations of the gas occurring in coordinate space. In this Section we focus on the dipole oscillation which can be excited by a sudden shift of the confining harmonic trap. According to Kohn's theorem the dipole oscillation in the absence of the lattice has no damping and its frequency is equal to the trap frequency. In a superfluid these oscillations exist also in the presence of the lattice thanks to the coherent tunneling of atoms through the barriers separating consecutive wells. The main consequence of the lattice is  a renormalization of the collective frequency determined by the effective mass of the superfluid. These oscillations have already been observed in Bose-Einstein condensates (Cataliotti {\it et al.}, 2001) and are expected to occur also in Fermi superfluids (Pitaevskii, Stringari and Orso, 2005).

The behavior of a non-interacting Fermi gas is  very different (Pezz\`{e} {\it et al.}, 2004). To understand the origin of the differences let us consider the simplest case of a one-dimensional Fermi gas characterized by the dispersion law $\epsilon_{p_z}=2\delta\sin^{2}\left(p_{z}d/2\hbar\right)$ and trapped by the harmonic potential $m\omega^2_zz^2/2$. Atoms with energy smaller than $2\delta$  can perform closed orbits in the $z-p_z$ phase plane.  These atoms oscillate around the center of the trap. Vice-versa, atoms with energy higher than $2\delta$ perform open orbits, being unable to fully transfer the  potential energy into the Bloch energy $\epsilon_{p_z}$. They consequently perform small oscillations in space, remaining localized on one side of the harmonic potential. As a consequence if $E_{F}>2\delta$ the cloud no longer oscillates around the new center of the trap but is trapped out of the center and performes small oscillations around an offset point, reflecting the insulating nature of the system.

In order to investigate a three-dimensional case, one can use the semiclassical collisionless kinetic equation for the distribution function with the Hamiltonian (\ref{Heff}). The results of the calculations show that, if $E_{F}>2\delta $, also in 3D the cloud is not able to oscillate around the new equilibrium position, but exhibits damped oscillations around an offset point, similarly to the 1D case. The damping is due to the fact that different atoms oscillate with different frequencies as a consequence of the non harmonic nature of the Hamiltonian. These phenomena were investigated experimentally using a Fermi gas of $^{40}$K atoms (Pezz\`{e} {\it et al.}, 2004). In Fig.~\ref{fig10.3} we show the observed time dependence of the $z$-coordinate of the cloud at $T=0.3\;T_F$, both without and in the presence of the lattice. One clearly sees the offset of the oscillations as well as their damping.

\begin{figure}[b]
\begin{center}
\includegraphics[width=8.5cm]{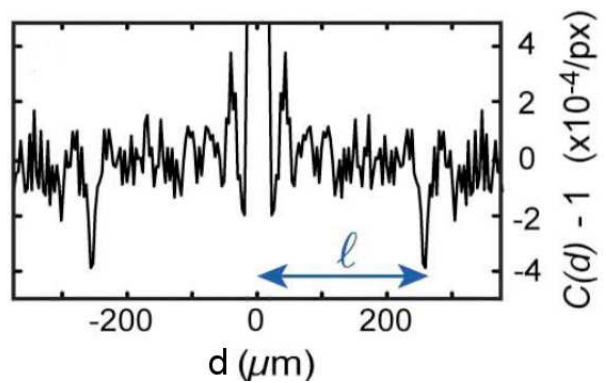}
\caption{Density correlation function $C(d)$ measured after expansion as a function of the distance $d=|z-z^\prime|$ in a Fermi gas of $^{40}$K atoms released from an optical lattice. The difference $C(d)-1$ of the correlation function with respect to its uncorrelated value shows a clear evidence of the anti-bunching effect at $d=\ell\equiv 2\hbar K t/m$. From Rom {\it et al.} (2006).}
\label{fig10.4}
\end{center}
\end{figure}

\subsubsection{Anti-bunching effect in the correlation function} \label{Sec10.1.4} 

The discussion has so far concerned one-body properties of the gas, like the momentum distribution and the density profiles. An interesting feature exhibited by spin-polarized Fermi gases is the occurrence of anti-bunching effects in the two-body correlation function [see discussion on $g_{\uparrow\uparrow}(r)$ in Sec.~\ref{Sec5.2.2}]. This suppression of the probability of finding two atoms within short distances is a direct consequence of the Pauli exclusion principle. 

In a recent experiment (Rom {\it et al.}, 2006)  the density correlation function $\left\langle n\left( z,t\right) n\left(z^{\prime},t\right)\right\rangle$ of a Fermi gas was measured after expansion from an optical lattice. The average is taken on different runs of the experiment and the distributions are integrated along the $x,y$ directions. The density correlation function, measured at large expansion times, is proportional to the momentum correlation function $\left\langle
n({p=mz/t})n({p^{\prime }=mz^{\prime }/t})\right\rangle$. Due to the periodicity  of the Bloch function $u_{q_zn}(z)$ an atom with quasi-momentum $p=\hbar q$ carries momenta with values $\hbar(q+2j K)$ for all integers $j$. As a consequence the same atom can be found at the points $z=\hbar\left(q+2jK\right) t/m$. Since the Pauli principle does not allow two fermions to occupy the same Bloch state, the correlation function $\left\langle n\left( z,t\right) n\left(z^{\prime},t\right)\right\rangle$ should vanish for relative distances which are integer multiples of $2\hbar Kt/m$. 

In the experiment of Rom {\it et al.} (2006) $^{40}$K atoms were confined at $T/T_{F}\approx 0.23$ in a 3D optical trap. In the conditions of the experiment the atoms filled the first energy band. The key experimental results are presented in Fig.~\ref{fig10.4}, where the anti-bunching effect is visible at $\left|z-z^{\prime }\right|=\ell\equiv 2\hbar Kt/m$. Notice that at short relative distances $|z-z^{\prime}|= 0$ the anti-bunching effect is masked by the positive contribution of the autocorrelation term to the density correlation function (\ref{gofr}).

\subsection{Interacting fermions in optical lattices} \label{Sec10.2}

The study of interacting Fermi gases in optical lattices is expected to become a growing field of research. The problem has been so far approached theoretically using two different perspectives. 

In a first approach one starts from the two-body Hamiltonian where interaction effects are accounted for in terms of a single parameter, the $s$-wave scattering length $a$. More microscopic details of the interatomic potential are unimportant provided that the lattice period is much larger than the effective range $|R^\ast|$ of the interaction.  The resulting many-body theories are in most cases well suited to treat the superfluid phase of the system, but have not so far extensively developed to investigate other states, like the Mott insulator or the antiferromagnetic phase. Basic applications of this approach concern the study of the dimer formation (see next Section) and the calculation of the BCS critical temperature (Orso and Shlyapnikov, 2005). 

A second approach is based on the development of more phenomenological models, like the Hubbard model, extensively employed in solid state physics. The Hubbard model is well suited to study the novel phases emerging in the presence of the optical lattice, like the Mott insulator and the antiferromagnetic phases (Werner {\it et al.}, 2005). Key questions are the identification of the parameters of the model in terms of the microscopic ingredients of the problem (the $s$-wave scatttering length and the intensity of the periodic potential) and its applicability under extreme conditions, for example at unitarity, where the scattering length is much larger than the period of the lattice. 

A separate class of problems finally concerns the physics of low dimensional systems, in particular of 1D systems, which can be experimentally produced using optical lattices and will be discussed in the next Section. 

From the experimental point of view first important results on the role of interaction have concerned the dimer formation in 3D tight lattices (St\"{o}ferle {\it et al.}, 2006) and the superfluid to Mott-insulator transition (Chin {\it et al.}, 2006).

\begin{figure}[b]
\begin{center}
\includegraphics[width=8.5cm]{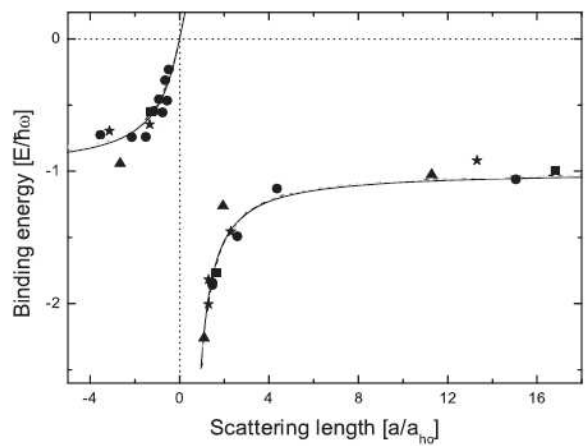}
\caption{Binding energy of molecules measured with $^{40}$K atoms in a 3D optical lattice. The data correspond to different intensities of the optical lattice: $s=6 E_R$ (triangles), $s=10 E_R$ (stars), $s=15 E_R$ (circles) and $s=22 E_R$ (squares). The solid line corresponds to Eq.~(\ref{harm}) with no free parameters. At the position of the Feshbach resonance ($a\to\pm\infty$) the binding energy takes the value $\epsilon_b=-\hbar\omega_{opt}$. From St\"oferle {\it et al.} (2006).}
\label{fig10.7}
\end{center}
\end{figure}

\subsubsection{Dimer formation in periodic potentials} \label{Sec10.2.1}

In this Section we discuss how the formation of a dimer is perturbed by the presence of a periodic potential. The problem is non trival because, differently from free space or harmonic trapping, the two-body problem can not be simply solved by separating the relative and center of mass coordinates in the Schr\"odinger equation. In particular, the center of mass motion affects the binding energy of the molecule.  

The problem of calculating the binding energy for arbitrary laser intensities of 1D lattices was solved by Orso {\it et al.} (2005) 
(the limit of tight lattices was previously considered by Fedichev, Bijlsma and Zoller, 2004). Differently from free space, where dimers are created only for positive values of the scattering length, in the presence of the periodic potential bound dimers exist also for negative values of the scattering length, starting from a critical value $a_{cr}<0$. When the laser intensity becomes very large the dimer enters a quasi-2D regime. In this limit the two interacting atoms are localized at the bottom of the same optical well where, in first approximation, the potential is harmonic  with frequency $\omega_{opt}$. Then the two-body problem can be solved analytically yielding, in particular, the value $\epsilon_b=-0.244\;\hbar\omega_{opt}=-0.488\sqrt{s}E_R$ at unitarity (Petrov, Holzmann and Shlyapnikov, 2000; Idziaszek and Calarco, 2006).  

The formation of dimers affects dramatically also the   tunneling of particles through the barriers produced by the optical lattice, particularly   the effective mass $M^\ast$ which is defined through the dispersion law $E(p_z)$ of a molecule as $1/M^\ast=[\partial^2 E(p_z)/\partial p_z^2]_{p_z=0}$. As a result the effective mass $M^*$ of the dimer is significantly larger than the value $2m^\ast$ where $m^\ast$ is the effective mass of a single atom in the presence of the same lattice potential (see Sec.~\ref{Sec10.1}). The difference is caused by the exponential dependence of the tunneling rate on the mass of a tunneling particle. Near  the threshold for the molecular formation $M^*$ approaches the non-interacting value $2m^\ast$.

Let us now discuss the case of a 3D lattice potential of the form (\ref{3D}). We restrict the analysis to a lattice of high intensity $s$. In this case the atomic pair is confined near one of the minima of the lattice, where the potential can be considered harmonic and isotropic  with  frequency $\omega_{opt}$ and the two-body problem can be solved analytically (Bush {\it et al.}, 1998). A bound state is found for any value of the scattering length, the binding energy being given by the solution of the equation 
\begin{equation}
\sqrt{2}\frac{\Gamma\left(-\epsilon_b/2\hbar\omega_{opt}\right)}{\Gamma\left(-\epsilon_b/2\hbar\omega_{opt}-1/2\right)}= \frac{a_{opt}}{a} \;,  
\label{harm}
\end{equation}
where $\Gamma$ is the Gamma function and $a_{opt}=\sqrt{\hbar /m\omega_{opt}}$. The resulting predictions are shown in Fig.~\ref{fig10.7}. For small and positive scattering lengths ($a\ll a_{opt}$) Eq.~(\ref{harm}) yields the   binding energy $\epsilon_b=-\hbar^2/ma^2$ relative to free space, while at unitarity one finds the result $\epsilon_{b}=-\hbar\omega_{opt}$.

The formation of molecules driven by the presence of the lattice was observed in the experiment of St\"{o}ferle {\it et al.} (2006), where the lattice was formed by three orthogonal standing waves and the binding energy was measured by radio-frequency spectroscopy. The radio-frequency pulse dissociates dimers and transfers atoms in a different hyperfine state which does not exhibit a Feshbach resonance. Therefore the fragments after dissociation are essentially non-interacting. The results are presented in Fig.~\ref{fig10.7} for different values of $s$ and show good agreement with theory. In particular, one can clearly see the existence of bound states for negative values of $a$ which would be impossible in the absence of the lattice.

\subsubsection{Hubbard model} \label{Sec10.2.3}

The  Hubbard model (Hubbard, 1963) provides a useful description for atoms in the lowest band of tight lattices. In the simplest version the Hamiltonian has the form: 
\begin{equation}
\hat{H}=-t\sum_{(i,j)}\left( \hat{c}_{i\uparrow}^{\dagger }\hat{c}_{j\uparrow }+\hat{c}_{i\downarrow }^{\dagger }\hat{c}_{j\downarrow}\right) +U\sum\limits_{i}\hat{n}_{i\uparrow }\hat{n}_{i\downarrow } \;,
\label{Hub}
\end{equation}
where the indices $i$ and $j$ run over the $N_{a}$ sites of the cubic lattice and correspond to first neighbor sites. The operators $\hat{c}_{i\sigma}^{\dagger }$ ($\hat{c}_{i\sigma}$) are the usual creation (annihilation) operators of particles with spin $\sigma=\uparrow, \downarrow$ on the site $i$,  while $\hat{n}_{i\sigma}=\hat{c}_{i\sigma}^{\dagger }\hat{c}_{i\sigma}$ is the corresponding number operator. The term in the Hamiltonian (\ref{Hub}) containing the parameter $t$ (hopping term) describes the tunneling of atoms between sites and plays the role of the kinetic energy operator. If $U=0$ this term gives rise to the dispersion law (\ref{tight3D}) with $\delta=2t$. The parameter $U$ describes instead the interaction between atoms and can have both positive and negative sign. Physically this term corresponds to the energy shift produced by the interaction when two atoms of opposite spin are localized in one of the lattice sites. The perturbative calculation of the shift using the pseudopotential (\ref{pseudo}) yields $U=(4\pi\hbar^2a/m)\int d{\bf r}|\psi_0|^4$ where $\psi_0$ is the ground-state wavefunction of an atom in the individual potential well. In the harmonic approximation for the local optical potential, holding for large laser intensities, one finds $U=\sqrt{8\over\pi}E_R s^{3/4}Ka$. The use of perturbation theory is justified if the shift is small compared to the optical oscillator energy $|U| \ll \hbar \omega_{opt}$,
or equivalently if $|a|\ll a_{opt}$. This condition is equivalent to the requirment that the energy $U$ is small compared to the gap between first and second band. For higher values of $U$ the applicability of the Hubbard model (\ref{Hub}) is questionable since in this case the Hamiltonian should account also for higher bands.

At zero temperature the model is characterized by two parameters: the ratio $u=U/t$ between the interaction and the hopping coefficients and the average occupancy $\rho =N/N_{a}$, which in the first band picture should be smaller or equal to $2$ due to Fermi statistics. The phase diagram predicted by the Hubbard model is very rich, including the superfluid, the Mott insulator and the antiferromagnetic phases. It is also worth noticing that in the strong coupling limit the Hubbard model is equivalent to the Heisenberg model (see, for example, Bloch, Dalibard and Zwerger, 2007). The detailed discussion of the various phases available in interacting Fermi gases trapped by a periodic potential lies outside the scope of this work and we refer to the recent reviews by Georges (2007) and Lewenstein {\it et al.} (2007).

\section{1D FERMI GAS} \label{Chap11}

The physical properties of 1D Fermi gases differ in many interesting aspects from the ones of their 3D counterparts.

In practice, to create a one-dimensional gas, atoms must be confined in a highly elongated harmonic trap, where the anisotropy parameter $\lambda=\omega_z/\omega_\perp$ is so small that the transverse motion is ``frozen'' to the zero point oscillation. At zero temperature this condition implies that the Fermi energy associated with the longitudinal motion of the atoms, $E_F=N\hbar\omega_z/2$, is much smaller than the separation between the levels in the transverse direction, $E_F\ll\hbar\omega_\perp$. This condition requires $\lambda\ll 1/N$. Such a configuration can be realized using a two-dimensional optical lattice formed by two perpendicular standing-wave laser fields. If the intensity of the beams is large enough, the tunneling between the minima of the lattice is absent and the atoms, confined in different minima, form an array of independent tubes. In the experiment by Moritz {\it et al.} (2005), the typical number of particles per tube is less than 100, while $\lambda\simeq 0.004$, thereby ensuring a reasonably safe 1D condition in each tube.

\begin{figure}[b]
\begin{center}
\includegraphics[width=8.5cm]{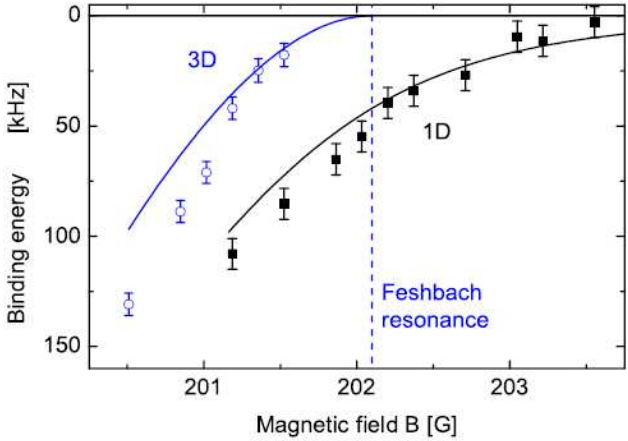}
\caption{(color online). Molecular binding energy measured with $^{40}$K in 1D [solid (black) symbols] and in 3D [open (blue) symbols] configurations. The lower (black) line corresponds to the theoretical prediction of Bergeman, Moore and Olshanii (2003) without free parameters. The upper (blue) line corresponds to the law $-\hbar^2/ma^2 +\, \text{finite-range corrections}$. The vertical dashed (blue) line represents the position of the Feshbach resonance. From Moritz {\it et al.} (2005).}
\label{fig10.10}
\end{center}
\end{figure}

\subsection{Confinement Induced Resonance} \label{Sec11.1}

At low energy the scattering process between two fermions with opposite spin colliding in a tightly confined waveguide ($\omega_z=0$) can be described by the effective 1D interaction potential 
\begin{equation}
V_{1D}(z)=g_{1D}\delta(z) \;,  
\label{delta1D}
\end{equation}
where the coupling constant $g_{1D}$ is expressed in terms of the 3D scattering length $a$ and of the transverse oscillator length  $a_\perp=\sqrt{\hbar/m\omega_\perp}$ (Olshanii, 1998)
\begin{equation}
g_{1D}=\frac{2\hbar^2a}{ma_\perp^2}\frac{1}{1-Ca/a_\perp} \;.
\label{g1}
\end{equation}
Here, $C=-\zeta(1/2)/\sqrt{2}\simeq1.0326$, with $\zeta(x)$ denoting the Riemann zeta function. The condition of validity for the effective 1D interaction (\ref{delta1D})-(\ref{g1}) is provided by
\begin{equation}
k_za_{\perp }\ll 1 \;,  
\label{cond1D}
\end{equation}
where $k_{z}$ is the longitudinal relative wavevector of the colliding atoms. The coupling constant $g_{1D}$ is resonantly enhanced for $a\to a_{cir}=a_\perp/C$, corresponding to the so called confinement induced resonance (CIR), while it remains finite at the position of the 3D resonance ($a\to\pm\infty$) where it takes the negative value $g_{1D}=-2\hbar^2/(Cma_\perp)$. Positive values of $g_{1D}$, corresponding to an effective 1D repulsive potential, are obtained only in the interval $0<a<a_{cir}$. Otherwise $g_{1D}<0$, corresponding to an effective attraction. If $|a|\ll a_\perp$, the coupling constant takes the limiting form $g_{1D}=2\hbar^2a/(ma_\perp^2)$, which coincides with the result of mean-field theory where the 3D coupling constant $g=4\pi\hbar^2a/m$ is averaged over the harmonic oscillator ground state in the transverse direction (see, {\it e.g.}, Pitaevskii and Stringari, 2003, Chap. 17).  

In the region where $g_{1D}$ is negative, two atoms can form a bound state. The wavefunction of the relative motion is obtained by solving the 1D Schr\"odinger equation with the potential (\ref{delta1D}) and is given by $\psi(z)=\sqrt{\kappa}e^{-\kappa\left|z\right|}$, where $\kappa =\sqrt{-m\epsilon_b}/\hbar$. For the binding energy $\epsilon_b$ one finds the result 
\begin{equation}
\epsilon_b=-\frac{m}{4\hbar^2}g_{1D}^{2} \;,  
\label{Eg1}
\end{equation}
yielding $\kappa =(m/2\hbar^2)\left|g_{1D}\right|$. Notice that $\epsilon_b$ is the energy of a dimer relative to the non-interacting ground-state energy $\hbar\omega_\perp$. The 1D result (\ref{Eg1}) for the binding energy is valid under the condition $\kappa a_\perp\ll 1$, or equivalently $\left|\epsilon_b\right|\ll\hbar\omega_\perp$. The general problem of calculating $\epsilon_b$ in a tightly confined waveguide has been solved by Bergeman, Moore and Olshanii (2003) using the pseudopotential (\ref{pseudo}). A molecular bound state exists for any value of the scattering length $a$. Its energy approaches the free-space result $-\hbar^2/ma^2$ for $a>0$ and $a\ll a_\perp$, and the 1D result (\ref{Eg1}) if $a<0$ and $|a|\ll a_\perp$. At the Feshbach resonance, $1/a=0$, these authors find the universal result $\epsilon_b\simeq-0.6\hbar\omega_\perp$.

These molecular bound states have been observed with $^{40}$K in the experiment by Moritz {\it et al.} (2005), where the binding energy $\epsilon_{b}$ was measured using radio-frequency spectroscopy. In Fig.~\ref{fig10.10} we show the experimental results obtained in highly elongated traps compared with the quasi-1D theoretical predictions of Bergeman, Moore and Olshanii (2003) (see also Dickerscheid and Stoof, 2005). The corresponding results for the molecular binding energy in 3D configurations are also reported in the figure, explicitly showing the existence of confinement induced molecules in the region of negative scattering lengths. 

The three- and four-body problems concerning the atom-dimer and the dimer-dimer scattering in quasi-1D configurations has been solved by Mora {\it et al.} (2004, 2005) using techniques similar to the 3D calculation by Petrov, Salomon and Shlyapnikov (2004) discussed in Sec.~\ref{Sec3.3}. In particular, one finds that the scattering process between 1D dimers with energy (\ref{Eg1}) can be described by the contact potential of the form (\ref{delta1D}) with the same atom-atom coupling constant $g_{1D}$. Since $g_{1D}<0$, the interaction between these dimers is attractive. Notice, however, that the fermionic nature of the atoms prohibits the formation of bound states with more than two particles.

\subsection{Exact theory of the 1D Fermi gas} \label{Sec11.2}

We consider a two-component Fermi gas with equal populations of the spin states ($N_\uparrow=N_\downarrow=N/2$) confined in a tight waveguide of length $L$. At zero temperature and in the absence of interactions, all single-particle states within the ``Fermi line'' $-k_F<k<k_F$ are occupied. The Fermi wavevector 
\begin{equation}
k_F=\frac{\pi}{2}n_{1D} 
\label{pF1D}
\end{equation}
is fixed by the linear density $n_{1D}=N/L$ and the corresponding Fermi energy is given by $E_F=(\pi\hbar n_{1D})^2/8m$. The condition (\ref{cond1D}), allowing for the use of the effective 1D interaction (\ref{delta1D})-(\ref{g1}), implies the requirement $n_{1D}a_{\perp }\ll 1$. In this case the many-body problem is completely determined by the Hamiltonian 
\begin{equation}
H_{1D}=-\frac{\hbar^2}{2m}\sum\limits_{i=1}^{N}\frac{d^{2}}{dz_{i}^{2}}+g_{1D}\sum\limits_{i=1}^{N_\uparrow}
\sum\limits_{i^\prime=1}^{N_\downarrow}\delta\left(z_{i}-z_{i^\prime}\right) \;, 
\label{H1D}
\end{equation}
which contains only one dimensionless parameter 
\begin{equation}
\gamma=\frac{mg_{1D}}{\hbar^2n_{1D}} \;.  
\label{gamma}
\end{equation}
Correspondingly the ground-state energy per atom can be written in the form 
\begin{equation}
\frac{E}{N}=\frac{\hbar^2n_{1D}^2}{2m} e\left( \gamma \right) \;,
\end{equation}
in terms of the dimensionless function $e(\gamma)$, and analogously for the chemical potential $\mu=d(n_{1D}E/N)/dn_{1D}$. From Eq.~(\ref{gamma}) one notices that the weak-coupling regime ($|\gamma|\ll 1$) corresponds to high densities $n_{1D}$, while the strong-coupling regime ($|\gamma|\gg 1$) is achieved at low densities. This is a peculiar feature of 1D configurations, resulting from the different density dependence of the ratio of kinetic to interaction energy in 1D ($\propto n_{1D}$) compared to 3D ($\propto n^{-1/3}$). 

The ground-state energy of the Hamiltonian (\ref{H1D}) has been calculated exactly using Bethe's ansatz both for repulsive, $g_{1D}>0$ (Yang, 1967), and attractive, $g_{1D}<0$ (Gaudin, 1967 and 1983), interactions. Some limiting cases of the equation of state at $T=0$ are worth discussing in detail. In the weak-coupling limit, $|\gamma|\ll 1$, one finds the perturbative expansion 
\begin{equation}
\mu=E_F(1+\frac{4\gamma}{\pi^2} +...) \;,
\label{muweak}
\end{equation}
where the first correction to the Fermi energy $E_F$ carries the same sign of $\gamma$. The above expansion is the 1D analogue of Eq.~(\ref{enexpansion}) holding in 3D. In the limit of strong repulsion, $\gamma\gg 1$, one instead finds (Recati {\it et al.}, 2003b)
\begin{equation}
\mu =4E_F\left( 1-\frac{16\ln 2}{3\gamma}+...\right) \;.
\label{mustrongrep}
\end{equation}
The lowest order term in the above expansion coincides with the Fermi energy of a single component non-interacting gas with a twice larger density ($N_\sigma=N$), consistently with the expectation that the strong atom-atom repulsion between atoms with different spins plays the role in 1D of an effective Pauli principle. 

The regime of strong attraction, $\left|\gamma\right| \gg 1$ with $\gamma<0$, is particularly interesting. In this case one finds the following expansion for the chemical potential (Astrakharchik {\it et al.}, 2004b)
\begin{equation}
\mu =\frac{\epsilon_b}{2}+\frac{E_F}{4}\left( 1-\frac{4}{3\gamma}+...\right) \;,
\label{mustrongattr}
\end{equation}
where $\epsilon_b$ is the binding energy (\ref{Eg1}) of a dimer. According to the discussion at the end of Sec.~\ref{Sec11.1}, in this regime the system is a gas of strongly attractive bosonic dimers. A crucial point is that the formation of bound states of these composite bosons is inhibited by the fermionic nature of the constituent atoms. In Eq.~(\ref{mustrongattr}) the leading term $E_F/4$, beyond the molecular binding energy $\epsilon_b$, coincides with half the chemical potential of a gas of impenetrable bosons (Tonks-Girardeau gas) with density $n_{1D}/2$ and mass $2m$. This peculiar behavior is an example of the exact mapping between bosons and fermions exhibited by 1D configurations (Girardeau, 1960). The consequences of the ``Fermi-Bose duality'' in 1D, including the regimes of intermediate coupling and for both $s$-wave and $p$-wave scattering, have been investigated in a series of papers (Cheon and Shigehara, 1999; Granger and Blume, 2004; Girardeau, Nguyen and Olshanii, 2004; Girardeau and Olshanii, 2004; Mora {\it et al.}, 2005).   

The strongly interacting regime, $|\gamma|\gg 1$, can be achieved by tuning the effective coupling constant $g_{1D}$. As in the experiment by Moritz {\it et al.} (2005) on 1D molecules, one makes use of a Feshbach resonance to tune the scattering lenth in the region around the value $a_{cir}$ corresponding to the confinement induced resonance where $\gamma=\pm\infty$ [see Eq.~(\ref{g1})]. This resonance connects a BCS-like weakly attractive regime ($\gamma$ negative and small), corresponding to weakly bound pairs with energy (\ref{Eg1}) and size $\kappa^{-1}>>n_{1D}^{-1}$, to a BEC-like regime ($\gamma$ positive and small) of tightly bound bosonic molecules with energies of order $\hbar\omega_\perp$ and size much smaller than the average distance between dimers. These dimers are expected to behave as a 1D gas of bosons interacting with a repulsive contact potential (Mora {\it et al.}, 2005). However, these tightly bound dimers can not be described by the Hamiltonian (\ref{H1D}) which, if $g_{1D}>0$, only describes the repulsive atomic branch. The occurrence of such a BCS-BEC crossover in 1D was first suggested by Tokatly (2004) and by Fuchs, Recati and Zwerger (2004). 

The properties at $T=0$ of the 1D Fermi gas with short-range interactions are considerably different from the ones of usual 3D Fermi liquids. Actually, this gas is an example of a Luttinger liquid (Luttinger, 1963). The low-energy properties of this liquid are universal and do not depend on the details of the interaction, on the specific Hamiltonian (lattice or continuum models) nor on the statistics of the atoms (Haldane, 1981). The only requirement is the existence of long-wavelength gapless excitations with linear dispersion. The Luttinger effective Hamiltonian is expressed in terms of the compressibility and of the velocity of propagation of these gapless excitations. For a review on the properties of Luttinger liquids see, {\it e.g.}, Voit (1995) and Giamarchi (2004). 

Hydrodynamic sound waves (phonons) are predicted by the Hamiltonian (\ref{H1D}). They propagate with the velocity $c$ determined by the compressibility through the general relation $mc^{2}=n_{1D}d\mu /dn_{1D}$. In this case, the knowledge of the equation of state allows for an exact determination of the effective Luttinger parameters (Recati {\it et al.}, 2003a). If $\gamma>0$, $c$ is larger than the Fermi velocity $v_F=\pi\hbar n_{1D}/2m$. For example, in the case of strong repulsion ($\gamma\gg 1$) the speed of sound takes the limiting value $c=2v_F(1-4\ln 2/\gamma)$. For small values of $\gamma$ the sound velocity tends to $v_F$, while for $\gamma<0$ it becomes smaller than the Fermi velocity. The inverse compressibility $mc^2$, however, remains positive indicating the stability of the gas even in the strongly attractive regime where one finds $c=v_F/2$. This is in sharp contrast with the behavior of a 1D Bose gas with attractive contact interactions, where the ground state is a soliton-like many-body bound state (McGuire, 1964).   

The presence of phonons in a Luttinger liquid affects dramatically the long-range behavior of the correlation functions, fixed by the dimensionless parameter $\eta=2\hbar k_F/mc$ (Luther and Peschel, 1974; Haldane, 1981). For example, the one-body density matrix behaves, for $\left|z-z^\prime\right|\gg 1/n_{1D}$, as
\begin{equation}
\left\langle\Psi_{\sigma }^{\dagger }\left(z\right)\Psi _{\sigma }\left(z^\prime\right)\right\rangle \propto \frac{n_{1D}\sin\left(\pi n_{1D}\left|z-z^\prime\right|\right)} {\left(n_{1D}\left|z-z^\prime\right|\right)^{\frac{1}{\eta}+\frac{\eta}{4}}}  \;.  
\label{decay}
\end{equation}
For a non-interacting gas $\eta=2$ and the one-body density matrix decays as $\sin\left(\pi n_{1D}|z-z^\prime|\right)/|z-z^\prime|$. This behavior reflects the presence of the jump from 1 to 0 in the momentum distribution at the Fermi surface $k=\pm k_F$. In the presence of interaction the correlation function (\ref{decay}) decreases faster. This implies that the jump at $k_F$ disappears and the momentum distribution close to the Fermi surface behaves as $n_k-n_{k_F}\propto sign(k_F-k)|k_F-k|^\beta$, with $\beta=\frac{1}{\eta}+\frac{\eta}{4}-1>0$.

The Hamiltonian (\ref{H1D}) supports also spin waves together with sound waves. For $\gamma>0$ these spin excitations also have a linear dispersion at small wavevectors. Their velocity of propagation $c_s$ tends to $v_{F}$ for small values of $\gamma$, but for finite interaction strengths it is different from the speed of sound $c$. In the strongly repulsive regime one finds the small velocity $c_s=v_F\pi^2/\gamma$, to be compared with the corresponding sound velocity $c=2v_F$ discussed above. This ``spin-charge separation'' is a peculiar feature of Luttinger liquids. The possibility of observing this phenomenon in ultracold gases has been investigated by Recati {\it et al.} (2003a). In the case of attractive interactions, $\gamma <0$, the spin-wave spectrum exhibits a gap $\Delta_{\text gap}$ (Luther and Emery, 1974). This ``spin-gap'' is defined according to Eq.~(\ref{Deltadef}) and is therefore analogous to the pairing gap in 3D Fermi superfluids. In the weak-coupling limit $|\gamma|\ll 1$, the gap is exponentially small being proportional to $\sqrt{|\gamma|} \exp\left(-\pi^2/2|\gamma|\right)$ (Bychkov, Gorkov and Dzyaloshinskii, 1966; Krivnov and Ovchinnikov, 1975).

So far we have considered uniform systems. In experiments the gas is confined in the longitudinal $z$-direction by a harmonic potential. If the average distance between particles is much smaller than the longitudinal oscillator length, $1/n_{1D}\ll a_z=\sqrt{\hbar/m\omega_z}$ (requiring small enough values of the trapping frequency $\omega_z$), one can use the local density approximation to calculate the properties of the trapped system (Astrakharchik {\it et al.}, 2004b). The density profile and the Thomas-Fermi radius can be obtained as a function of the interaction strength from the solution of Eq.~(\ref{LDA}). One can
also calculate the frequency of the lowest compression mode using the hydrodynamic theory of superfluids of Sec.~\ref{Sec7.3}.
Since $\mu\propto n_{1D}^2$ both in the weak- and strong-coupling limit, independent of the sign of interactions [see Eqs.~(\ref{mustrongrep}) and (\ref{mustrongattr})], the frequency of the mode tends to $2\omega_z$ in these limits. The study of the breathing-mode frequency for intermediate couplings has been carried out by Astrakharchik {\it et al.} (2004b).  

Another interesting application of the local density approximation to trapped 1D configurations is provided by the study of spin polarized systems (Orso, 2007). The ground-state energy of the Hamiltonian (\ref{H1D}) can be calculated exactly also with unequal spin populations $N_\uparrow \neq N_\downarrow$ (Yang, 1967; Gaudin, 1967). In the case of attractive interactions, $g_{1D}<0$, the $T=0$ phase diagram always consists of a single phase corresponding to: i) a fully paired state with a gap in the spin excitation spectrum if $N_\uparrow=N_\downarrow$, ii) a non-interacting fully polarized state if $N_\downarrow=0$ and iii) a gapless partially polarized state if $N_\uparrow>N_\downarrow$ (Oelkers {\it et al.}, 2006; Guan {\it et al.}, 2007). This latter state is expected to be a superfluid of the FFLO type (Yang, 2001). In harmonic traps spin unbalanced configurations result in a two-shell structure: a partially polarized phase in the central region of the trap and either a fully paired or a fully polarized phase in the external region depending on the value of the polarization (Orso, 2007). This structure is in sharp contrast with the behavior in 3D configurations, where the unpolarized superfluid phase occupies the center of the trap and is surrounded by two shells of partially polarized and fully polarized normal phases (see Sec.~\ref{Sec9.3}). One can understand this behavior by noticing that the larger densities occurring in the center of the trap correspond in 1D to a weak-coupling regime and, consequently, pairing effects are smaller in the central than in the external region of the trap. The opposite situation takes place instead in 3D configurations.

\section{CONCLUSIONS AND PERSPECTIVES} \label{Chap12}

In this review we have discussed some relevant features exhibited by atomic Fermi gases from a theoretical perspective. The discussion has pointed out a general good agreement between theory and experiment, revealing that the basic physics underlying these quantum systems is now reasonably well understood. The most important message emerging from these studies is that, despite their diluteness, the role of interactions in these quantum gases is highly non trivial, revealing the different facets of Fermi superfluidity in conditions that are now accessible and controllable experimentally. The possibility of tuning the value and even the sign of the scattering length is actually the key novelty of these systems with respect to other Fermi superfluids available in condensed matter physics. The long sought BCS-BEC crossover, bringing the system into a high $T_c$ superfluid regime where the critical temperature is of the order of the Fermi temperature, can be now systematically investigated and many theoretical approaches are available to explore the different physical properties. The situation is particularly well understood at zero temperature where  important properties of these Fermi gases in harmonic traps, like the equilibrium density profiles, the values of the release energy and of the collective frequencies, can be calculated in an accurate way using the equation of state of uniform matter available through Quantum Monte Carlo simulations and employing the local density approximation. Other important features that can be now considered reasonably well understood theoretically and confirmed experimentally are the momentum distribution along the crossover, the collisional processes between pairs of fermions on the BEC side of the resonance and the basic properties of Fermi gases in optical lattices such as Bloch oscillations, the structure of the Fermi surface and the binding of molecules. In general the most interesting regime emerging from both the theoretical and experimental investigations is the unitary limit of infinite scattering length, where the dilute gas becomes strongly correlated in conditions of remarkable stability. In this regime there are no length scales related to interactions entering the problem, which consequently exhibits a universal behavior of high interdisciplinary interest.  

Many important issues remain nevertheless to be addressed or to be explored in a deeper and systematic way. A brief list is presented below:

- {\bf Thermodynamics}. More theoretical work needs to be done to determine the transition temperature along the crossover and the thermodynamic functions below and above $T_c$. Very little is known about the temperature dependence of the superfluid density (Taylor {\it et al.}, 2006; Akkineni, Ceperley and Trivedi, 2006; Fukushima {\it et al.}, 2007) and its role on physically observable quantities like, for example, the propagation of second sound (Taylor and Griffin, 2005; Heiselberg, 2006; He {\it et al.}, 2007). Another open question remains the identification of a good thermometry in these ultracold systems, where the experimental value of the temperature is often subject to large uncertainties.  

- {\bf Collective modes and expansion}. The transition from the hydrodynamic to the collisionless regime on the BCS side of the resonance and the consequences of the superfluid transition on the frequencies and the damping of the collective oscillations, as well as on the behavior of the aspect ratio during the expansion, still requires a clear understanding. This problem raises the question of the inclusion of mesoscopic effects in the theoretical description which become important when the pairing gap is of the order of the harmonic oscillator energy. Also the question of the transition at finite temperature between the hydrodynamic and the collisionless regime in the normal phase near resonance requires more theoretical and experimental work. Another important question is the temperature dependence of the viscosity of the gas, which is in principle measurable through the damping of the collective modes. The problem is particularly interesting at unitarity where a universal behavior is expected to occur (Son, 2007). 

- {\bf Diabatic transformation of the scattering length}. The experimental information on the condensation of pairs (Regal, Greiner and Jin, 2004b; Zwierlein {\it et al.}, 2004) and on the structure of vortices (Zwierlein {\it et al.}, 2005b) along the crossover is so far based on the measurement of the density profiles following the fast ramping of the scattering length before expansion. The theoretical understanding of the corresponding process is only partial and requires more systematic investigations through the implementation of a time-dependent description of the many-body problem. 

- {\bf Coherence in Fermi superfluids}. Although the experimental measurement of quantized vortices has proven the superfluid nature of these ultracold gases, coherence effects have not yet been explored in an exhaustive way. For example, the implications of the order parameter on the modulation of the diagonal one-body density in interference experiments remain to be studied. Other relevant topics are the interference effects in the two-body correlation function (Carusotto and Castin, 2005) and the Josephson currents (Spuntarelli, Pieri and Strinati, 2007).

- {\bf Rotational properties}. The physics of interacting Fermi gases under rotation in harmonic traps is a relatively unexplored subject of research, both in the absence and in the presence of quantized vortices. Questions like the difference between the  superfluid and the collisional hydrodynamic behavior in rotating configurations, the Tkachenko modes of the vortex lattice (Watanabe, Cozzini and Stringari, 2007) and the nature of the phase diagram of the rotating gas at high angular velocity are topics requiring both theoretical and experimental investigation.

- {\bf Spin imbalanced Fermi gases}. This subject of research has attracted a significant amount of work in the last few years. Many important questions still remain open, like the nature of the superfluid phases caused by the polarization at zero and at finite temperature in the different regimes along the crossover. Recent experiments have also raised the  question of the role of surface tension effects at the interface between the normal and the superfluid component (De Silva and Mueller, 2006b; Partridge {\it et al.}, 2006b) and of correlation effects in the spin polarized normal phase (Schunck {\it et al.}, 2007)

- {\bf Fermi-Fermi and Fermi-Bose mixtures}. A growing activity, both on the experimental and on the theoretical side, is expected to characterize future studies of mixtures of different atomic species. The quantum phases of Fermi-Bose mixtures in optical lattices (Albus, Illuminati and Eisert, 2003; Lewenstein {\it et al.}, 2004; G\"unter {\it et al.}, 2006), the formation of dipolar gases (Ospelkaus {\it et al.}, 2006c; Modugno, 2007) and the superfluid behavior of Fermi-Fermi mixtures of different atomic masses are important topics for future research.

- {\bf RF transitions}. RF transitions provide valuable information on the gap parameter and on pairing effects in interacting Fermi gases (Chin {\it et al.}, 2004; Shin {\it et al.}, 2007). The proper inclusion of final-state interactions in the calculation of the spectral response is a crucial ingredient to make the analysis of the experimental findings conclusive on a quantitive basis.  

- {\bf Quantum impurities}. The investigation of the motion of impurities added to a trapped quantum gas can open interesting possibilities for the determination of the viscosity coefficients and for the study of the Landau criterion of superfluidity.

- {\bf Phase transitions in optical lattices}. The recent experiments on the superfluid to Mott-insulator phase transition (Chin {\it et al.}, 2006) have already stimulated first theoretical work involving the use of a multi-band Hubbard model (Zhai and Ho, 2007). An important issue is the behavior of the transition at unitarity where the scattering length is much larger than the lattice spacing of the optical potential. Further important topics concern the study of the various magnetic phases, which can be implemented by playing with the geometry and the dimensionality of the lattice, as well as the role of disorder (Lewenstein {\it et al.}, 2007).

- {\bf $p$-wave superfluidity}. The recent experiment by Gaebler {\it et al.} (2007) on the production and detection of molecules in a single-component $^{40}$K gas using a $p$-wave Feshbach resonance and the measurement of the binding energy and lifetime of these $p$-wave molecules opens new exciting possibilities of realizing $p$-wave superfluids with ultracold gases. First theoretical investigations of the many-body properties of these systems as a function of the interaction strength predict a much richer phase diagram compared to $s$-wave superfluids, including quantum and topological phase transitions (Gurarie, Radzihovsky and Andreev, 2005; Cheng and Yip, 2005; Iskin and S\'a de Melo, 2006a; Gurarie and Radzihovsky, 2007).

- {\bf Low-dimensional configurations}. Two- and one-dimensional configurations of ultracold gases can be easily produced in laboratories. The experiment by Moritz {\it et al.} (2005) on 1D molecules provides the first example of a low-dimensional arrangement combined with the use of a Feshbach resonance. Many interesting features are expected to take place when the scattering length becomes larger than the characteristic length of the confinement. The properties of the BCS-BEC crossover in 2D and in 1D as well as the physics of Luttinger liquids in 1D can be addressed by further investigations.

\section*{ACKNOWLEDGEMENTS}

This review paper has been the result of many fruitful collaborations  with the members of the CNR-INFM Center on Bose-Einstein Condensation in Trento. In particular we like to thank Grigori Astrakharchik, Iacopo Carusotto, Marco Cozzini, Franco Dalfovo,  Carlos Lobo, Chiara Menotti, Giuliano Orso, Paolo Pedri, Sebastiano Pilati and Alessio Recati. Stimulating collaborations and discussions with D\"orte Blume, Jordi Boronat, Aurel Bulgac, Joe Carlson, Joaquin Casulleras, Frederic Chevy, Roland Combescot, Antoine Georges, Rudy Grimm, Jason Ho, Murray Holland, Randy Hulet, Massimo Inguscio, Debbie Jin, Wolfgang Ketterle, Kathy Levin, Giovanni Modugno, Pierbiagio Pieri, Nikolai Prokof'ev, Leo Radzihovsky, Cindy Regal, Christophe Salomon, Henk Stoof, Giancarlo Strinati, John Thomas, Sungkit Yip and Martin Zwierlein are also acknwledged.

\end{document}